\documentclass[longauth]{aa}  

\usepackage{newtxtext,newtxmath}
\usepackage{orcidlink}

\usepackage{multirow}
\usepackage[english]{babel}
\usepackage[T1]{fontenc}
\usepackage[dvipsnames]{xcolor}
\usepackage{caption}

\DeclareRobustCommand{\VAN}[3]{#2}
\let\VANthebibliography\thebibliography
\def\thebibliography{\DeclareRobustCommand{\VAN}[3]{##3}\VANthebibliography}

\usepackage{graphicx}
\usepackage{amsmath}
\usepackage{booktabs}
\usepackage{longtable}
\usepackage{xcolor}
\usepackage{adjustbox}
\usepackage{fix-cm}

\definecolor{blazeorange}{rgb}{1.0, 0.4, 0.0}
\definecolor{seagreen}{rgb}{0.18, 0.55, 0.34}
\definecolor{rufous}{rgb}{0.66, 0.11, 0.03}
\definecolor{royalfuchsia}{rgb}{0.79, 0.17, 0.57}
\definecolor{scarlet}{rgb}{1.0, 0.13, 0.0}
\definecolor{royalpurple}{rgb}{0.47, 0.32, 0.66}
\definecolor{darkblue}{rgb}{0, 0, 0.66}
\definecolor{violet}{rgb}{0.5,0.,0.5}

\begin{document}
\title{Multi-epoch afterglow rebrightenings in GRB~250129A: Evidence for successive shock interactions}

\author{D.~Akl\,\orcidlink{0009-0006-4358-9929}\inst{1,2,3},
S.~Antier\,\orcidlink{0000-0002-7686-3334}\inst{1,4},
H.~Koehn\,\orcidlink{0009-0001-5350-7468}\inst{5}, 
P.T.H.~Pang,\orcidlink{0000-0001-7041-3239}\inst{6,7}$\star$,
J.J.~Geng\,\orcidlink{0000-0001-9648-7295}\inst{8}$\star$,
R.~Gill\,\orcidlink{0000-0003-0516-2968}\inst{9}$\star$,
E.~Abdikamalov\,\orcidlink{0000-0001-5481-7727}\inst{10,11}, 
C.~Adami\,\orcidlink{0000-0002-8904-3925}\inst{12},
V.~Aivazyan\,\inst{13},
L.~Almeida\,\orcidlink{0000-0001-8179-1147}\inst{14,15}
% ...
          }
\institute{Universit\'e de la Côte d'Azur, Nice, France 
        \and 
             New York University Abu Dhabi, PO Box 129188, Saadiyat Island, Abu Dhabi, UAE
        \and 
             Center for Astrophysics and Space Science (CASS), New York University Abu Dhabi, Saadiyat Island, PO Box 129188, Abu Dhabi, UAE
        \and 
             IJCLab, Univ Paris-Saclay, CNRS/IN2P3, Orsay, France
        \and 
             Institute of Physics and Astronomy, Theoretical Astrophysics, University Potsdam, Haus 28, Karl-Liebknecht-Str. 24/25, 14476
        \and 
             Nikhef, Science Park 105, 1098 XG Amsterdam, The Netherlands
        \and 
             Institute for Gravitational and Subatomic Physics (GRASP), Utrecht University, Princetonplein 1, 3584 CC Utrecht, The Netherlands
        \and 
             Purple Mountain Observatory, Chinese Academy of Sciences, Nanjing 210023, China
        \and 
             Instituto de Radioastronomía y Astrof\'isica, Universidad Nacional Aut\'onoma de M\'exico, Antigua Carretera a P\'atzcuaro \# 8701,\\ Ex-Hda. San Jos\'e de la, Huerta, Morelia, Michoac\'an, M\'exico C.P. 58089, Mexico
        \and 
             Physics Department, Nazarbayev University, 53 Kabanbay Batyr Ave., Astana 010000, Kazakhstan
        \and 
             Energetic Cosmos Laboratory, Nazarbayev University, 53 Kabanbay Batyr Ave., Astana 010000, Kazakhstan
        \and 
             Université Paris-Saclay, Université Paris Cité, CEA, CNRS, AIM, 91191, Gif-sur-Yvette, France
        \and 
             E.Kharadze Georgian National Astrophysical Observatory, Mt. Kanobili, Abastumani, 0301 Adigeni, Georgia
        \and 
             SOAR Telescope/NSF's NOIRLab, Avda Juan Cisternas 1500, 1700000, La Serena, Chile
        \and 
             Laboratório Nacional de Astrofísica - LNA, Rua Estados Unidos, 154 Itajubá - MG CEP 37504-364, Brazil
% ...
}
\date{Received ...; accepted ...}

\abstract
   {Most long gamma-ray bursts (GRBs) exhibit afterglows broadly consistent with external forward-shock emission, typically described by smooth broken power-law decays in the multiband light curve. However, a minority of well-sampled GRBs deviate from this behavior, including the GRB we are investigating in this article, GRB~250129A. This object shows multiple late-time rebrightenings at X-ray and optical wavelengths.
   Rebrightenings are often attributed to energy injection from prolonged central engine activity, refreshed shocks from delayed shell collisions, density jumps in the ambient medium, or angular jet structure and viewing-angle effects. 
}
   {After a comprehensive analysis of the prompt emission of the GRB observed in gamma-rays and the near-infrared, we aim to investigate the physical origin of the multiple X-ray and optical flaring episodes observed in GRB~250129A.
   }
   {We conduct comprehensive temporal and spectral multiband analyses of GRB~250129A. The physical processes at play in the afterglow light curves were investigated using several methods, ranging from empirical fitting to Bayesian inference. The high-quality monitoring of the flare episodes, together with the connection between the prompt emission and the afterglow, enables us to test the consistency of the fireball model and to constrain, reject, or confirm alternative scenarios.}
  {GRB 250129A is an interesting GRB with multiwavelength prompt and afterglow emission. Conducting the prompt and time-resolved analyses, we obtained an isotropic-equivalent energy of $E_{\mathrm{iso,\gamma}} = (1.35 \pm 0.12) \times 10^{53}$~erg. By modeling the afterglow using an agnostic Bayesian framework (\textsc{nmma}), we rule out both a single external-shock evolution and a one-time energy-injection scenario. By further performing numerical calculations, we demonstrate that the rebrightening episodes are consistent with refreshed shocks arising from delayed collisions between relativistic shells, in agreement with evolving outflow dynamics.
} 
  {Based on the consistency between our analyses of the prompt and afterglow GRB 250129A data, together with prior general knowledge on microphysical properties, we find that two statistically significant rebrightening episodes occurred within 1.1 days post trigger and can be explained by a sequence of refreshed shocks. We stress that the availability of temporally and spectrally rich GRB datasets, such as the one presented in this work, provides a powerful means to test current modeling frameworks.}

\keywords{gamma-ray burst: individual: GRB 250129A --
                transients: gamma-ray bursts}

\titlerunning{GRB 250129A} 
\authorrunning{D.~Akl et al.}
\maketitle

\section{Introduction}

\begingroup
\renewcommand{\thefootnote}{}
\footnotetext{$\star$:\footnotesize{thopang@nikhef.nl,~jjgeng@pmo.ac.cn,~r.gill@irya.unam.mx}}
\endgroup

\label{intro}
Gamma-ray bursts (GRBs)\ are among the most luminous and violent electromagnetic phenomena observed in the universe~\citep[e.g.,][]{Piran04physics, Kumar2015}. 
Based on their $T_{90}$ duration, defined as the time interval over which the central 90\% of the gamma-ray fluence is accumulated, they are typically classified into two categories, short ($T_{90}<2$\,s) or long ($T_{90}>2$\,s) GRBs \citep[e.g.,][]{1993ApJ...413L.101K, 2023ApJ...945...67S}. A large fraction of long GRBs is suggested to originate from the collapse of massive stars, as confirmed through associations with broad-line Type Ic supernovae \citep[e.g.,][]{Woosley1993,MacFadyen1999,Woosley_2012,10.1093/mnras/stae503,Srinivasaragavan_2024}, although a subset may originate from alternative progenitors \citep{2022Natur.612..223R,Luo_2023,2025arXiv250922792N}. In contrast, short GRBs are mostly caused by the merging of compact objects \citep[e.g.,][]{Paczynski1986,Eichler1989,Paczynski1991,1992ApJ...395L..83N}, as evidenced by their joint detections of associated kilonovae and gravitational waves \citep{2017ApJ...850L..40A,2017AdAst2017E...5C}. 

One of the most favored explanations for GRBs is the \textit{fireball model} scenario \citep{1993ApJ...405..278M, 1994ApJ...433L..85W, Piran99, Kumar2015}, in which the prompt gamma-ray emission arises from energy dissipation within a narrow relativistic jet due to internal shocks produced by collisions between discrete shells with different velocities. The jet is decelerated by its interaction with the external medium, driving an external forward shock that produces the long-lasting, broadband (radio to X-rays) synchrotron afterglow emission. Following deceleration, the afterglow emission typically declines over time as a power law in the X-ray and optical bands when they originate from the same power-law segment of the synchrotron spectrum. Although fireball dynamics are often approximated as spherical at early times, the outflow is in fact collimated into a jet of half-opening angle $\theta_{\rm j}$. As the blast wave decelerates, the relativistic beaming angle (with size $1/\Gamma$, where $\Gamma$ is the Lorentz factor) increases and eventually exceeds $\theta_{\rm j}$, causing the jet edge to become visible and allowing lateral expansion to set in. This leads to the characteristic steepening of the afterglow light curve known as the jet break \citep{1999ApJ...525..737R,1999ApJ...519L..17S}. 

The light curve of GRB afterglows is expected to follow a smooth temporal evolution; however, intense, rapid rebrightenings (or flares), gradual ones lasting over minutes to days, and other irregularities in the multiband afterglow light curve are often revealed in a fraction of high-cadence GRBs with well-sampled multiband observations \citep[e.g.,][]{Nousek2006, Evans2009, Liang13Comprehensive, Kumar2015, Postigo2018, Busmann25curious}. Between 20\% and 30\% of long GRBs exhibit optical rebrightenings in their afterglow~\citep[e.g.,][]{Li2012,Becerra2023}. 

One possible explanation includes energy injection into the blast wave due to the expulsion of multiple shells by the central engine that later collide and produce a brightening of the afterglow~\citep{Panaitescu:1998km, Rees:1997nx, Kumar-Piran-00, Sari2000,Zhang-Meszaros-02, Uhm:2012yg, Laskar15, Fraija:2022btg, 10.1093/mnras/stad2594}. The GRB central engine is expected to be highly variable, and the expulsion of late shells may be associated with fallback of material onto the central engine~\citep{2001ApJ...550..410M}. Late-time energy injection has been used to explain observed plateaus and rebrightenings in a substantial number of GRBs~\citep{Nousek2006, 2007ApJ...671..637Y, 2010MNRAS.402...46M, Laskar15}.

Alternatively, energy can be added to the blast wave gradually and continuously. This can occur when the ejecta have a radial velocity gradient~\citep{Granot:2005ye, Zhang2006, vanEerten:2018amz, Ryan2020}, with progressively slower-moving inner shells trailing behind the faster moving outer shell, or if the central engine is a magnetar that injects energy continuously into the blast wave via a magnetohydrodynamic (MHD) pulsar-type wind \citep{Dai98,Zhang-Meszaros-01,Zhang-Meszaros-02,Geng16}. In both scenarios, the afterglow light curve shows an achromatic bump or rebrightening as the energy is injected, with the post-injection light curve following the same temporal decay as that of the pre-injection one in the absence of spectral-break passage. Multiple rebrightening episodes are generally not expected in this scenario, since the energy injection occurs smoothly and continuously.

These multiple episodes can instead be explained by having an asymmetric structured jet, where different azimuthal jet components, with distinct energies and Lorentz factors, sequentially dominate the afterglow emission, producing multiple peaks \citep{Li_2025}. 
 
Another possible explanation is a nonuniform circumburst environment in which the blast wave encounters density enhancements that could lead to rebrightenings~\citep{WangLoeb2000, Dai:2001fn, Uhm:2014qta}. These density structures may be organized as spherical shells formed by unstable wind episodes of the progenitor or the clumpy remnant of a previous supernova explosion, a scenario consistent with constraints from multiwavelength imaging and spectroscopy ~\citep{WangLoeb2000, Lazzati2002, Huang_2006}. However, analytic calculations \citep{Nakar07} and numerical simulations \citep{vanEerten09} show that even large density enhancements will only produce a mild rebrightening in the light curve. 

In this paper, we present multiwaveband observations of GRB 250129A, which was detected by the Burst Alert Telescope (BAT) on the \emph{Neil Gehrels Swift Observatory (Swift)} at $T_0=$ 2025 January 29 04:45:09 UTC~\citep{2025GCN.39066....1B} (trigger number \href{https://gcn.gsfc.nasa.gov/other/1285812.swift}{1285812}, signal-to-noise ratio (SNR) in the image trigger of 10.13, duration of the trigger 64~s in the 15--50 keV energy band). It was simultaneously detected by \textit{Konus-Wind} from $T_0-66$~s to $T_0+208$~s in the 20--400 keV range with detection significance above $6\sigma$~\citep{2025GCN.39116....1F}. The GRB was classified as a long burst with a duration of $T_{90} = 262.25\pm 23.71$~s as determined by the online analysis in the 15--350~keV band. \textit{Swift} initiated an automatic slew and performed follow-up observations \citep{2025GCN.39085....1S} with the X-Ray Telescope (XRT) and the Ultra-Violet and Optical Telescope \citep[UVOT;][]{Roming2005} commencing at $T_0+158.6$~s. Both X-ray and optical counterparts were found: the first observation revealed an uncataloged X-ray source located $54''$ from the BAT position~\citep{2025GCN.39066....1B} and a candidate optical afterglow with a magnitude of 17.88, positioned within $2.7''$ of the XRT localization, in the white filter~\citep{2025GCN.39066....1B}.

Various follow-up observations were initiated rapidly, beginning as early as 1.9 min after $T_0$, while the prompt emission was ongoing, and observations continued over several weeks. These efforts involved about 30 ground-based telescopes and instruments. For instance, early optical imaging was conducted by the OHP-T193 telescope, 30 min after the BAT trigger~\citep{2025GCN.39078....1S} and the Las Cumbres Observatory Global Telescope~\citep{2025GCN.39077....1G} 1.17~hr post-trigger. Extended monitoring was carried out by the 2~m robotic Liverpool Telescope and the SAO RAS 1~m telescope, $\sim24.1$~hr and $\sim43.298$~hr after the trigger, respectively~\citep{2025GCN.39099....1B,2025GCN.39107....1M} (see :\href{https://gcn.nasa.gov/circulars/events/grb-250129a}{the GCN reports}).
Observations were also conducted by the COLIBRI telescope~\citep{2025GCN.39104....1W, 2025GCN.39106....1A, 2025GCN.39136....1W}, as well as the Global Rapid Advanced Network Devoted to Multi-messenger Addicts (GRANDMA)~\citep{GRANDMA,2025GCN.39096....1A, 2025GCN.39246....1A} and its citizen science program, \href{http://kilonovacatcher.in2p3.fr/}{Kilonova-Catcher (KNC)} (\href{https://skyportal-icare.ijclab.in2p3.fr/source/2025aji}{See the GRB 250129A SkyPortal Public Page} to retrieve results submitted by various teams).

The first observations showed a steep optical and X-ray rebrightening within the first day, a feature that cannot be accounted for by a standard fireball afterglow model~\citep{Piran99,Zhang2006}, and thus motivated long-term monitoring that revealed additional rebrightening episodes. The redshift of GRB~250129A  was also measured to be $z = 2.151$ from VLT/X-shooter at $T-T_0 = 1.315$ hr~\citep{2025GCN.39071....1S}; the observations revealed Ly$\alpha$ absorption at $\sim 3830$~\AA, along with absorption lines from ions such as Si~II, C~IV, and Fe~II. This redshift was later confirmed by the Nordic Optical Telescope \citep[NOT;][]{2025GCN.39073....1I} and the MISTRAL spectro-imager at Observatoire de Haute-Provence~\citep{2025GCN.39078....1S}.

We performed a multistage modeling analysis of the GRB 250129A afterglow to probe the physical origin of its rebrightening episodes, investigating mechanisms linked to central energy reactivation. We focus on scenarios involving energy injection into the forward shock and a series of refreshed shocks produced by collisions between shells with different bulk Lorentz factors launched during the prompt phase.

The paper is structured as follows. \S\ref{sec:observations} presents the observational dataset, including the prompt emission and multi-wavelength afterglow data from gamma-ray to near-infrared observations. In \S\ref{environment}, we perform an analysis of the dust environment and the possible host galaxy. \S\ref{sec:interpretation} discusses theoretical modeling and physical interpretation of the event based on various jet-structure scenarios. Our conclusions are summarized in \S\ref{sec:discussion}.

\section{Observational dataset}
\label{sec:observations}
\subsection{Prompt Observations}

\paragraph*{Gamma-Rays --} 
 
\begin{figure}[h]
    \centering
    \includegraphics[width=0.8\linewidth]{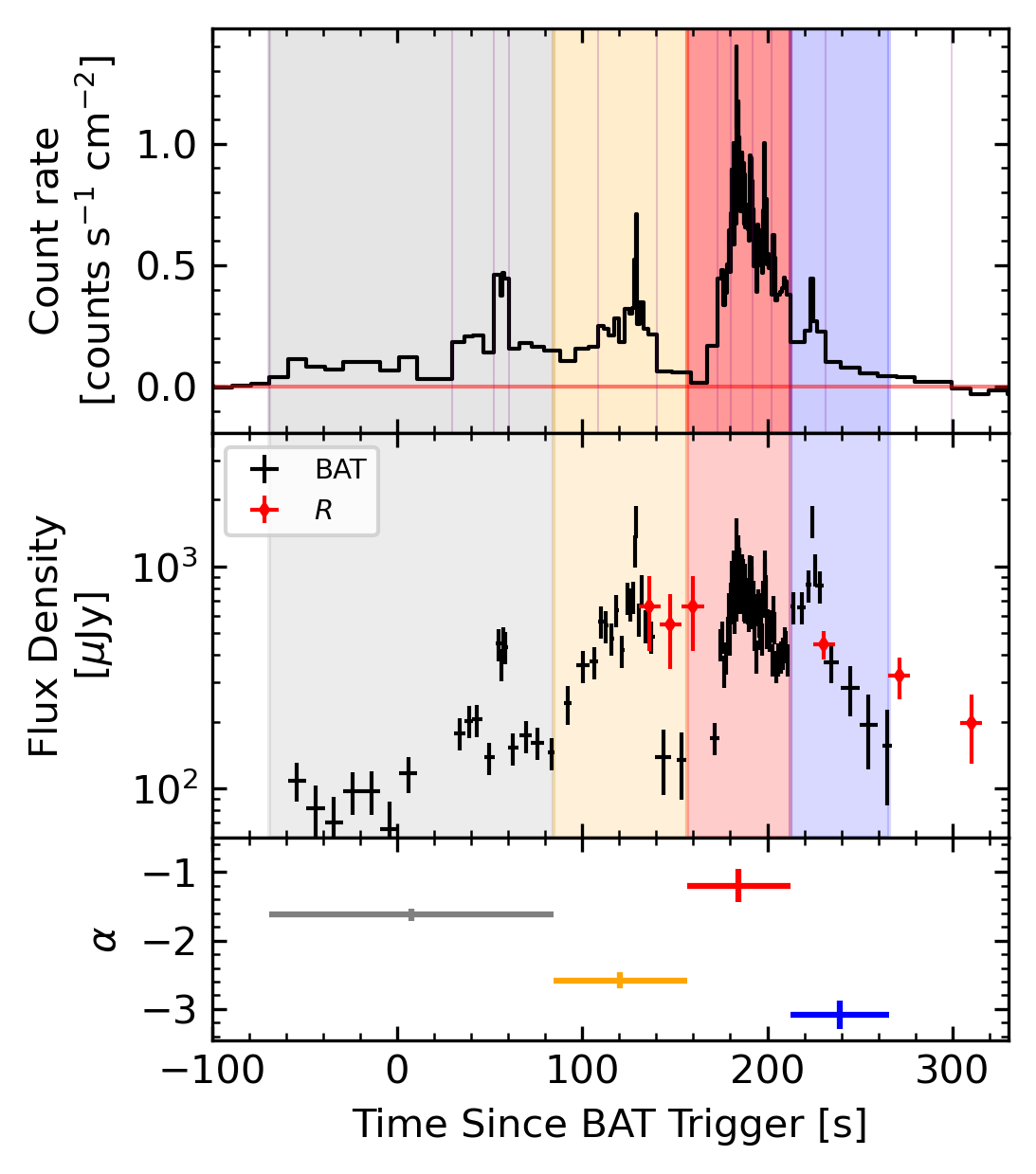}    
    \caption{{\it Top:} BAT mask-weighted count-rate light curve with Bayesian block edges. {\it Middle:} BAT (10 keV) and TAROT $R$-band flux light curves. {\it Bottom:} Best-fit spectral indices.}
    \label{fig:bat-lc}
\end{figure}

We analyzed the temporal and spectral characteristics of the GRB prompt emission observed by BAT. Mask-weighted light curves (binned to SNR = 5) and spectra were constructed using \texttt{batbinevt}. We defined four burst intervals based on the edges obtained from the Bayesian block algorithm \citep{2013ApJ...764..167S} and the rise and decay slopes of the pulses, as shown in the top panel of Figure~\ref{fig:bat-lc}. 

\begin{figure}[h]
    \centering
    \includegraphics[width=0.8\linewidth]{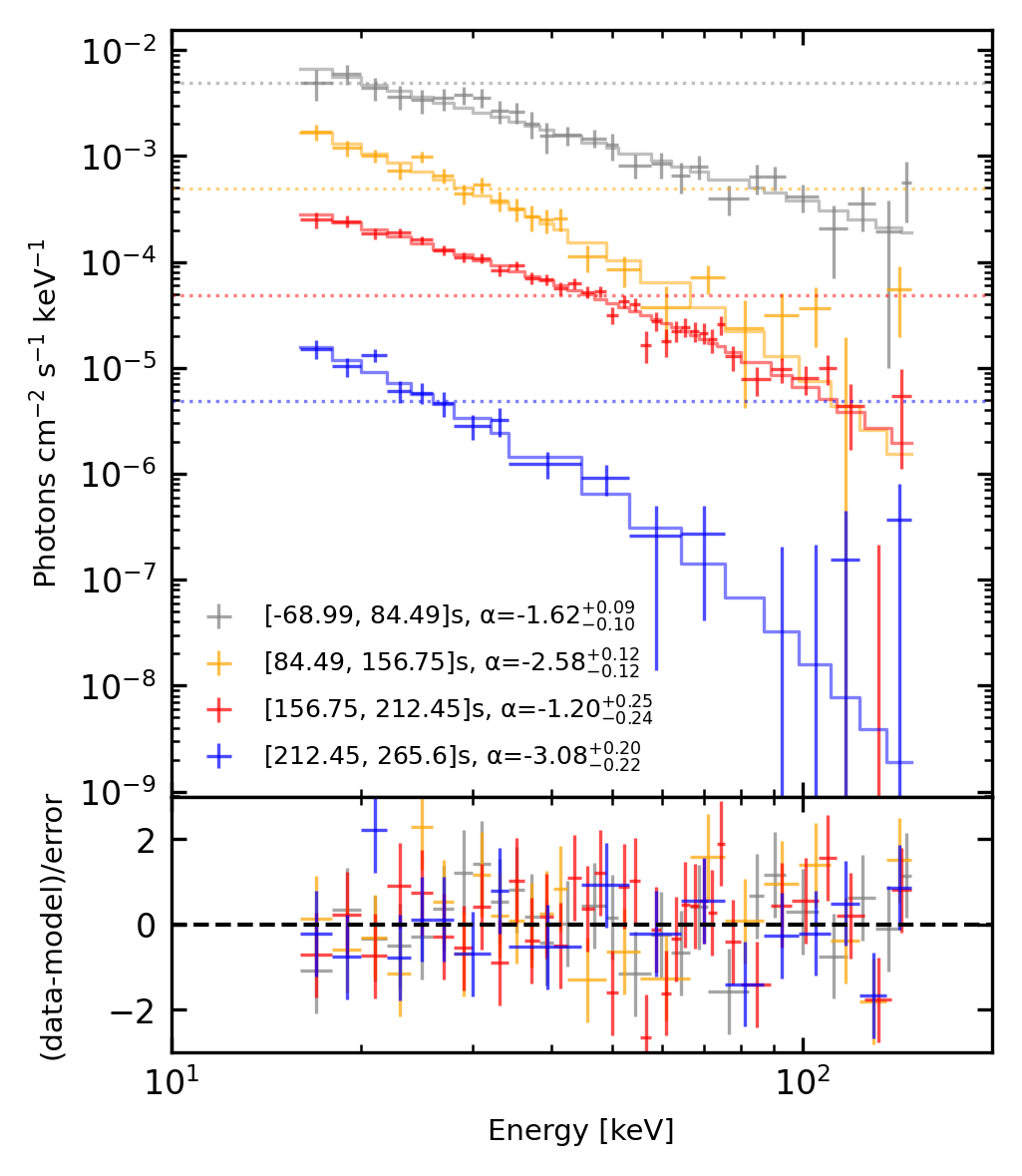}
    \caption{BAT 15--150 keV time-resolved spectra and best-fit models for GRB 250129A. For clarity, the fluxes of successive intervals are scaled by a constant factor of 0.1 relative to the previous interval; the horizontal dashed lines indicate the reference flux levels of the first interval. The fit results are reported with 90\% confidence intervals. ``0'' corresponds to the time of the trigger $T_0$.}
    \label{fig:bat-spec}
\end{figure}

Time-resolved spectra were extracted for the four intervals, and the spectral models were selected between a single power law (SPL) and a cutoff power law (CPL) using Bayesian Information Criterion (BIC) tests, requiring an improvement of $\Delta \mathrm{BIC} = 4$. Intervals~1, 2, and~4 are best fitted by SPL models, while Interval~3 is best fitted by a CPL model with $\alpha = -1.20^{+0.25}_{-0.24}$ and $E_{0} = 42^{+17}_{-10}$~keV, consistent with the result reported by \textit{Konus--Wind} \citep{2025GCN.39116....1F}. The time-resolved spectra as well as corresponding intervals are shown in Figure~\ref{fig:bat-spec}. Based on the best-fit spectral models, the photon counts were subsequently converted to the flux density at 10~keV, as shown in Figure~\ref{fig:bat-lc}. Accounting for the cosmological K-correction \citep{2001AJ....121.2879B}, we calculate an isotropic-equivalent energy of $E_{\mathrm{iso,\gamma}} = (1.35 \pm 0.12) \times 10^{53}$~erg by summing the fluence over the four intervals. 

\paragraph*{Optical Observations During the Prompt Emission --}
The Télescopes à Action Rapide pour les Objets Transitoires (\href{http://tarot.obs-hp.fr/}{TAROT}) associated with the GRANDMA collaboration started observing the GRB about 106~s after $T_0$ in drift mode and without a filter (e.g., the tracking of the hour-angle motor was adapted to a drift of 0.30 pixel~s$^{-1}$), enabling a record of the flux during 60 s without dead time~\citep{klotztrail}. This sequence, covering the prompt emission during about 70 s, is shown in Figure~\ref{fig:bat-lc}. 

\subsection{X-ray Afterglow}
\label{x-ray-analysis}
\paragraph*{Light Curve --} We retrieved the XRT unabsorbed flux density light curves resampled for the 10 keV band from the \href{https://www.swift.ac.uk/burst_analyser/01285812/}{\texttt{Burst Analyser}}. The light curve was rebinned by grouping data points into 16 manually defined, noncontiguous clusters based on temporal proximity to improve the SNR while preserving the time evolution. Among these, 4 bins contained clusters of 2--5 closely spaced observations, while the remaining bins contained a single point each. For each bin, we computed the mean magnitude and standard deviation.
We then converted the flux density from 10 keV to 2 keV (see Appendix~\ref{app:detailX-ray}), to ensure our analysis is grounded in the energy range where the XRT is most sensitive. The light curve can be seen in Figure~\ref{fig:LC}, and all the data are available online in SkyPortal in section~\ref{data:photo}.

\paragraph*{Time Resolved Spectra --} We extracted the spectral characteristics in the 0.3--10 keV range of XRT data in various epochs (see Figure~\ref{fig:xrtsourcesfits} in Appendix~\ref{app:detailX-ray}). The data reduction and spectral analysis were conducted with the same approach for all epochs using \texttt{HEASoft v6.34}, and processed using the \texttt{XRT tool xrtpipeline (v0.13.7)}. The spectra (available \href{https://www.swift.ac.uk/xrt_spectra/01285812/}{at this link}) were fitted with an absorbed power-law model, leaving only the photon index ($\Gamma$) and the normalization as free parameters, with the Galactic column density ($N_{\rm H} =2.4 \times 10^{20}$ cm$^{-2}$) fixed (see Appendix~\ref{app:detailX-ray}). 

We initially investigated potential spectral evolution within the first day following the discovery from 2025-01-29T05:04:40 to 2025-01-29T18:49:13, as the low count rate after the first day prevented the fitting for properties such as the photon index. The best-fit photon indices for the three source spectra within the first day shared the same characteristics with the photon index $\Gamma = 1.99 \pm 0.23$, $\Gamma = 2.15 \pm 0.15$, and $\Gamma = 1.87 \pm 0.18$, respectively. We then analyzed one epoch more carefully, which corresponds to high variation in the optical. We split the period [$0.061$–$0.197$] post $T_0$ at $T=0.061$, which coincides with a transition in the optical light curve from a rising to a decaying phase. Separate spectral fits were performed for each using the method outlined in Appendix~\ref{app:detailX-ray}. The resulting photon indices were $\Gamma = 2.22 \pm 0.18$ for the rising phase and $\Gamma = 2.11 \pm 0.34$ for the decaying phase. While the values differ slightly, the uncertainties overlap substantially, indicating no statistically significant spectral evolution during this period. This is further discussed by fitting the spectral energy distribution (SED), as detailed in Sec.~\ref{sec:sed}.

\subsection{UV-Optical-Infrared Afterglow}
\label{uvot}

\paragraph*{Space --} We performed \textit{Swift} aperture photometry in $UVW1$, $UVW2$, $UVM2$ (ultraviolet), and $U$, $B$, $V$ (optical) to extract the UVOT light curves using a 5\arcsec-radius circular source region and a 15\arcsec-radius background region away from the source. For individual exposures, we employed the \texttt{uvotmaghist} task. In addition, we selectively co-added exposures that were close in time within a range of a few minutes in the same band, identified based on their temporal distribution on a log-time plot. These subsets were summed using \texttt{uvotimsum}, and photometry was then performed using \texttt{uvotsource}. This approach allowed us to preserve time resolution where needed and improve the SNR, as well as decrease the uncertainty, particularly with the exposures that were clustered closely in time.

Very early-time exposures initially excluded by \texttt{uvotmaghist} due to failed aspect correction (i.e., \texttt{ASPCORR} keyword set to \texttt{NONE}) were manually corrected using \texttt{uvotaspcorr} to allow for their inclusion. The initial $V$-band settling image was also examined for detector ramp-up effects. To assess its reliability, we compared the photometry of field stars in the settling image with that from later stacked $V$-band exposures. Additionally, we considered nondetection when SNR $< 2$, which is typically associated with contamination from known internal dust regions within the telescope. All the UVOT data are available in Section~\ref{data:photo} of the Appendix (online material) and shown in Figure~\ref{fig:LC}.

\begin{figure}
    \centering
    \includegraphics[width=0.45\textwidth]{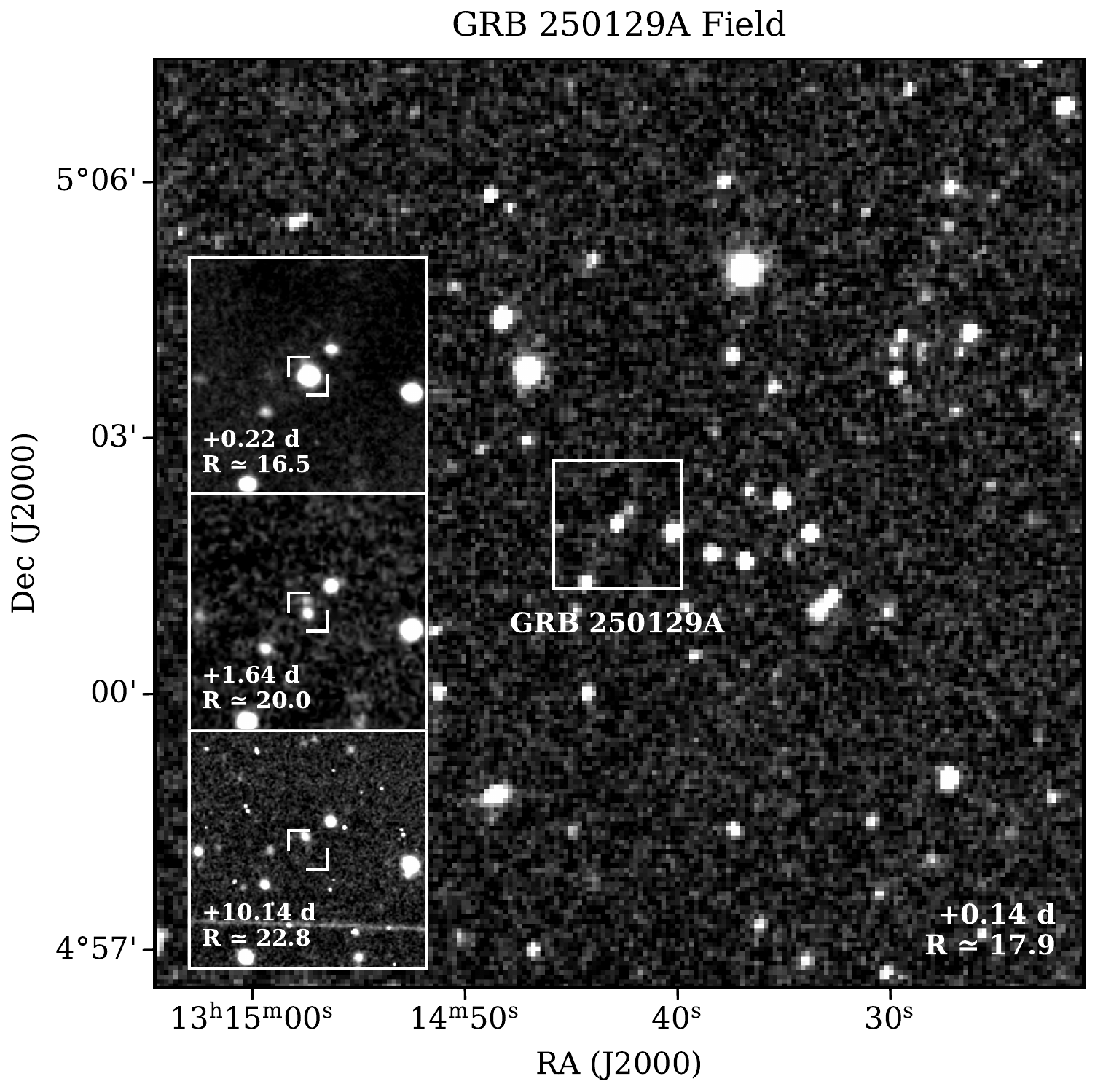}
    \caption{
    TAROT-TCH telescope image of GRB~250129A taken at $T-T_0 = 0.14$ d. The insets show the zoomed-in region of three images from the KAIT, Abastumani-T70, and Euler telescopes centered on the GRB location, taken at phases of $T-T_0 = 0.22$, $T-T_0 = 1.64$, and $T-T_0 = 10.14$ d, respectively, relative to the \textit{Swift} GRB trigger (MJD 60704.198).}
    \label{fig:tarot}
\end{figure}

\label{sec:photometric_methods}
\paragraph*{Ground --} The processing of all optical observations of the 30 different instruments at play followed the same methodologies to maintain consistency and data quality. The pre-processing after data acquisition started with bias and dark subtraction, as well as flat-field correction according to each telescope's specifics. We then manually masked areas within each image with visible imaging artifacts or areas that were not entirely corrected during pre-processing (especially at the edges). We verified or derived astrometric solutions using the \texttt{astrometry.net} service~\citep{2010astrometrynet}, on either individual images or stacked images. We performed dynamical image stacking using the SWarp software~\citep{2010ascl.soft10068B} to retain high temporal resolution and SNR $> 3$ at the location of the transient.

Forced photometry of all the stacked images ($\sim290$ total) was performed at the transient position (RA, Dec. = 198.676728$^\circ$, 5.030631$^\circ$) using \texttt{STDPipe}~\citep{stdpipe,Karpov2025}, a home-made Python-based set of codes for performing astrometry, photometry, and transient detection tasks on optical images, originally created as part of the GRANDMA Collaboration. We utilized the associated web interface \href{http://stdweb.favor2.info}{\texttt{STDWeb}} to conduct our photometry and seek anomalies that need further adjustments (e.g., high background noise, calibration issues). Catalogs for the photometric calibration were chosen based on the filters used to acquire the images. Instrumental magnitudes were calibrated using the Gaia DR3 Synthetic Photometry Catalog~\citep{2023A&A...674A..33G} and  Pan-STARRS Data Release 1 catalog~\citep{2016arXiv161205560C} for images obtained in filters close to the Johnson-Cousins filter system ($U, B, V, Rc, Ic$) and Sloan filters ($g, r, i, z$), respectively. Images obtained using nonstandard filters (not part of the Johnson-Cousins or Sloan systems) were calibrated using the filter and catalog, which minimized the color term to below 0.1: for example, $g-r$ for Sloan-like filters in order to assess how much the individual photometric system of the image deviates from the catalog one. Unfiltered images are particularly challenging since they cover the full wavelength range.

Specific treatment of several images was required to correctly calibrate them using standard catalogs. For example, a first nonlinear color term was needed to adjust the data of the early afterglow ($T-T_0 < 0.2$ day): the color-term estimation had to be modified into a parabolic color term, which was then used to fit both datasets from the TAROT/TCH and FRAM-Auger telescopes. Specifically, the correction for the color we obtained is $B-V=1.46$ mag ($\sim g-r=1.27$ mag). Hence, the color term was applied to TAROT/TCH as $[R + 0.44(B-V) - 0.29(B-V)^{2}]$ and to FRAM-Auger data as $[R + 0.06(B-V)]$ (as shown in Figure~\ref{fig:LC}). We additionally used data from Skynet telescopes in the $B$, $V$, $R$, and $I$ bands, and we applied a joint color correction of $(B-V) = 0.4$ mag to Skynet, TAROT/TCH, and FRAM-Auger to minimize the offsets introduced by the distinct filter responses of the telescopes. This correction indicated that the transient exhibited a relatively blue color of $(B-V) = 0.4$ mag during the early-time window, thereby superseding our earlier interpretation that the transient was red at early times.

At later times, from $0.2 < T-T_0 < 1$ days, the analysis of KAIT data taken without a filter was challenging owing to poor background quality and high noise levels in certain images; no color-term corrections were applied as their inclusion led to further instability in the fit. To mitigate these issues, the number of free parameters was minimized: the background mesh size, or the size of the grid cells used for global background estimation, was reduced to 32 pixels, with a constant zero-point, and a reduced annulus size was constrained to apertures of 2 and 4 times the FWHM (full width at half-maximum intensity) for the sky inner and outer annuli, respectively. Finally, when the afterglow reached a magnitude above 21.5, we subtracted a template image from the Legacy DESI Survey \href{www.legacysurvey.org/dr10}{Data Release 10} to isolate the flux contribution using High Order Transform of Psf and the template-subtraction \texttt{HOTPANTS}~\citep{2015ascl.soft04004B} algorithm for image subtraction.  Besides, late time observations also included one infrared $J$-band image obtained by the 1m of Pic Du Midi.

\begin{figure*}
    \centering
    \includegraphics[clip, width=0.90\textwidth]{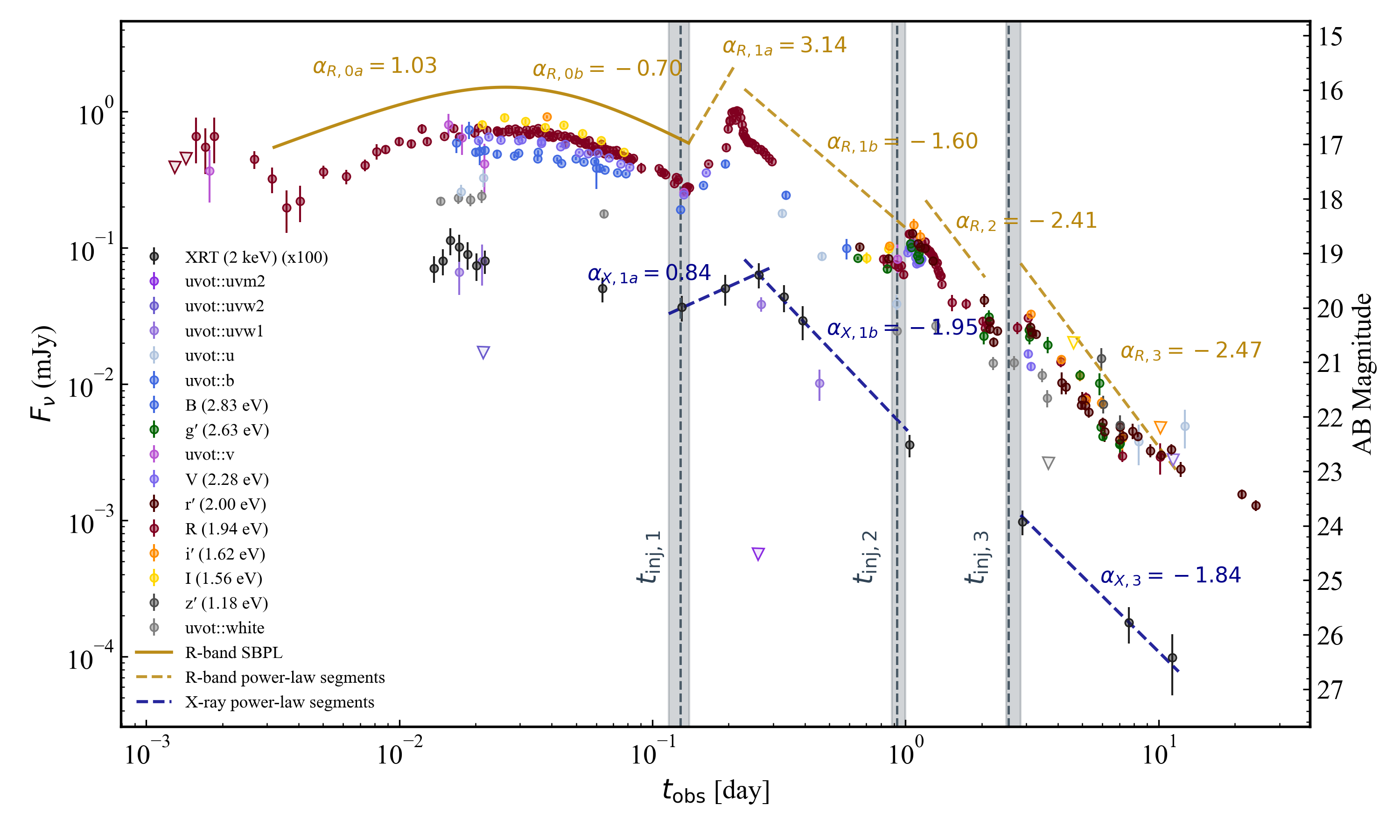}
    \caption{Observations from this work. Apparent magnitudes corrected for Milky Way extinction with the calibration of the Milky Way dust maps from \citep{Schlafly_2011}. Here, $t_{\rm inj,1}$, $t_{\rm inj,2}$, and $t_{\rm inj,3}$ correspond to the time of the first ($T-T_0 = 0.129^{+0.011}_{-0.013}$), second ($T-T_0 = 0.926^{+0.074}_{-0.042}$), and third ($T-T_0 = 2.55^{+0.301}_{-0.05}$) rebrightenings in days (see Section~\ref{tab:posterior_nmma_analysis}). For clarity, XRT flux densities are scaled by a factor of 100, and the AB-magnitude axis reflects the corresponding scaled flux densities \citep{Oke83Secondary} The dashed lines represent the power-law fits to the afterglow, and the corresponding temporal slope values are listed in Table~\ref{tab:empirical_slopes}. The solid curve shows the smooth broken power-law (SBPL) fit to the early $R$-band evolution. 
    The $R$-band temporal decay fits are shown with a +0.35 dex vertical offset to aid visualization.}
    \label{fig:LC}
\end{figure*}

\section{Environment}
\label{environment}

\subsection{Host Galaxy Line-of-Sight Extinction} 
\label{sec:sed}

The line-of-sight host-galaxy dust extinction has been estimated by creating an X-ray-to-optical spectral energy distribution (SED) at a mid-time of $T-T_\mathrm{0} \approx 7.04$ days after the \textit{Swift} trigger, since no further spectral evolution of the afterglow is expected at this later epoch. The SED was built using XRT and optical data ($g'r'i'z'$), which were corrected from the Galactic reddening of $E(B-V) = 0.03$ mag \citep{Schlafly_2011}.

Compared to Section~\ref{x-ray-analysis}, we retrieved the time-sliced X-ray spectrum, covering the 2.22--12.74 days interval, keeping high timing resolution and directly from the automated data products provided by the public \textit{Swift}/XRT archive~\citep{evans_online_2007}. Photometric data in the $g'r'i'z'$ bands from Section~\ref{uvot} were interpolated to the common reference time of the $z'$ band (MJD 60711.237) at $\sim 7.04$ days after the trigger. The SED fit was performed by following a standard routine using the average extinction curves of the Milky Way (MW), Large Magellanic Cloud (LMC), and Small Magellanic Cloud (SMC) \citep{pei1992}, and the intrinsic X-ray-to-optical spectrum was modeled with a single or broken power law. Furthermore, the slope of the X-ray wavelengths was assumed to be 0.5 steeper than the spectral slope below the cooling break of the intrinsic spectra, $\Delta\beta = \beta_\mathrm{X} - \beta_\mathrm{o} = 0.5$, as predicted by the standard afterglow model~\citep{Sari1998}. We fixed the redshift and the Galactic foreground absorption of $N_{\mathrm{H}}^{{\mathrm{Gal}}} = 2.43 \times 10^{+20} \; \text{cm}^{-2}$~\citep{willingale,bekhti2016}.
\begin{figure}
    \centering
    \includegraphics[width=0.48\textwidth]{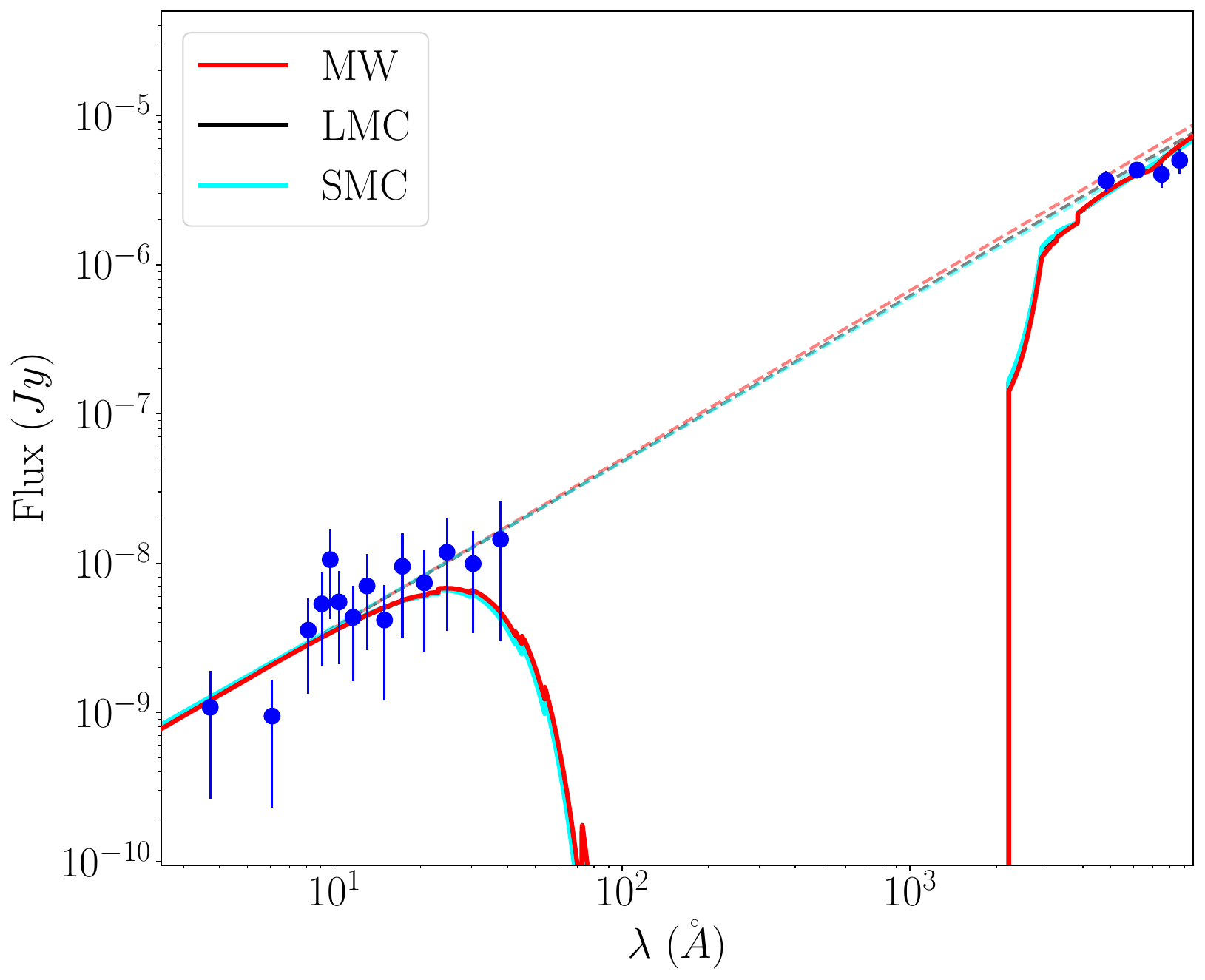}
    \caption{X-ray-to-optical SED of GRB 250129A at $T-T_\mathrm{0} \approx 7.04$ days using the MW (red), LMC (black), and SMC (cyan) extinction curves. \textit{Dashed lines}:  intrinsic simple power-law model of the afterglow. \textit{Solid lines}: Best fits to the data, including the X-ray absorption and the optical extinction. The three curves overlap because the negligible best-fit host extinction makes the models indistinguishable.}
    \label{fig:sed250129}
\end{figure}

The simple power-law model with the MW extinction curve provides a fit with the parameters $\beta = 1.15^{+0.18}_{-0.17}$, $E(B-V) = 0.05^{+0.14}_{-0.05}$ mag, $N_\mathrm{H,X} = 0.89^{+8.72}_{-0.88} \times 10^{+22}$ cm$^{-2}$ ($\chi^2_\text{MW}/\text{d.o.f.}$ = 12.12/14). Uncertainties are given at the 3$\sigma$ confidence level. Additionally, the fitted parameters remain consistent across the other extinction curves, with the LMC and SMC yielding statistically similar fits of $\chi^2_\text{LMC}/\text{d.o.f.}$ = 12.45/14 and $\chi^2_\text{SMC}/\text{d.o.f.}$ = 12.57/14 (see Fig. \ref{fig:sed250129}). Given that our estimated extinction is consistent with zero at the 3$\sigma$ confidence level, we assumed the no-host dust extinction scenario in subsequent analyses and applied no further corrections.

\subsection{Host Galaxy}
\label{sec:host}

\begin{table}
	\centering
	\caption{Multi-segment temporal (\(\alpha\)) and spectral (\(\beta\)) indices of the afterglow. The temporal indices in the \(R\) band and X-ray are derived from power-law fits within the time intervals listed in the first column. The onset bump is fitted with a smoothly broken power-law, while single power-law fits are applied to the remaining segments. Spectral indices are obtained from multi-band SED fitting over the corresponding epochs.}
	\label{tab:empirical_slopes}
 \begin{tabular}{cl}
		\hline
Time (days) & Slopes \\
\hline
               & Temporal index     \\  
               \hline
0.004-0.032    &    $\alpha_{\rm{R},0a}$ = 1.03$\pm$0.05      \\  
0.032-0.139    &    $\alpha_{\rm{R},0b}$ = -0.70$\pm$0.07     \\  
0.139-0.211    &    $\alpha_{\rm{R},1a}$ = 3.14 $\pm$0.40     \\  
0.231-1.020    &    $\alpha_{\rm{R},1b}$ = -1.60 $\pm$0.02    \\  
1.200-2.060    &    $\alpha_{\rm{R},2}$  = -2.41 $\pm$0.15    \\  
2.850-12.000   &    $\alpha_{\rm{R},3}$  = -2.47 $\pm$0.10    \\ 
              \hline
              & X-ray band                                  \\  
0.116-0.291    &    $\alpha_{\rm{X},1a}$ = 0.84$\pm$0.43      \\  
0.231-1.020    &    $\alpha_{\rm{X},1b}$= -1.95$\pm$0.04     \\  
2.850-12.000   &    $\alpha_{\rm{X},3}$  = -1.84 $\pm$0.51    \\  
\hline
             &  Spectral index                                \\
             \hline
$\sim$0.08    &     $\beta_{\rm0b}$ = -0.75$\pm$0.07      \\
$\sim$0.2     &     $\beta_{\rm1a}$ = -0.71$\pm$0.01      \\  
0.30-0.33      &     $\beta_{\rm1b}$ = -0.91$\pm$0.15      \\
1.2-2.0       &     $\beta_{\rm2}$ = -1.14$\pm$0.79       \\  
3.0-4.0       &     $\beta_{\rm3}$ = -1.25$\pm$0.83       \\
\hline
		\hline
	\end{tabular}
\label{empirical}
\end{table}


We searched for the host galaxy of GRB~250129A using the \texttt{Galclaim} tool \citep{galclaim} to identify catalogued galaxies in Pan-STARRS DR2 \citep{Pan-STARRS}. The nearest candidate, PSO J198.6769+05.0319 (SDSS J131442.44+050154.9; RA = 198.6769$^\circ$, Dec = 5.0319$^\circ$), lies 4.695 arcseconds from the afterglow position, corresponding to a chance coincidence probability named $P_{\rm cc} = 0.034$. $P_{\rm cc}$, defined in \citet{galclaim}, expresses the chance alignment of the afterglow position with the field galaxies, using catalogued galaxies in Pan-STARRS DR2 \citep{Pan-STARRS}. It quantifies the likelihood that a galaxy of comparable or greater brightness would be found within a circle of radius defined by the angular separation between the GRB and the candidate host purely by chance. Although this value is relatively low, it exceeds the typical threshold adopted to confirm host associations for long GRBs ($P_{\rm cc} \sim 0.01$). Moreover, the galaxy has a photometric redshift of $0.443 \pm 0.0812$ in SDSS DR19 \citep{SDSSdr19}, which is clearly inconsistent with our GRB’s redshift ($z = 2.151$). Furthermore, adopting \citet{Planck18}'s cosmological parameters, a 4.695 arcsecond offset at $z = 2.151$ would correspond to a projected physical separation of $\sim 39$ kpc, well beyond the known distribution of offsets for long GRBs (e.g., \citealt{Blanchard2016}). We therefore conclude that this galaxy cannot be the host.

No other nearby objects in Pan-STARRS yield a $P_{\rm cc}$ value suitable for a plausible association. In the Legacy Survey DR10 \citep{LegacySurvey}, only one object (RA = 198.6776$^\circ$, Dec = 5.0292$^\circ$) not detected in Pan-STARRS is found near the afterglow at an angular separation of $\sim 6$ arcseconds. This source is significantly fainter than the previously discussed galaxy, resulting in a significantly higher $P_{\rm cc}$. 

We therefore conclude that the host galaxy is not detected in the available catalogs, with a $3\sigma$ upper limit of $r > 24.13$ mag, obtained from our reduction of the Legacy Survey DR10 images following the procedure described in Section 2.3. The non-detection of a host galaxy down to this magnitude limit is not unusual for GRB host galaxies at comparable redshifts (e.g., \citealt{Cenko2008,Greiner2015,Perley2016,Perley2016b}).

\section{Interpretation}
\label{sec:interpretation}

To model the afterglow emission before the rebrightening episodes seen in the optical and X-rays, we first consider the simple and standard afterglow model of a thin and ultrarelativistic spherical shell. This serves as a good approximation for a uniform top-hat jet viewed on-axis at times before the jet-break time. 
We assume that the number density distribution of the circumburst medium could be described as $n_{\rm ext}(\mathcal{R})=n_0(\mathcal{R}/\mathcal{R}_0)^{-k} = A\mathcal{R}^{-k}$ with $\mathcal{R}$ the distance from the explosion center, $k$ the index of the density profile, $n_0$ the density normalization defined at a fixed $\mathcal{R}_0$, and $A=n_0\mathcal{R}_0^k$. Considering a shell propagating in such a medium with an initial bulk Lorentz factor (LF) of $\Gamma_0 \gg 1$, it will start to decelerate when the swept-up mass $M_{\rm sw}(\mathcal{R}) = 4\pi m_{\rm p} n_{\rm ext}(\mathcal{R}) \mathcal{R}^3/(3-k) = M_0/\Gamma_0$, where $M_0 = E_{\rm k,iso}/\Gamma_0c^2$ is the baryon load of the shell and $E_{\rm k,iso}$ is its isotropic-equivalent kinetic energy after the prompt phase, and $m_{\rm p}$ and $c$ are the proton mass and speed of light. 
The corresponding deceleration radius is given by $\mathcal{R}_{\rm dec} = [(3-k)E_{\rm k,iso}/4\pi m_{\rm p} c^2A\Gamma_0^2]^{1/(3-k)}$ \citep[e.g.,][]{Panaitescu-Kumar-00}, where the bulk of the kinetic energy of the shell is given to the shocked swept-up medium that causes the FS afterglow emission to peak. At $\mathcal{R}>\mathcal{R}_{\rm dec}$, the dynamical evolution of the shell follows the \citet{Blandford76} solution where its bulk LF declines as a power law in radius with $\Gamma(\mathcal{R})\propto \mathcal{R}^{-(3-k)/2}$, as dictated by energy conservation.

We model the emission coming from the FS using the standard afterglow theory~\citep{Sari1998}. The shock accelerates a large fraction of the electrons in the swept-up medium into having a power-law energy distribution, with comoving number density $n_{\rm e}(\gamma)\propto\gamma^{-p}$ for $\gamma>\gamma_{\rm m}$ and where $\gamma$ is the LF of the electrons and $\gamma_{\rm m}$ is the LF of the minimal energy electrons. These electrons hold a fraction $\epsilon_{\rm e}$ of the total internal energy in the shocked medium, while a fraction $\epsilon_{\rm B}$ goes into generating and/or amplifying small-scale magnetic fields. The accelerated electrons then cool behind the shock front by emitting synchrotron radiation. The power-law temporal index ($\alpha$) and the spectral index ($\beta$) of the afterglow emission ($F_\nu\propto t^{\alpha} \nu^{\beta}$) are related through the \textit{closure relations}, where slow-cooling emission produced at a frequency $\nu_\mathrm{m}<\nu<\nu_\mathrm{c}$ (with $\nu_\mathrm{m}$ the peak synchrotron frequency of minimal-energy electrons and  $\nu_\mathrm{c}$ the cooling-break frequency ~\citep{2002ApJ...568..820G})  has $\alpha=-3(p-1)/4$ and $\beta=-(p-1)/2$ when $k=0$. We exploit these relations below to check for consistency and infer the value of $p$.

\subsection{Probing External Shock Scenario with Empirical Fitting}
\label{sec:empirical}
 
Figure~\ref{fig:LC} shows the optical and X-ray light curves. The optical emission exhibits an onset bump peaking at $\sim 0.03$~d, followed by three rebrightening episodes at $\sim 0.2$, $\sim 1$, and $\sim 2.5$~d. The first two rebrightenings are clearly visible, while the third is less pronounced. The X-ray light curve, although sparsely sampled, follows the overall trend of the optical emission; however, the limited data make it difficult to clearly identify distinct rebrightening episodes. To characterize this morphology, we derive the temporal and spectral slopes. Temporal slopes are obtained by fitting power laws over the time intervals listed in Table~\ref{tab:empirical_slopes}, which also summarizes the results.

The onset bump has rising and decaying slopes of $1.03 \pm 0.05$ and $-0.70 \pm 0.07$, respectively. Assuming a constant-density interstellar medium (ISM), a slow-cooling synchrotron regime, and no energy injection in this phase, we infer an electron power-law index of $p = 1.93 \pm 0.09$. This is somewhat smaller than typically reported for GRBs \citep[e.g.,][]{Kumar2015}. However, we measure a spectral index of $-0.75 \pm 0.07$ at $0.08$~d, which, under the slow-cooling assumption, implies $p = 2.5 \pm 0.14$. This value is consistent with the commonly observed range \citep{Kumar2015}. The discrepancy between the values of $p$ inferred from the temporal decay versus the spectrum likely stems from the short duration of the first decay phase. Measuring a slope over such a limited interval near the peak is prone to curvature effects, leading to an artificially shallow decay measurement. Indeed, the fits obtained with the \textsc{NMMA}/\textsc{afterglowpy} code (Section~\ref{sec:tophat-jet-nmma}) are consistent with the assumed physical scenario without requiring early energy injection during the onset bump (i.e., before the first reebrightening). We therefore attribute the low temporal $p$-value to uncertainties in the slope measurement rather than intrinsic physics.

The first rebrightening after the onset bump rises with a temporal slope of $3.14 \pm 0.40$, after which it declines with a slope of $-1.60$. The second and third rebrightening episodes are followed by more rapid declines of $-2.41 \pm 0.15$ and $-2.47 \pm 0.10$. This steepening is indicative of a jet break. Owing to limited data, we are unable to calculate the rising slopes of the second and third rebrightening episodes.

The spectral indices measured at about 0.08, 0.2, 0.3, 1.2--2, and 7 days after the trigger are respectively $-0.75 \pm 0.07$, $-0.71 \pm 0.01$, $-0.91 \pm 0.15$, $-1.14 \pm 0.79$ , and $-1.25 \pm 0.83$ (cf. Table~\ref{tab:empirical_slopes}). The steepening of the spectral index at late times suggests the passage of the cooling frequency, while the steep late-time temporal decays are consistent with a jet-break scenario.

\subsection{Agnostic Approach}
\label{sec:tophat-jet-nmma}

\paragraph{Framework and Analysis Period --} We aim to extract the physical parameters of the jet and emission properties by analyzing the light-curve data with the \textsc{afterglopwy} model~\citep{Ryan2020} using the Bayesian inference framework \textsc{nmma}~\citep{Pang:2022rzc}. 
The \textsc{afterglowpy} package models the GRB afterglow as the result of synchrotron emission from accelerated electrons. Here we make the explicit assumption that the jet propagates inside a constant-density ISM with $n_{\rm ext}=n_{\rm ISM}$. The jet structure is assumed to be a top-hat jet for which the isotropic-equivalent kinetic energy, $E_{\rm k,iso}$, is constant across the jet aperture with half-opening angle $\theta_{\rm j}$. Moreover, the dynamical evolution of the jet in \textsc{afterglowpy} does not consider a coasting phase and instead makes the simplifying assumption that it has already decelerated and follows the asymptotic power-law decline of its bulk LF with radius. For this reason, when analyzing the data from GRB~250129A with \textsc{afterglowpy}, we exclude early data within the first 0.05~days (4320~s) post-burst, as these points show a rising light curve that is likely due to a jet coasting phase and cannot be properly modeled in \textsc{afterglowpy}.

\paragraph{Energy Injections in a Phenomenological Approach --} In order to capture the rebrightening epochs from GRB~250129A's afterglow, we set an ad-hoc prescription that allows the jet energy to vary with time. Specifically, we introduce an energy injection parameter $\Delta \log_{10}E$ that, starting from time $t_{\rm inj}$, linearly increases the jet log energy within a given time interval (see Appendix~\ref{nmma:injection} for details). 
This parametrization can be extended to include multiple energy injections.
We emphasize that this prescription represents a phenomenological approach and does not account for the mechanism underlying the increase in jet energy. 
Despite these limitations, the ad hoc increase of the jet energy still allows us to phenomenologically recover the rebrightening in the afterglow light curve and statistically investigate the number of rebrightenings and their approximate duration. 

To infer the posterior $p(\vec{\theta}|d)$ from the light-curve data $d$, we sample the parameter space of the model using the nested sampling algorithm as implemented in \textsc{pymultinest}~\citep{Feroz:2008xx,Buchner:2014nha}.
In particular, we use the likelihood function  $\ln \mathcal{L}(\vec{\theta}|d)$, which compares the observed magnitudes to the model predictions from \textsc{afterglowpy}. 
The model parameters and priors are listed in Table~\ref{NMMAprior}.
In addition to the model parameters, \textsc{nmma} also samples a nuisance parameter $\sigma_{\text{sys}}$ that accounts for systematic uncertainty in the modeling. Throughout the analysis, we fix the redshift at $z = 2.151$ (Sec.~\ref{intro}), corresponding to a luminosity distance of $d_L = 17.3$~Gpc.
We begin by establishing a reference baseline to evaluate the statistical significance of the observed rebrightening epochs. 
Specifically, we select data from the time interval between 0.05 and 0.2 days, which we refer to as the ``early-time'' dataset. 
The posterior light curves estimated using the data from this interval serve as a proxy for a standard GRB afterglow in our subsequent analysis (see Table~\ref{NMMAprior}). 
Using posterior light curves derived from the ``early-time'' data, we compute the average chi-squared statistic, $\langle\chi_j^2\rangle$, for each data point (see Appendix~\ref{nmma:injection} for details). 
In the $R$ band, the computed $\langle\chi_j^2\rangle$ values reveal three distinct rebrightening phases spanning the intervals [0.15, 1] days, [1, 3] days, and [3, 30] days (see Figure~\ref{fig:LC}). 
To investigate these features, we apply the ad hoc energy injection model described in Eq.~\eqref{eq:energy_injection} and analyze the full dataset beyond 0.05 days. 
Parameter priors are listed in Table~\ref{tab:prior_nmma_analysis}. 

In total, four models are considered: one without energy injection, and others with 1, 2, and 3 energy injection epochs. The posterior distributions for the four models considered are summarized in Table~\ref{tab:posterior_nmma_analysis} (and visualized in Fig.~\ref{fig:nmma_corner} in the Appendix). 
As described by~\cite{Pang:2022rzc}, the parameter $\sigma_{\rm sys}$ is included to account for the systematic uncertainty within the light-curve  modeling. Comparing the inferred systematic uncertainty across models with varying numbers of energy-injection epochs to the baseline model without injections, we observe a general decreasing trend. This suggests that including energy injections improves the model's ability to fit the data.

\begin{figure}
    \centering
    \includegraphics[width=0.5\textwidth]{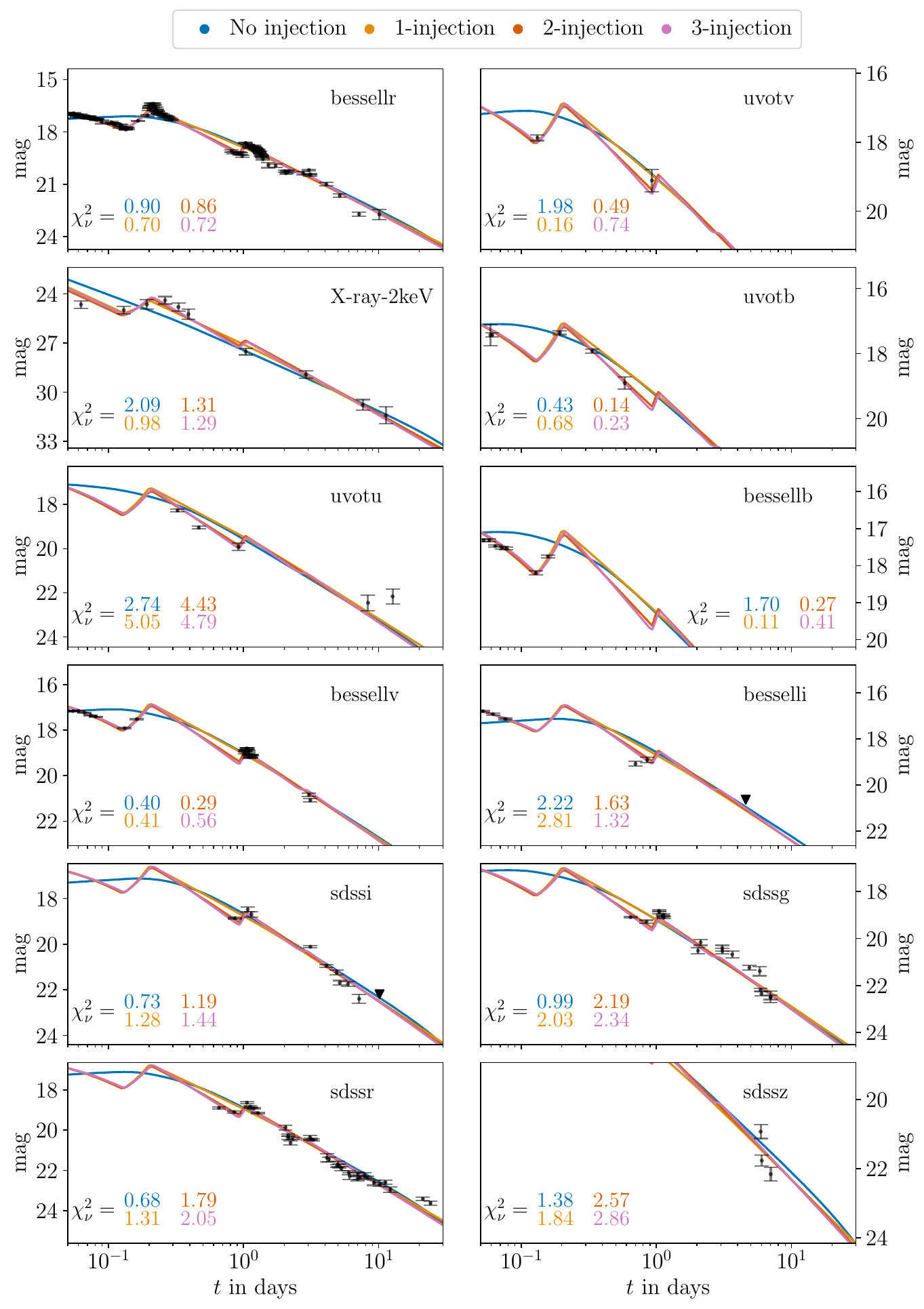}
    \caption{Best-fit light curves of the GRB afterglow, together with the reduced $\chi^2$ of each filter, for each of the models considered.}
    \label{fig:nmma_bestfit}
\end{figure}

We also compare the maximum likelihood values and compute the Bayes factors between the models, as reported in Table~\ref{tab:posterior_nmma_analysis}. Our analysis shows that models with one, two, or three injection epochs yield higher maximum likelihoods and are favored by the Bayes factor. However, in the case of three injections, the (ln-) Bayes factor relative to the two-injection model is modest ($\ln\mathcal{B} \approx 6$). Furthermore, the estimated fractional energy injected associated with the third injection is notably small. This is consistent with the best-fit light curves, as there is minimal difference between the two-injection and three-injection light curves shown in Fig.~\ref{fig:nmma_bestfit}, leading us to conclude that only the first two rebrightenings are statistically significant. This conclusion is further supported by visual inspection of the light curve (cf. Fig.~\ref{fig:LC}): the first and second rebrightenings are clearly identifiable, whereas the third episode is less distinct.

The energy injection scenario provides a physically motivated explanation for afterglow rebrightenings. This scenario is to some extent equivalent to the refreshed-shock scenario, in which the ejecta possess a distribution of LFs rather than a single characteristic value. As the FS decelerates, slower but energetically significant ejecta catch up with the blast wave and re-energize it, leading to a rebrightening of the afterglow light curve \citep{Sari2000}. This framework is also adopted by \citet{Laskar15}, who model energy injection as arising from a continuous distribution of Lorentz factors released over a time interval short compared to the afterglow timescale. Despite the apparent consistency between the model and data for each individual rebrightening phase when energy injection is included, the inferred physical parameters, such as the electron distribution power-law index $p$, vary significantly across the different epochs.
The variation between the inferred parameters for GRB~250129A indicates that such energy injection alone is insufficient to capture the underlying physics in a self-consistent manner.

Although energy injection could in principle explain the rebrightening, fully accounting for the observed temporal and spectral evolution before and after the injection remains challenging, requiring a detailed treatment of the underlying physics, as presented in the next section.

\begin{table}[h]
    \caption{Parameters employed and the associated prior bound in our Bayesian inferences. Three sets of priors are used for each of the re-brightening peaks. $\epsilon_{\rm e}$ refers to the energy fraction in electrons, $\epsilon_{\rm B}$ to the energy fraction in magnetic field, $p$ to the electron distribution power-law index, $\theta_{\rm j}$ to the jet core opening angle $\sigma_{\rm sys}$ to the systematic error,  $t_{\rm inj}$ to the start time of energy injection,  $\Delta t_{\rm inj}$ to the duration of energy injection and $\Delta \log E_{\rm k,iso}$ to the fractional energy injected.  }\label{tab:prior_nmma_analysis}
    \centering
    \begin{tabular}{lc}
    \hline
    Parameter & Prior bound \\
    \hline
    (log$_{10}$-) $E_{\rm k,iso}$ [erg] & $[47, 57]$ \\
    (log$_{10}$-)  $n_{\rm ISM}$ [cm$^{-3}$] & $[-6, 2]$\\
    (log$_{10}$-) $\epsilon_{\rm e}$ & $[-4, 0]$\\
    (log$_{10}$-) $\epsilon_{\rm B}$ & $[-8,0]$ \\
    $p$ & $[2.01,3]$ \\
    Viewing angle [$^\circ$] & $[0,36]$\\
    $\theta_{\rm j}$ [$^\circ$] & $[0.6,20]$ \\
     $\sigma_{\rm sys}$ [mag] & $[0,2]$\\
    \hline
   $t_{\rm inj}$ [day] & $[0.07,0.3]$, $[0.7,1.3]$, $[2.5,3.5]$\\
   $\Delta t_{\rm inj}$ [day] & $[0.001,0.3]$, $[0.001,0.4]$, $[0.001,1]$\\
    (log$_{10}$-) $\Delta \log E_{\rm k,iso}$ & $[0,3]$, $[0,4]$, $[0,4]$\\
    \hline
    \end{tabular}
\label{NMMAprior}    
\end{table}

\begin{figure}
    \centering
    \includegraphics[width=0.5\textwidth]{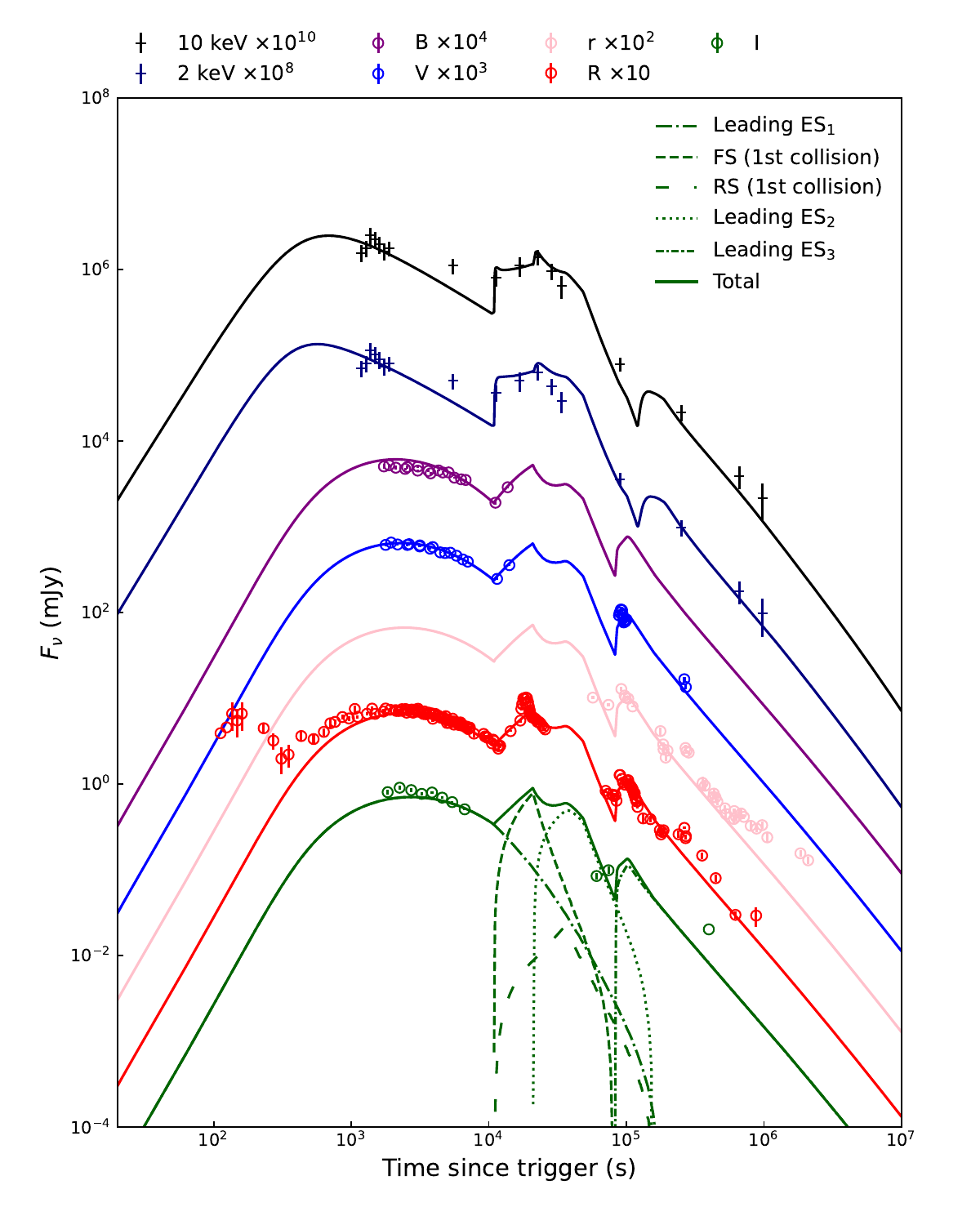}
    \caption{Fit to the GRB afterglow within the scenario of shock collision. The dash-dotted line is the emission of the outermost external shock (marked as ES$_1$) propagating in the environment for the first outflow. The dashed and loosely dashed line represents emission from the forward shock (FS) and the reverse shock (RS) during the collision, respectively. The dotted line is the emission of the external shock at the second stage (marked as ES$_2$) after the FS crossing the leading shell, which is refreshed again and marked as ES$_3$ at the third stage (densely dashed-dotted lines) by the second collision near 1~day.
    Since the crossing timescales of the FS/RS during the second collision are relatively shorter and their flux contribution are subdominant, they are not shown by individual lines for clarity. The total flux from all emission components is given by the solid line for each band. 
    }
    \label{fig:collision}
\end{figure}

\begin{table*}[h]
    \caption{Parameters employed in our Bayesian inferences. We report maximum-likelihood values at 95\% credibility for various subdatasets considered.}
    \label{tab:posterior_nmma_analysis}
    \centering
    \begin{tabular}{lcccc}
    \hline
    Parameter & No injection & 1-injection & 2-injection & 3-injection \\
    \hline
    (log$_{10}$-) On-axis isotropic equivalent energy $E_{\rm k,iso}$ [erg] & $52.53^{+0.43}_{-0.02}$ & $53.72^{+0.95}_{-0.58}$ & $53.85^{+0.80}_{-0.18}$ & $54.12^{+0.93}_{-0.36}$ \\
    (log$_{10}$-) Ambient medium’s density $n_{\rm ISM}$ [cm$^{-3}$] & $-0.31^{+2.3}_{-0.22}$ & $-0.32^{+2.32}_{-3.45}$ & $0.82^{+0.81}_{-3.8}$ & $-0.84^{+1.53}_{-4.65}$ \\
    (log$_{10}$-) Energy fraction in electrons $\epsilon_{\rm e}$ & $-0.01^{+-0.0}_{-0.44}$ & $-0.93^{+0.45}_{-0.84}$ & $-0.61^{+0.14}_{-0.81}$ & $-0.89^{+0.29}_{-0.96}$ \\
    (log$_{10}$-) Energy fraction in magnetic field $\epsilon_{\rm B}$ & $-1.86^{+1.66}_{-0.35}$ & $-3.71^{+2.21}_{-1.23}$ & $-5.26^{+2.31}_{-0.48}$ & $-4.25^{+2.78}_{-0.9}$ \\
    Electron distribution power-law index $p$ & $2.29^{+0.09}_{-0.06}$ & $2.68^{+0.09}_{-0.12}$ & $3.00^{+0.00}_{-0.03}$ & $3.00^{+0.00}_{-0.06}$ \\
    Viewing angle [degree] & $7.15^{+8.85}_{-0.5}$ & $13.44^{+3.99}_{-9.94}$ & $0.13^{+9.61}_{-0.13}$ & $14.79^{+4.38}_{-10.59}$ \\
    Jet core opening angle $\theta_{\rm j}$ [degree] & $9.75^{+10.25}_{-0.2}$ & $16.62^{+3.38}_{-10.76}$ & $16.66^{+3.34}_{-5.88}$ & $15.93^{+4.07}_{-10.92}$ \\
    \hline
    Time of $1^{\rm st}$ energy injection $t_{\rm inj,1}$ [day] & - & $0.128^{+0.021}_{-0.015}$ & $0.126^{+0.012}_{-0.015}$ & $0.129^{+0.011}_{-0.013}$ \\
    Duration of $1^{\rm st}$ energy injection $\Delta t_{\rm inj,1}$ [day] & - & $0.070^{+0.024}_{-0.033}$ & $0.075^{+0.019}_{-0.021}$ & $0.075^{+0.014}_{-0.018}$ \\
    (log$_{10}$-) $1^{\rm st}$ Fractional energy injected $\Delta \log E_{1}$ & - & $0.51^{+0.057}_{-0.065}$ & $0.512^{+0.059}_{-0.039}$ & $0.515^{+0.059}_{-0.033}$\\
    Time of $2^{\rm nd}$ energy injection $t_{\rm inj,2}$ [day] & - & - & $0.974^{+0.031}_{-0.077}$ & $0.926^{+0.074}_{-0.042}$\\
    Duration of $2^{\rm nd}$ energy injection $\Delta t_{\rm inj,2}$ [day] & - & - & $0.007^{+0.102}_{-0.006}$ & $0.083^{+0.037}_{-0.082}$\\
    (log$_{10}$-) $2^{\rm nd}$ Fractional energy injected $\Delta \log E_{2}$ & - & - & $0.166^{+0.008}_{-0.043}$ & $0.204^{+0.013}_{-0.034}$\\
    Time of $3^{\rm rd}$ energy injection $t_{\rm inj,3}$ [day] & - & - & - & $2.55^{+0.301}_{-0.05}$ \\
    Duration of $3^{\rm rd}$ energy injection $\Delta t_{\rm inj,3}$ [day] & - & - & - & $0.282^{+0.294}_{-0.281}$ \\
    (log$_{10}$-) $3^{\rm rd}$ Fractional energy injected $\Delta \log E_{3}$ & - & - & - & $0.053^{+0.02}_{-0.044}$ \\
    \hline
    Systematic error $\sigma_{\rm sys}$ [mag] & $0.38^{+0.03}_{-0.04}$ & $0.24^{+0.03}_{-0.02}$ & $0.19^{+0.03}_{-0.01}$ & $0.17^{+0.03}_{-0.01}$ \\
    \hline
    (ln-) likelihood ratio $\ln\Lambda$ & $-128.52$ & $-41.35$ & reference & $15.63$ \\
    (ln-) Bayes factor $\ln\mathcal{B}$ & $-110.44 \pm 0.24$ & $-31.56 \pm 0.33$ & reference & $6.51\pm 0.34$ \\
    \hline
    \end{tabular}
\end{table*}

\setlength{\tabcolsep}{3pt}
\begin{table}[h]
\centering
\caption{Parameters employed in the afterglow fitting within the framework of multiple shell collisions.}
\label{tab:fit_collision}
\begin{tabular}{ccccccccccc}
\toprule
\multirow{2}{*}{\small Param.} & \multicolumn{3}{c}{\small{Shells}} & 
\multicolumn{3}{c}{\small{1st Collision}} &
\multicolumn{3}{c}{\small{2nd Collision}} \\
\cmidrule(lr){5-7} \cmidrule(lr){8-10}
 &\small{1st} & \small{2nd} & \small{3rd} & \small{FS} & \small{RS} & \small{ES} & \small{FS} & \small{RS} & \small{ES} \\
\midrule
$E_{\rm k,iso}$  & \multirow{2}{*}{$1.70^{+0.12}_{-0.11}$} & \multirow{2}{*}{10.0} & \multirow{2}{*}{8.0} & & & & & & \\
\small{($10^{53}$~erg)} & & & & & \\
\midrule
$\Gamma_0$ & $97.72^{+2.28}_{-4.40}$ & 39.0 & 24.5 & & & & & & \\
\midrule
$\theta_{\rm j}$ \small{(rad)} & \multicolumn{9}{c}{$0.10^{+0.02}_{-0.02}$} \\
\midrule
$n_{\rm ISM}$  & \multicolumn{9}{c}{\multirow{2}{*}{$1.05^{+0.27}_{-0.16}$}} \\
\small{(cm$^{-3}$)} & & & \\
\midrule
$\epsilon_{\rm e}$ & $0.09^{+0.01}_{-0.00}$ & & & 0.18 & 0.10 & 0.11 & \multicolumn{2}{c}{0.10}  & 0.12 \\
\midrule
$\epsilon_{\rm B}$ & $2.51^{+0.31}_{-0.32}$ & & & \multicolumn{2}{c}{\multirow{2}{*}{0.10}}  & \multirow{2}{*}{2.0} & \multicolumn{2}{c}{\multirow{2}{*}{0.10}} & \multirow{2}{*}{2.0} \\
\small{($10^{-3}$)} & & & & \\
\midrule
$p$ & $2.07^{+0.00}_{-0.00}$ & & & \multicolumn{5}{c}{2.3} & 2.4 \\
\midrule
$\eta$ & \multicolumn{2}{c}{0.12} &  & & & & & & \\
\bottomrule 
\end{tabular}
\end{table}

\subsection{Refreshed Shocks Scenario: Shell Collisions}
\label{sec:refreshed-shocks}

The multiple pulses in the prompt emission phase suggest that there are multiple mass shells launched by the central engine. We consider that the earliest launched shell (outermost in the radial direction, or the external shock) has a bulk LF of $\Gamma_{0,1}\gg1$, which starts to decelerate at $\mathcal{R} \ge \mathcal{R}_{\rm dec}$. Later shells ejected with bulk LFs $\Gamma_{0,i} \gg 1$, with $i=(2,3,4)$, propagate almost unimpeded and may collide with the outer shell in succession, thereby dissipating their kinetic energy and causing potential rebrightening episodes~\citep[e.g.,][]{Kumar-Piran-00,Sari2000,Zhang-Meszaros-02}. The correspondence between the prompt pulses and shells, and the timing analyses on the temporal features, is crucial to identify whether collisions could occur, which has been applied in the explanation of GRB 140304A~\citep{Laskar18}, GRB 240529A~\citep{Sun24}, GRB 060729~\citep{Geng25a}, and (most recently) GRB 250221A~\citep{Angulo-Valdez+25}. When the two shells collide, another set of forward and reverse shocks develops, where the FS heats the already relativistically hot material of the outer shell, while the RS heats the inner shell.

The first optical rebrightening episode starts at $T_1 \approx 1.1 \times 10^4$\,s. Since the rebrightening occurs after the peak of the light curve, it is clear that the outermost shell has already decelerated. In general, for 
an ultrarelativistic shell with dynamics $\Gamma^2\propto \mathcal{R}^{-m}$, where $m=0$ for a coasting shell at $\mathcal{R}<\mathcal{R}_{\rm dec}$ and 
$m=3-k$ for a decelerating shell with self-similar evolution at $\mathcal{R} > \mathcal{R}_{\rm dec}$, the on-axis arrival time of radiation is $T_z\equiv T/(1+z) \approx \mathcal{R}/2(1+m)\Gamma^2c$. Then, given $\hat T_1 \equiv T_1/T_{\rm dec}$ ($T_{\rm dec} \approx 2 \times 10^3$~s if we treat the peak time of the first onset bump as the deceleration time), 
we can calculate the radius at which shocks develop in the two colliding shells, such that (e.g., \citealt{Laskar18,Angulo-Valdez+25})
\begin{equation}
\label{eq:R1hat}
    \mathcal{\hat R}_1 \equiv \frac{\mathcal{R}_1}{\mathcal{R}_{\rm dec}} = [(4-k) \hat T_1]^{1/(4-k)} \simeq 2.2 \quad (k=0)\,,
\end{equation}
with the assumption that the material that makes the dominant contribution to the rebrightened emission is moving with LF $\approx \Gamma_1$. To reach radius $\mathcal{R}_1$, the outer shell takes lab-frame time of $t_1 = t_{\rm em,1} + \mathcal{R}_1 / c + (\mathcal{\hat R}_1^{4-k} - 1) T_{\mathrm{dec},z}/(4-k)$, where $t_{\mathrm{em},i}$ marks the launching time of the $i$-th shell. To reach the same radius, the inner shell takes lab-frame time $t_1 = t_{\rm em,2} + \mathcal{R}_1/\beta_{0,2}c \approx t_{\rm em,1} + \delta t_{\rm em, 12} + (1+1/2\Gamma_{0,2}^2)\mathcal{R}_1/c$, where $t_{\rm em,1} < t_{\rm em,2} < T_{90}/(1+z) \approx 83$\,s and $\delta t_{\rm em, 12}$ is the waiting time between the ejection of the first two mass shells. Since $\delta t_{\rm em,12}(1+z)/T_{\rm dec}\ll1$, it can be neglected when equating the lab-frame arrival times of the two shells at $\mathcal{R}_1$, which yields an estimate of the contrast between the initial LFs of the two shells,
\begin{equation}
\label{eq:Gamma_ratio}
    \frac{\Gamma_{0,1}}{\Gamma_{0,2}}\approx\sqrt{\frac{\mathcal{\hat R}_1^{4-k}-1}{(4-k)\mathcal{\hat R}_1}} 
    \approx\frac{[(4-k)\hat T_1]^{(3-k)\over2(4-k)}}{\sqrt{4-k}}
    \approx 1.6 \quad (k=0)\,,
\end{equation}
where the second approximation assumes that $\mathcal{\hat R}_1\gg1$. During the collision, the bulk LF of the FS can be approximated as the velocity of the merged two shells (e.g., \citealt{Piran99}),
\begin{equation}
\Gamma_{\rm F} \simeq \left(\frac{\Gamma_{1} m_1+\Gamma_{0,2} m_2}{m_1/\Gamma_{1} + m_2/\Gamma_{0,2}} \right)^{1/2} \in [\Gamma_1, \Gamma_{0,2}],
\end{equation}
according to the conservation of energy and momentum, where $m_1$ and $m_2$ are the masses of the two shells. 
The comoving thickness of the leading shell can be approximated as $\Delta_1 \approx \eta \mathcal{R}_1/\Gamma_1 \approx 2 \eta \Gamma_{0,2}^2 c T_{1,z} / \Gamma_1$ at $T_1$, where $\eta$ is a geometry factor that could be taken as $\eta=1/4(3-k)$ from the consideration of particle number conservation for a homogeneous shell.

Let $\beta_{\rm F1}$ denote the relative velocity between the FS and the leading shell.
We may expect that the optical flare would reach its peak luminosity when the FS has crossed the entire leading shell after an observational duration of~\citep{Geng25a}
\begin{eqnarray}
\label{eq:time_rise}
T_{\rm flare,1} - T_1 &\approx& (1+z) \Gamma_{\rm F} (1-\beta_{\rm F} \cos(\min[\theta_{\rm j},1/ \Gamma_{\rm F}])) \frac{\Delta_1}{\beta_{\rm F1} c} \\ \nonumber
&\approx& 2 \eta \Gamma_{0,2}^2 /(\Gamma_{\rm F} \Gamma_1 \beta_{\rm F1}) T_1 \\ \nonumber
& \in & [\frac{\Gamma_{0,2}}{\Gamma_{0,1}} \hat{\mathcal{R}}_1^{(3-k)/2}, \frac{\Gamma_{0,2}^2}{\Gamma_{0,1}^2} \hat{\mathcal{R}}_1^{3-k}] \frac{2 \eta}{\beta_{\rm F1}} T_1
\end{eqnarray}
for $\theta_j > 1/\Gamma_{\rm F}$, where the last range is obtained by taking the two extremes for $\Gamma_{\rm F}$ in Equation (3). As observation of the $R$-band data gives a dimensionless timescale of $\delta \hat t = (T_{\rm flare,1} - T_1)/T_1 \approx 0.5$ for the rising phase of the first optical flare, Equations (\ref{eq:R1hat}--\ref{eq:time_rise})
suggest that $\eta \le 0.12$ since $\beta_{\rm F1} < 1$. 

The analyses on the shock dynamics above show the conditions for the shell collision. We further conduct more accurate numerical calculations to verify that this scenario can well explain the multiple peaks in GRB~250129A with a set of reasonable parameters.  
Before collision, the dynamics of the external shock of the first outflow is described by a generic model based on energy conservation~\citep{Huang99,Peer12}. During collision, the dynamics of the FS/RS system is described by the mechanical model that incorporates the conservation of energy and momentum~\citep{Beloborodov06,Geng25b}.

We calculate the time-dependent electron spectrum heated by each shock by solving the continuity equation in energy space \citep{Geng18}, and derive the resulting synchrotron and synchrotron self-Compton emission using standard formulae~\citep{Rybicki79}.
In our calculations, the environment is assumed to be ISM-type for simplicity, and the accelerated post-shock electron spectrum is generalized as a Maxwellian component and a power-law component for each shocked/emitting region.
This hybrid spectrum is characterized by an energy fraction parameter defined as the nonthermal energy fraction of the total, and we set it to a typical value of $\sim 0.4$ following earlier studies for simplification~\citep{Giannios09,Gao24}.
We fit the early afterglow within $10^4$~s, produced by the outermost shell, using the standard Bayesian approach to derive its parameters: $E_{\rm k,iso}$, $\Gamma_{0,1}$, $\theta_{\rm j}$, $n_{\rm ISM}$, $\epsilon_{\rm e}$, $\epsilon_{\rm B}$, and $p$. 
We assume that subsequently launched shells share the same jet opening angle $\theta_{\rm j}$ and circumburst density $n_{\rm ISM}$ as the outermost shell.
To account for the observed optical flares, we introduce two additional shells that are launched after the outermost shell. Since their initial bulk Lorentz factors are only of order several tens, deceleration is not yet significant; these shells therefore move at nearly constant speed until they catch up and collide with the preceding outermost shell, producing the optical flares.
During each collision, the equation of state of the material ahead of the FS is determined by calculating the averaged LF of material accumulated by the leading shock with adiabatic cooling accounted~\citep[e.g.,][]{Nava13}.
The parameters of the subsequent two shells and the relevant collisions are chosen according to the discussions above (e.g., $p$ in Sec.~\ref{sec:empirical}) and several trials around typical values (cf. Table~\ref{tab:fit_collision}).

As shown in Fig.~\ref{fig:collision}, the multiwavelength afterglow and two significant rebrightenings could be well fitted with a set of parameters listed in Table~\ref{tab:fit_collision}.  
For the first collision that generates the intense optical flare, the adopted $\Gamma_{0,1} \approx 100$ and $\Gamma_{0,2} \approx 40$ generally align with Equation~\ref{eq:Gamma_ratio}.  
The optical flare is mainly contributed by the FS as the FS sweeps and re-accelerates the material collected from the environment by the leading shell. After each collision, the FS crosses the entire leading shell, and the merged shell evolves into a new leading/outmost external shock propagating in the environment.
The shell collision scenario provides details of the shock interaction in comparison with the traditional energy injection scenario, while the fitting results of both scenarios are consistent with each other.

\label{sec:density-enhancements}






\section{Conclusion and discussion}
\label{sec:discussion}

We explored whether the light curve of GRB~250129A can be reproduced by a physically consistent set of microphysical parameters commonly adopted in the literature, and a sequence of refreshed shocks produced by collisions between shells with different bulk Lorentz factors ejected during the prompt emission phase (see Sec.~\ref{sec:refreshed-shocks}). To this end, we adopted a kinematic approach to calculate the dynamical evolution and radiation from such shell collisions, using a numerical code that will be publicly released by Geng et al. (in preparation). 

While the present analysis requires the introduction of additional microphysical parameters to describe the separate emission components and does not involve a full Bayesian inference, the observed flares can be broadly reproduced using a reasonable set of plausible shell parameters and typical microphysical parameters ($\epsilon_{\rm e} \approx 0.1$, $\epsilon_{\rm B} \approx 10^{-3}$–$10^{-2}$). This supports the physical plausibility of the refreshed-shock interpretation. While the present analysis demonstrates the physical plausibility of the refreshed-shock scenario using reasonable parameter values, future studies combining informative priors from particle-in-cell simulations with high-cadence and broader multiwavelength coverage (e.g., radio and GeV bands) could provide tighter constraints on the system parameters.

Several lines of evidence support the refreshed-shock scenario considered here. First, the prompt emission of GRB~250129A displayed four distinct emission episodes in gamma-rays and at least one in optical, suggesting the ejection of multiple relativistic shells. Second, the optical rebrightenings exhibit a steep temporal index, in contrast to the shallower behaviors seen in some GRBs~\citep{Filgas11,Nardini14} at comparable epochs ($\sim 10^4$~s). Such steepness supports the interpretation that these are new emission components associated with shell collisions, rather than a gradual energy injection process arising from a continuous Lorentz-factor stratification or prolonged central-engine activity.  
The latter scenario is difficult to reconcile with the data, as it remains unclear how a central engine --- whether a magnetar or an accreting black hole --- could directly modulate the blast-wave energetics at such large radii. 
If multiple jets arise from accretion onto the central compact object, the properties of these shells (e.g., velocities, kinetic energies) could provide insights into the system’s accretion history and jet-launching mechanisms. However, the kinetic energy of the rear shells may not be well constrained owing to the lack of simultaneous, high-quality X-ray data and the degeneracy of microphysical parameters during the collision. 

An alternative explanation not explored in detail in this work involves variations in the external density profile encountered by the relativistic blast wave. Such density structures are plausible in the vicinity of massive progenitors, where strong stellar winds and eruptive mass-loss episodes can shape a complex circumburst environment \citep{Ramirez-Ruiz10}. However, previous studies indicate that even substantial density jumps are generally insufficient to reproduce sharp and intense optical flares such as the one observed at $\sim 10^{4}$ s in GRB~250129A \citep{Nakar07, vanEerten09, Gat13, Geng14}.
Similarly, off-axis viewing scenarios are disfavored, as they generally predict a smooth evolution and cannot support a sequence of multiple, rapidly rising rebrightening episodes \citep{Beniamini20Afterglow, Abdikamalov25Reverse, Wang25Forward}. 
As observational coverage continues to improve in a timely manner, the detection of more GRBs exhibiting similar behavior could provide deeper insight into the prolonged activity and complex dynamics of GRB central engines.

\section*{Data Availability}
The raw measurements are all public and can be retrieved at\href{https://skyportal-icare.ijclab.in2p3.fr/source/2025aji)}{~the GRB~250129A SkyPortal Public Page}. The images are available upon request.

\bibliographystyle{mnras}
\bibliography{references}

\appendix

\section{Contributions} 

D. Akl and S. Antier are the main contributors to the work, including the analysis and writing across all sections, coordination, and organization of contributors related to observational and high-energy results. M. Pillas was the chair of the GRB program and manages the article on behalf of GRANDMA.

Z. Wang, S. Antier, and A. Klotz are the main authors of Sec.~2.1. They received help from A. Lien. 

D. Akl and M. Molham are the main contributors and authors of Sec.~2.2. D. Akl and M. Molham are the authors of Sec.~2.3 related to Space observations with the help of S. Oates and A. de Ugarte Postigo, and D. Akl is the main author related to ground-based observations, with the coordination and contribution of S. Antier and S. Karpov.

N. Rakotondrainibe is the main contributor of Sec.~3.1. 
J.-G. Ducoin is the main contributor of Sec.~3.2.

J. Mao, R. Strausbaugh, E. Abdikamalov, and D. Berdikhan are the main authors of Sec.~4.1. P. Pang and H. Koehn are members of the paper-writing team and wrote Sec.~4.2. J.J.~Geng is the main contributor of Sec.~4.3. R. Gill and T. Laskar contributed to reviewing Sec.~4 as well as guiding the modeling of the rebrightening episodes.


D.~Akl, S.~Antier, C.~Adami, C.~Angulo-Valdez, V.~Aivazyan, L.~Almeida, C.~Andrade, Q.~André, S. Antier, V.~Aivazyan, J.-L.~Atteia, K.~Barkaoui, S. Basa, R. L. Becerra, P.~Bendjoya, E.~Bernaud, S.~Boissier, S.~Brunier, A.Y.~Burdanov, N. R. Butler, J.~Chen, F.~Colas, W. Corradi,  D.~Darson, D. Dornic, C.~Douzet, C.~Dubois, J-G Ducoin, A.~Durroux, D.~Dutton, P.-A.~Duverne, F.~Dux, E.G.~Elhosseiny, A.~Esamdin, A.V.~Filippenko, F. Fortin, M.~Freeberg, J.J.~Geng, M.~Gillon, N.~Globus, P.~Gokuldass, R.~Hellot, Y.H.M.~Hendy, Y.L.~Hua, R.~Inasaridze, A.~Iskandar, M.~Jelínek, S.~Karpov, A. Klotz, N. Kochiashvili, T.~du~Laz, A.~Le~Calloch, W.H.~Lee, S.~Leonini, X.Y.~Li, C.~Limonta,  J.~Liu, D.~L\'opez-C\'amara, F. Magnani, M.~Mašek, B.M.~Mihov, M. Molham, E.~Moreno Méndez, W.~Mercier, M.~Odeh, M. Pereyra, D.~Reichart, J.-P.~Rivet, F.~Romanov, F.~Sánchez-Álvarez, N.~Sasaki, B. Schneider, D.~Schlekat,  L.~Slavcheva-Mihova, A.~Simon, T.R.~Sun, A. Takey, D. Turpin, A.~de~Ugarte Postigo, L.T.~Wang, X.F.~Wang, A.M.~Watson, Y.S.~Yan, J.~de~Wit, S.~Z\'u\~niga-Fern\'andez, and W.~Zheng contributed to this work via observations and measurements taken by various GRANDMA/KNC observatories, their partners such as KAIT and Skynet, and the COLIBRÍ observatory. 

N. Globus, N. Guessoum, E. Abdikamalov, N. Kochiashvili contributed to the review and editing of the full article to improve its quality with the help of M. Coughlin, R. Becerra, J-G Ducoin, A.V. Filippenko, L. García-García.

\section{Acknowledgements (extended)}

D. Akl is supported by Tamkeen under the NYU Abu Dhabi Research Institute grant CASS.

E.A.'s and D.B.'s work was funded by the Science Committee of the Ministry of Science and Higher Education of the Republic of Kazakhstan (Grant No. AP26103591). E.A. acknowledges support by the Nazarbayev University Faculty Development Competitive Research Grant Program (No. 040225FD4713).

We would like to thank the Pierre Auger Collaboration for the use of its facilities.

This work made use of data supplied by the UK \textit{Swift} Science Data Centre at the University of Leicester. 

C.A.V. acknowledges support from a SECIHTI fellowship. 

D.G.S. acknowledges support from a NASA North Carolina Space Grant Undergraduate Research Scholarship.

B.M.M. and L.S.-M. research was carried out with the help of infrastructure renovated under the National Roadmap for Research Infrastructure (2020-2027), financially coordinated by the Ministry of Education and Science of Republic of Bulgaria (agreement D01-326/04.12.2023).

J.G. is supported by the National Natural Science Foundation of China (grant Nos. 12273113, 12393812, 12393813, and 12321003), the Strategic Priority Research Program of the Chinese Academy of Sciences (grant No. XDB0550400), and the Youth Innovation Promotion Association (2023331).

J.M. is financially supported by the National Key R\&D Program of China (2023YFE0101200), Natural Science Foundation of China 12393813, and the Yunnan Revitalization Talent Support Program (YunLing Scholar Project). 

S.K. acknowledges financial support from the European Union and the Czech Ministry of Education, Youth and Sports (Project No. CZ.02.01.01/00/22\_008/0004632 -- FORTE).

H.K. and T.D. acknowledge funding from the EU Horizon under ERC Starting Grant No. SMArt-101076369.

P.T.H.P. is supported by the research program of the Netherlands Organization for Scientific Research (NWO) under grant number VI.Veni.232.021. 

N.G. and L.G.G. gratefully acknowledge the support of the Simons Foundation (MP-SCMPS-00001470, N.G., L.G.G).

A.M.W. is grateful for financial support from UNAM/DGAPA/PAPIIT project IN109224.

F.D.R. is grateful to iTelescope.Net for giving him some complimentary points for observing time to use their remote telescopes; the AAVSO for granting him a complimentary membership.

Members of TNOT acknowledge financial support from the Natural Science Foundation of Xinjiang Uygur Autonomous Region under No. 2024D01D32; Tianshan Talent Training Program grant 2023TSYCLJ0053. X. Wang is supported by NSFC (12288102, 12033003), the Tecent Xplorer Prize, the Ma Huateng Foundation, and the New Cornerstone Science Foundation through the XPLORER PRIZE.

A.V.F.'s research group at UC Berkeley acknowledges financial assistance from  Gary and Cynthia Bengier, Clark and Sharon Winslow, Alan Eustace and Kathy Kwan (W.Z. is a Bengier-Winslow-Eustace Specialist in Astronomy), and numerous other donors.  KAIT and its ongoing operation were made possible by donations from Sun Microsystems, Inc., the Hewlett-Packard Company, AutoScope Corporation, Lick Observatory, the U.S. National Science Foundation, the University of California, the Sylvia \& Jim Katzman Foundation, and the TABASGO Foundation. Research at Lick Observatory is partially supported by a generous gift from Google. 


TAROT was built with the support of the Institut National des Sciences de l'Univers, CNRS, France. TAROT is funded by the CNES and thanks to the help of the technical staff of the Observatoire de Haute-Provence, OSU-Pytheas.

We thank the staff of the Observatorio Astron\'omico Nacional on Sierra San Pedro M\'artir.

The AbAO team acknowledges Shota Rustaveli National Science Foundation of Georgia (SRNSFG). This work was supported by SRNSFG grant FR-24-7713.

Based in part on observations made at Observatoire de Haute Provence (CNRS), France, with MISTRAL. This research has made use of the MISTRAL database, operated at CeSAM (LAM), Marseille, France. 

The Virgin Islands Robotic Telescope (VIRT) is located at the University of the Virgin Islands' (UVI's) Etelman Observatory on St. Thomas, U.S. Virgin Islands. UVI operates VIRT and uses it for robotic optical observations in support of time-domain and transient astronomy, including GRANDMA follow-up campaigns. The observatory acknowledges support in part from NASA EPSCoR 80NNSC22M0063, NSF AST 2319415, and NASA EPSCoR 80NSSC24M0112.

FRAM-Auger telescope operation is supported by the Czech Ministry of Education, Youth and Sports (projects MEYS LM2018105, LM2023047, and EU/MEYS CZ.02.01.01/00/22\_008/0004632).

The KAO-NRIAG team acknowledges financial support from the Egyptian Science, Technology \& Innovation Funding Authority (STDF) under grant number 45779. M. Molham acknowledges the support provided by the Women for Africa Foundation, whose contribution enabled the tools and methodologies applied in this work. KAO thanks M.~Abdelkareem for conducting observations.

The AST3-3 and YAHPT team would like to express their sincere thanks to the staff of the Yaoan observation station.

M.G. and E.J. are FNRS-F.R.S. Research Directors. J.d.W. and MIT gratefully acknowledge financial support from the Heising-Simons Foundation, Dr. and Mrs. Colin Masson, and Dr. Peter A. Gilman for Artemis, the first telescope of the SPECULOOS network situated in Tenerife, Spain. The ULiege's contribution to SPECULOOS has received funding from the European Research Council under the European Union's Seventh Framework Programme (FP/2007-2013) (grant Agreement No. 336480/SPECULOOS), from the Balzan Prize and Francqui Foundations, from the Belgian Scientific Research Foundation (F.R.S.-FNRS; grant No. T.0109.20), from the University of Liege, and from the ARC grant for Concerted Research Actions financed by the Wallonia-Brussels Federation. 

Some of the data used in this paper were acquired with the DDRAGO instrument on the COLIBRÍ telescope at the Observatorio Astronómico Nacional on the Sierra de San Pedro Mártir. COLIBRÍ and DDRAGO are funded by the Universidad Nacional Autónoma de México (CIC and DGAPA/PAPIIT IN109418 and IN109224), and CONAHCyT (1046632 and 277901). COLIBRI received financial support from the French government under the France 2030 investment plan, as part of the Initiative d’Excellence d’Aix-Marseille Université-A*MIDEX  (ANR-11-LABX-0060 -- OCEVU and AMX-19-IET-008 -- IPhU), from LabEx FOCUS (ANR-11-LABX-0013), from the CSAA-INSU-CNRS support program, and from the International Research Program ERIDANUS from CNRS. COLIBRÍ and DDRAGO are operated and maintained by the Observatorio Astronómico Nacional and the Instituto de Astronomía of the Universidad Nacional Autónoma de México.

Views and opinions expressed are those of the authors only and do not necessarily reflect those of the European Union or the European Research Council. Neither the European Union nor the granting authority can be held responsible for them.

\section{X-ray Data Reduction: Additional Details}
\label{app:detailX-ray}

We rebinned the XRT light curve into noncontiguous segments. Each of these segments was treated as a distinct temporal window, within which we calculated the mean flux and its standard deviation. This approach allows us to combine temporally proximate measurements, hence increasing the SNR of temporally close data points and retaining the overall time evolution of the light curve. For each bin, we calculated the average time in the bin, average flux density, and the uncertainty of the flux density, which was calculated as the larger of the standard deviation of the flux values or the average propagated error. 

To convert the flux density from 10 keV to 2 keV, and derive the AB magnitudes \citep{Oke83Secondary}, we used the standard energy scaling relation for flux densities in frequency space,
\[
F_{\nu}(E_2) = F_{\nu}(E_1) \left( \frac{E_2}{E_1} \right)^{\Gamma - 1},
\]
where $E_1 = 10~\mathrm{keV}$, $E_2 = 2~\mathrm{keV}$, and \( \Gamma - 1 = 0.94 \) for the photon index \( \Gamma = 1.94 \), obtained from the online Swift Time-Averaged Spectrum available \href{https://www.swift.ac.uk/xrt_spectra/01285812/}{here}.

This conversion is motivated by the fact that \textit{Swift}-XRT's effective area, and hence its sensitivity, declines significantly at higher energies, making 10~keV flux measurements noisier and less reliable. We selected 2~keV as a reference energy for spectral scaling and magnitude conversion, as it lies near the logarithmic midpoint of the \textit{Swift}-XRT energy band (0.3--10~keV). 

\begin{figure}[h!]
    \centering
    \includegraphics[width=1\linewidth]{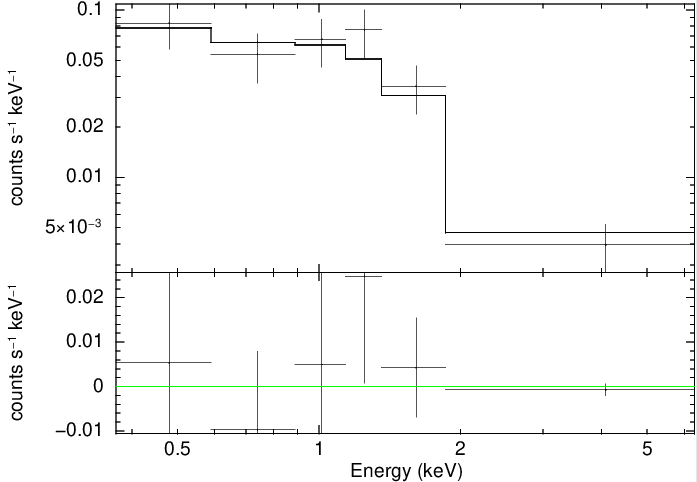}  
    \includegraphics[width=1\linewidth]{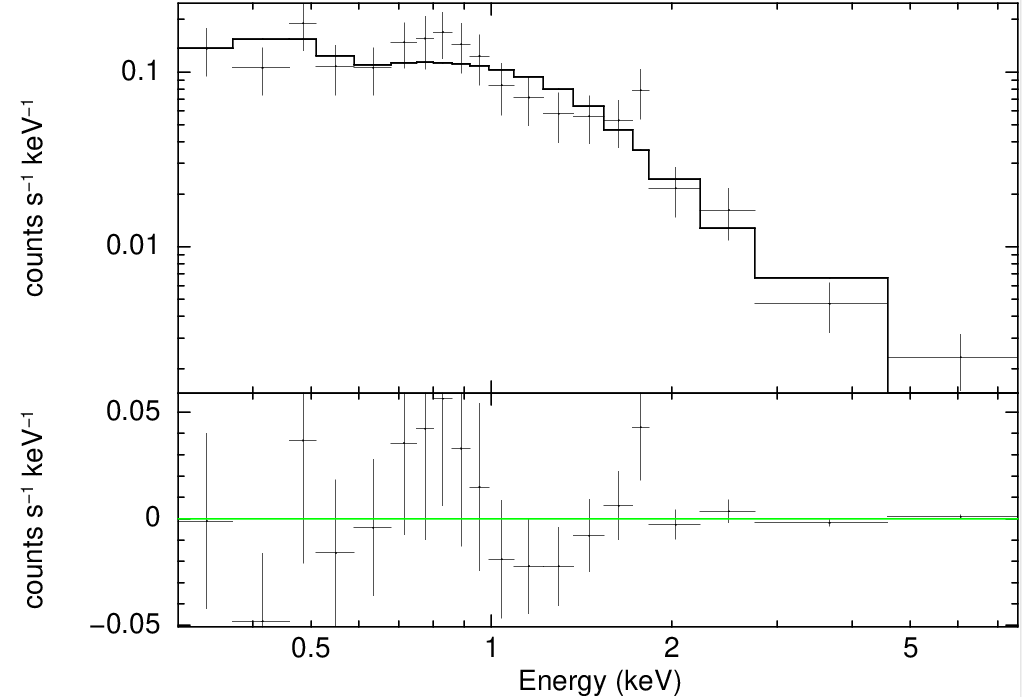}  
    
    \caption{X-ray spectra fitted with a power-law (PL) model. \textit{Top panel:} Spectrum from the decay interval of observation ID \texttt{01285812001}, corresponding to $T - T_0 = 0.1355$ to $0.1969$ days. \textit{Bottom panel:} Spectrum from the rise interval of the same observation, spanning $T-T_0 = 0.061$ to $0.1355$ days. 
    }
    \label{fig:xrtsourcesfits}
\end{figure}

The output flux values were subsequently converted to AB magnitudes using the standard relation 
\[
m_{AB} = -2.5 \log_{10} \left(\frac{F_{\nu}}{3631 \ \text{Jy}}\right)\, ,
\]
where $F_{\nu}$ is the flux at 2 keV in $Jy$ and \( 3631~\text{Jy} \) corresponds to the flux density of an object with an AB magnitude of 0. 

The data reduction and spectral analysis were conducted using \texttt{HEASoft v6.34} using the \textit{XRT tool xrtpipeline (v0.13.7)} on the three XRT spectra on 2025-01-29, with the latest updated calibration data files (\texttt{CALDB}, released on 2024 February 28). Spectral modeling and parameter estimation were performed using \texttt{XSPEC v12.14.1}, employing \textit{c}-statistics as the fitting method (\texttt{Fit.statMethod = "cstat"}), which is particularly suited for low-count Poisson-distributed X-ray data as it provides a more reliable parameter estimation compared to the traditional $\chi^2$ fitting, which assumes Gaussian errors.

XRT spectra (0.3--10 keV) were extracted from circular regions ($25''$) around the source center and background regions ($50''$) away from the source, grouped such that the source spectrum contained at least 1 count per bin. The spectra were fitted with an absorbed power-law model, with the Galactic column density ($N_{\rm H} =2.4 \times 10^{20}$ cm$^{-2}$) (Dickey \& Lockman 1990) fixed, using the TBabs model of Wilms, Allen \& McCray(2001), leaving only the photon index and the normalization as free parameters. Spectra were fit in the 0.3--10.0 keV energy range. Using the photon index from the fitting and the normalization, the spectral model used for fitting is defined as \texttt{tbabs*powerlaw}, where \texttt{tbabs} represents the Tübingen-Boulder absorption model, accounting for interstellar photoelectric absorption characterized by the neutral hydrogen column density ($N_{\rm H}$), while \texttt{powerlaw} describes the X-ray emission as a simple power-law function. 

The best-fit photon indices for the three source spectra within the first day were $\Gamma = 1.99 \pm 0.23$, $\Gamma = 2.15 \pm 0.15$, and $\Gamma = 1.87 \pm 0.18$, respectively. The corresponding flux at 2 keV was $2.75 \times 10^{-11}$ erg cm$^{-2}$ s$^{-1}$, $8.51 \times 10^{-11}$ erg cm$^{-2}$ s$^{-1}$, and $6.53 \times 10^{-11}$ erg cm$^{-2}$ s$^{-1}$, obtained from the fitted spectral parameters. 

\section{Detailed of the observations}
\label{app:Optical Observations}




In this section, we detail observations for GRB 250129A by GRANDMA and its associated partners. Early observations of the optical afterglow of GRB 250129A started on MJD 60704.20 using the TAROT-TCA and TAROT-TCH telescopes, two robotic 25 cm aperture telescopes located at the Calern Observatory (Observatoire de la Côte d’Azur) and La Silla Observatory, respectively; they detected the afterglow simultaneously at $T-T_0 = 106$~s.

The full observational campaign lasted 24.23 days. We use for this work the data obtained in $g'$, $r'$, $i'$, $z'$, $B$, $V$, $R$, $I$ bands for extracting the physical properties of the event. We provide in Table~\ref{tab:studyobservations} a list of all telescopes involved in the observational campaign, including the start and end times (relative to $T_0$) of the first and last observations made by each telescope. We also provide the filters/bands used during the entire campaign. 

The telescopes that contributed to this campaign are as follows: 
AbAO T-70 Telescope at Abastumani Observatory in Georgia, the ground-based ARTEMIS telescope of the SPECULOOS Northern Observatory (SNO) in Spain, Colibri at San Pedro Martir, C2PU at Calern Observatory, Euler telescope at La Silla, FRAM-Auger telescope at Pierre Auger Observatory in Argentina, KAIT at Lick Observatory in California, KAO at Kottamia Observatory in Egypt, T193/MISTRAL at Haute-Provence Observatory in France, NAO-2m at Rozhen National Astronomical Observatory in Bulgaria, OPD-0.6m at the  Pico dos Dias's Observatory in Brazil, Pic du Midi 1~m Telescope at Pic du Midi Observatory in France, Skynet Network, TAROT-TCA located at the Calern Observatory in France, TAROT-TCH located at La Silla, TNOT telescope located at the Nanshan Station of Xinjiang Astronomy Observatory in China, and YAHPT/AST3-3 in Yunnan Province in China. Our preliminary results were reported publicly through the General Coordinates Network (\href{https://gcn.nasa.gov/}{GCN}) \citet{2025GCN.39091....1S}, \citet{2025GCN.39096....1A}, \citet{2025GCN.39104....1W}, \citet{2025GCN.39106....1A}, \citet{2025GCN.39110....1S}, and \citet{2025GCN.39246....1A}. 

\paragraph*{KNC--}In addition to the professional network, GRANDMA activated its Kilonova-Catcher (KNC) citizen science program for further observations with amateurs’ telescopes. Seven amateur telescopes observed GRB 250129A, including nicknames MONTARRENTI, T-CAT, MLC, AITP, T11, T30, CH!CMOS, and CDK. Overall, the KNC telescopes started observing from $T-T0=$ 0.857 days up to $T-T_0 = 3.108$ 
days. Photometry of KNC images followed the methodology detailed in Section~\ref{sec:photometric_methods}. 

\paragraph*{Abastumani T-70--} AbAO T-70 Telescope at Abastumani Observatory in Georgia, with the main mirror diameter of 100 cm located on  Mt. Kanobili (altitude 1610 m above sea level; 41$^{\circ}$ 45$^{'}$ 17$^{''}$ N, 42$^{\circ}$ 49$^{'}$ 20$^{''}$ E) and equipped with a pl4240-ccd-camera-back-illum-63-5mm-shutter-grade-1 with $BVR_cI_c$ filters provides a field of view (FOV) of $30'$, with a readout of 10 s. The limiting magnitude in the $R_c$ filter with an exposure time of 1 min is 18.2 mag (AB system, 3$\sigma$).

\paragraph*{Artemis} SPECULOOS-North/Artemis \citep{Burdanov2022} is a 1.0~m Ritchey-Chr\'etien telescope located at the Teide Observatory (Tenerife, Spain). It is equipped with a thermoelectrically cooled 2k$\times$2k Andor iKon-L BEX2-DD CCD camera with a  scale of $0.35\arcsec$/pixel and a total FOV of $12\arcmin\times12\arcmin$. SPECULOOS-North is a twin of the SPECULOOS-South \citep{Jehin2018Msngr,Delrez2018,Sebastian_2021AA} and SAINT-EX \citep{demory2020} telescopes. 

\paragraph*{C2PU--} \href{https://www.oca.eu/fr/c2pu-accueil}{C2PU} is a two 1.04~m telescopes facility of Observatoire de la Côte d'Azur located at the Calern observing station in Southern France (longitude $06^\circ\,55'\,22.7''$~E, latitude  $43^\circ\,45'\,13.2''$~N, elevation  $1274$~m, IAU observatory code  $R87$) . For this photometric follow-up, the West ``Omicron'' telescope of C2PU has been used in its $f/3.17$ optical configuration (parabolic prime focus with a three-lens Wynne corrector). The camera used was a QHY600 from QHYCCD (CMOS sensor Sony IMX455 with $9600\times6422$ pixels of $3.76\times3.76$~$\mu$m) in binning $2\times2$. The resulting scale and FOV were $0.47''$/pixel, and $37.6'\times25.2'$. 

\paragraph*{COLIBRÍ--} \href{https://www.colibri-obs.org/}{COLIBRÍ} is a Franco-Mexican fast, robotic 1.3~m telescope located at the Observatorio Astronómico Nacional (OAN) in the Sierra de San Pedro Mártir, Baja California \citep{Basa2022}. COLIBRI used the blue channel of the DDRAGO science imager \citep{Langarica2024} and the OGSE camera.

OGSE is an ON Semi KAF-16803 front-illuminated CCD in an FLI ML 16803 package. The CCD is $4\mathrm{k}\times4\mathrm{k}$ with 9~$\mu$m pixels. The scale is $0.20''$/pixel, and the field is 13.6~arcmin square. OGSE is equipped with a fixed Baader red filter that transmits from 590 to 690 nm. This filter is narrower and redder than the standard SDSS/Pan-STARRS $r$ filter; the color term in the transformation is $-0.10(g-r)$.

DDRAGO is a two-channel image with the blue channel working in $g,r,i$ and the red in $z,y$ \citep{Langarica2024}. During COLIBRI observations, the red channel was not available. The blue channel uses a backside-illuminated, deep-depleted e2v 231-84 CCD in a Spectral Instruments 1110S package. The CCD is $4\mathrm{k} \times 4\mathrm{k}$ with 15~$\mu$m pixels. The scale is $0.38''$/pixel, and the FOV is 25.9~arcmin square. All of our observations were performed with the $r$ filter, which closely approximates SDSS/Pan-STARRS $r$; the color term in the transformation is smaller than $0.01(g-r)$.



\paragraph*{Euler--} The 1.2~m Euler Telescope is located at La Silla,  built and operated by the Geneva Observatory, Université de Genève, Switzerland. The Euler Telescope is equipped with three complementary instruments: the CORALIE spectrograph, the EulerCam (ECAM), and PISCO, a smaller telescope mounted piggyback on the Euler Telescope.

\paragraph*{FRAM-Auger--}FRAM-Auger is a fully robotic 30 cm $f/6.8$ telescope located at the Pierre Auger Observatory, Malargue, Argentina. The telescope is equipped with Moravian Instruments G4-16000 CCD, and $B$, $V$, $R$, and $I$ filters, and has a field of view of $60' \times 60'$, with a scale of $0.92''$/pixel.

\paragraph*{KAIT--} The 0.76~m Katzman Automatic Imaging Telescope (KAIT), located at the Lick Observatory, California (as part of the Lick Observatory Supernova Search \citep[LOSS;][]{Filippenko2001}), observed the field of GRB 250129A from $T-T_0=0.196$~days to $T-T_0=1.393$~days. Observations were performed in the $Clear$-band (close to the $R$ band; see \citealt{Li2003}) with a set of $60$~s exposure images. 

\paragraph*{KAO--} The KAO data presented in this study were acquired using the 1.88~m telescope at the Kottamia Astronomical Observatory (KAO), operated by the National Research Institute of Astronomy and Geophysics (NRIAG), Egypt \citep{10.1007/978-3-642-03325-4_16}. These observations utilized the Kottamia Faint Imaging Spectro-Polarimeter (KFISP), an instrument capable of imaging, spectroscopy, and polarimetry \citep{2022ExA....53...45A}. KFISP is mounted at the Cassegrain focus of the telescope and provides a field of view of approximately $8.2'\times 8.2'$. It is equipped with a $2048 \times 2048$ pixel liquid nitrogen-cooled CCD camera integrated within the KFISP optics, yielding a scale of $0.24''$/pixel.

\paragraph*{NAO-2m--} NAO-2m denotes the 2~m Ritchey-Chr\'etien telescope of the Rozhen National Astronomical Observatory, Bulgaria. For the present observations, the telescope was equipped with the multimode, two-channel focal reducer FoReRo-2 \citep{2000KFNTS...3...13J}, 2k~$\times$ 2k Andor iKon-L CCD cameras, and a Sloan set of filters. The gain is 1.0\,$\rm e^{-}\,ADU^{-1}$ for the blue channel CCD and 1.1\,$\rm e^{-}\,ADU^{-1}$ for the red channel one. For both channels, the pixel size is 0\farcs497 on the sky, and the FOV is $17' \times 17'$. Ten or more frames were acquired of the GRB 250129A field in each of the $g'r'i'$ bands with exposure times of 300 or 600\,s.

\paragraph*{OHP/MISTRAL--} \href{https://ohp.osupytheas.fr/mistral-spectro-imager/MISTRAL}{MISTRAL} (Multi-purpose InSTRument for Astronomy at Low resolution, \citealt{2024A&A...687A.198S}) is a Faint Object Spectroscopic Camera mounted at the folded Cassegrain focus of the 1.93 m telescope of the Haute-Provence Observatory (OHP). Present observations were made in the blue mode (400–800 nm) during four observing slots. We used $r'$ and $i'$  filters. The CCD is an ANDOR deep-depletion 2k$\times$2k CCD camera (iKon-L DZ936N BEX2DD CCD-22031) with 13.5~$\mu$m pixels. The cooling is made by a five-layer Peltier device. The operating temperature is $-90^\circ$C to $-95^\circ$C. The dark current is lower than 3 electrons/hour/pixel. 

\paragraph*{Pic Du Midi--}Pic du Midi 1~m Telescope at Pic du Midi Observatory in France. The telescope is equipped with two cameras. The first one is an ANDOR 2k$\times$2k CCD camera (iKon-L DZ936) with a spatial sampling of $0.5''$/pixel. The second camera is a \href{https://lytid.com/wp-content/uploads/2024/06/SIRIS-brochure-June-2024.pdf}{LYTID Siris} InGaAs sensor FPA 640$\times$512 pixels, with a 
readout noise of 5 electrons, operated at 77~K.

\paragraph*{Skynet--}
\href{https://skynet.unc.edu/}{The Skynet Robotic Telescope Network} (Skynet) is a fully automated global network of optical and radio telescopes used for both education and research. Founded and operated out of the University of North Carolina at Chapel Hill, Skynet is comprised of the  Panchromatic Robotic Optical Monitoring and Polarimetry Telescopes (PROMPT; \citealp{reichart_prompt_2005}), which were originally dedicated to rapid-response observations of GRBs \citep{reichart_unc-chapel_2006}, as well as various other optical telescopes from participating institutions across the globe. 

Approximately one minute after the \textit{Swift}-BAT trigger time (2025-01-29 4:45:09), Skynet's \texttt{Campaign Manager} software \citep{dutton_skynets_2022} automatically scheduled observations of the field of GRB 250129A on all available Skynet telescopes. The field of the GRB was subsequently observed using the 40 cm PROMPT-2, PROMPT-5, and PROMPT-6 telescopes at the Cerro Tololo Inter-American Observatory, the 40 cm PROMPT-MO-1 telescope at Meckering Observatory, the 40 cm MLC-RCOS16 telescope at the Montana Learning Center, the 50 cm OAUJ-CDK500 telescope at the Astronomical Observatory of the Jagiellonian University, and the 40 cm APUS-CDK24 telescope \citep{albin_look_2018} at the American Public University
System Observatory. 

\paragraph*{TNOT--} The Tsinghua-Nanshan Optical Telescope is a 80~cm reflector located at Nanshan Station of Xinjiang Astronomy Observatories (XAO), Chinese Academy of Sciences (CAS). The field of view was designed to $26.5'' \times 26.5''$, and the Bessel $BV$ system and Sloan $ugri$ system were chosen as the main filter set. The imaging system utilizes an Andor iKon-L DZ936 CCD camera with a 2048 $\times$ 2048 pixel array, and the pixel scale is $0.78''$. 
  
\paragraph*{YAHPT/AST3-3--}
The YaoAn High Precision Telescope (YAHPT) is an 80~cm automatic telescope manufactured by ASA and deployed at the YaoAn Astronomical Station in Yunnan Province, China. Equipped with a PIXIS 2024B camera, it provides a field of view of $11'' \times 11''$ and is dedicated to high-precision astrometric measurements of natural satellites and asteroids. The third Antarctic Survey Telescope (AST3-3, 68 cm Schimdt) was also temporarily deployed at the YaoAn station for commissioning before its final deployment to Dome A, Antarctica. AST3-3 is now equipped with a QHY411 camera featuring a Sony IMX-411 sensor, offering a field of view of $1.65^\circ \times 1.23^\circ$ with a 14k $\times$ 10k pixel array. The basic data processing for both YAHPT and AST3-3 follows the standard CCDPROC procedures, with astrometric calibration based on the Gaia DR3 catalog.

\begin{table}
    \centering
    \resizebox{0.95\linewidth}{!}{
    \begin{tabular}{|c|c|c|c|}
        \hline
        Instrument & $T-T_{0,{\rm start}}$ & $T-T_{0,{\rm end}}$ & Bands\\
        \hline
        \hline
        TAROT/TCH & $<$ 0.001 & 0.140 & $R$ \\
        FRAM-Auger & 0.008 & 0.090 & $R$ \\
        Skynet-PROMPT-5 & 0.019 & 3.165 & $B,V,R$ \\
        Skynet-PROMPT-6 & 0.019 & 3.166 & $B,V,R,I$ \\
        OHP/MISTRAL & 0.038 & 7.992 & $r',i'$ \\
        KAIT & 0.196 & 1.393 & $R$ \\
        Skynet-MLC & 0.201 & 0.233 & $R$ \\
        TNOT & 0.646 & 4.816 & $g',r'$ \\
        Skynet-OUAJ & 0.806 & 1.835 & $R$ \\
        KAO & 0.847 & 0.874 & $g',r',i',z'$ \\
        KNC-Montarrenti & 0.857 & 0.8570 & $R,I$ \\
        KNC-AITP & 1.034 & 3.092 & $V,g',r',i'$ \\
        Skynet-APUS & 1.061 & 1.128 & $R$ \\
        COLIBRI & 1.141 & 24.234 & $r'$ \\
        KNC-T11 & 1.317 & 1.329 & $R,V$ \\
        KNC-T30 & 1.529 & - & $R$ \\
        AbAO-T70 & 1.730 & 2.767 & $R$ \\
        KNC-CH!CMOS & 2.126 & 3.108 & $g',r'$ \\
        Skynet-PROMPT-MO & 2.500 & 2.507 & $R$ \\
        Skynet-PROMPT-2 & 3.011 & 3.168 & $I$ \\
        KNC-CDK & 3.018 & - & $V$ \\
        Euler & 3.115 & 10.158 & $R,i'$ \\
        AST3-3 & 3.648 & 4.573 & $g'$ \\
        YAHPT & 3.571 & 4.595 & $I$ \\
        NAO-2m & 4.884 & 4.951 & $g',r',i'$ \\
        T-CAT & 5.836 & - & $V,g',r'$ \\
        C2PU & 5.884 & 5.970 & $g'$, $r'$, $i'$, $z'$ \\
        Artemis & 6.009 & 7.039 & $g',r',z'$ \\
        PicDuMidi-T1M & 6.994 & 7.885 & $g',r',J$ \\
    
        \hline
    \end{tabular}
    }
    \caption{All observational instruments whose data have been analyzed and incorporated into the results presented in this study. The $T-T_0$ provided is in days. The filters listed correspond to the reference-system (like-)filters after cross-calibration and may differ from the instrumental filters used during acquisition. The transformation between systems is performed using color-term relations  (see section \ref{sec:photometric_methods}).}
    \label{tab:studyobservations}
\end{table}

\section{Energy injection and posterior distribution}
\label{nmma:injection}

 \paragraph{Energy Injection --} In \textsc{nmma} and \textsc{afterglowpy}, we implement energy injection prescription that linearly increases the log of energy from $t_{\rm inj}$ to $t_{\rm inj}+\Delta t_{\rm inj}$ as described by 
\begin{equation}
\label{eq:energy_injection}
\begin{aligned}
    \log_{10}E_{\rm k,iso}(t) =
    \begin{cases}
    \log_{10}E_{\rm k,iso} & t < t_{\rm inj}\\
    \log_{10}E_{\rm k,iso} +\Delta \log_{10}E\frac{t - t_{\rm inj}}{\Delta t_{\rm inj}} & t_{\rm inj} < t < t_{\rm inj} + \Delta t_{\rm inj}\\
    \log_{10}E_{\rm k,iso} +\Delta \log_{10}E & t > t_{\rm inj} + \Delta t_{\rm inj}.\\
    \end{cases}
\end{aligned}
\end{equation}

 \paragraph{Likelihood function definition --} To infer the posterior $p(\vec{\theta}|d)$ from the light-curve data $d$, we sample the model parameter space using the nested sampling algorithm as implemented in \textsc{pymultinest}~\citep{Feroz:2008xx,Buchner:2014nha}.
In particular, we use the likelihood function  $\ln \mathcal{L}(\vec{\theta}|d)$,
\begin{align}
\begin{split}
    &\ln \mathcal{L}(\vec{\theta}|d) =\\
    &\sum_{t_j} \biggl(-\frac{1}{2}  \frac{(m(t_j) - m^{\star}(t_j, \vec{\!\theta}\,))^2}{\sigma(t_j)^2 + \sigma_{\text{sys}}^2} + \ln(2\pi (\sigma(t_j)^2 + \sigma_{\text{sys}}^2)) \biggr)\ .
    \label{eq:likelihood}
\end{split}
\end{align}
This function compares the observed magnitudes $m(t_j)$ at times $t_j$ to the model predictions $m^{\star}(t_j, \vec{\theta})$ from \textsc{afterglowpy}. Here, $\sigma(t_j)$ represents the measurement uncertainties, and $\sigma_{\text{sys}}$ denotes the systematic uncertainty.

 \paragraph{Average chi-squared statistic --} Using posterior light curves derived from the ``early time'' data, we compute the average chi-squared statistic, $\langle\chi_j^2\rangle$, for each data point $d_j$,
\begin{equation}
\langle\chi_j^2\rangle = \int \left[\frac{(m(t_j) - m^{\star}(t_j, \vec{\!\theta}))^2}{\sigma(t_j)^2 + \sigma_{\text{sys}}^2}\right] p(\vec{\theta} | d) d\vec{\theta}\, .
\end{equation}

\begin{figure*}
    \centering
    \includegraphics[width=0.8\textwidth]{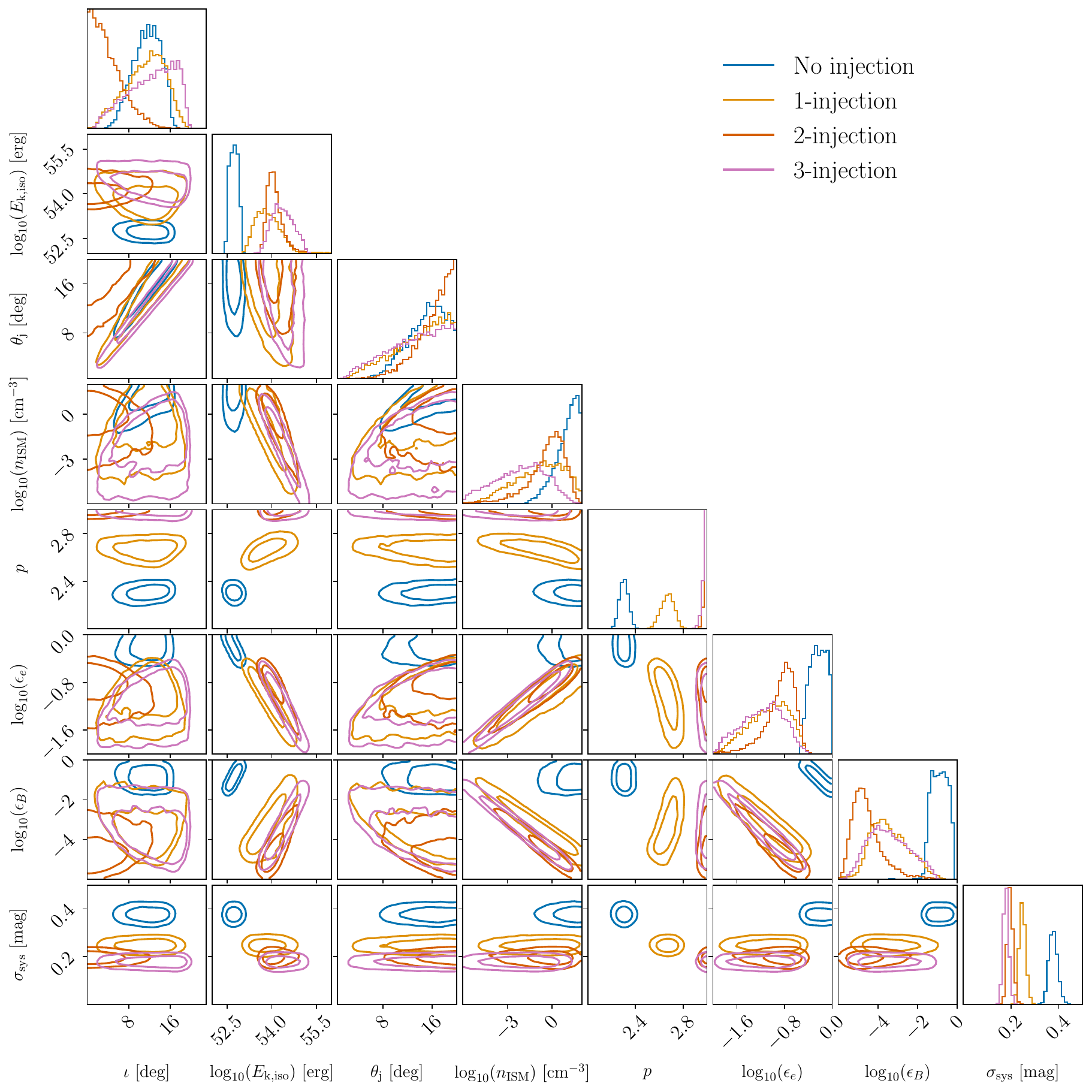}
    \caption{Posterior of the GRB afterglow for each of the models (see Section~\ref{sec:tophat-jet-nmma}) considered using \textsc{NMMA}. Priors are given in Table~\ref{NMMAprior}. The different times of rebrightening (e.g., energy injections) are consistent with the light curve (see Figure~\ref{fig:LC}).}
    \label{fig:nmma_corner}
\end{figure*}



\clearpage

\onecolumn

\section{Online Material (Data tables)}
\label{data:photo}
All the data used in this work can be found online at the GRB 250129A SkyPortal Public Page: \url{https://skyportal-icare.ijclab.in2p3.fr/source/2025aji} and in Tables~\ref{tab:xrt_data} and \ref{tab:all_observations}.

\small
\begin{longtable}{|cc|cc|c|ccc|c|}
\caption{X-ray data used in this work. ``Delay'' is the time interval between the start of the observation ($T_{\rm start}$) and the {\it Swift}
GRB trigger time (2025-01-29T04:45:09). We display both the unabsorbed flux densities and the corresponding computed AB magnitudes.} 
\label{tab:xrt_data} \\
\hline
\multicolumn{2}{|c|}{$T_{\rm start}$}  & \multicolumn{2}{c|}{Delay} & Band & \multicolumn{3}{c|}{Flux} & Instrument  \\
UT & MJD & (day) & (s)  & Central frequency  & AB Magnitude & Flux density ($\mu$Jy) & Error ($\mu$Jy)  & \\
\hline
\multicolumn{9}{|c|}{X-ray bands}\\
\hline
\endfirsthead

\hline
\multicolumn{2}{|c|}{$T_{\rm start}$}  & \multicolumn{2}{c|}{Delay} & Band & \multicolumn{3}{c|}{Flux} & Instrument  \\
UT & MJD & (day) & (s)  & Central frequency  & AB Magnitude & Flux density ($\mu$Jy) & Error ($\mu$Jy)  & \\
\hline
\multicolumn{9}{|c|}{X-ray bands}\\
\hline
\endhead

2025-01-29T05:04:53  & 60704.212 & 0.014 & 1185 & 2keV & 24.27 $\pm$  0.24 & 0.711 & 0.16 & Swift XRT \\
2025-01-29T05:06:32  & 60704.213 & 0.015 & 1284 & 2keV & 24.14 $\pm$  0.24 & 0.804 & 0.18 & Swift XRT \\
2025-01-29T05:08:05  & 60704.214 & 0.016 & 1376 & 2keV & 23.76 $\pm$  0.25 & 1.140 & 0.26 & Swift XRT \\
2025-01-29T05:10:00  & 60704.215 & 0.017 & 1491 & 2keV & 23.88 $\pm$  0.24 & 1.019 & 0.23 & Swift XRT \\
2025-01-29T05:11:54  & 60704.217 & 0.019 & 1605 & 2keV & 24.02 $\pm$  0.25 & 0.896 & 0.20 & Swift XRT \\
2025-01-29T05:14:10  & 60704.218 & 0.020 & 1742 & 2keV & 24.23 $\pm$  0.25 & 0.741 & 0.17 & Swift XRT \\
2025-01-29T05:16:28  & 60704.220 & 0.022 & 1879 & 2keV & 24.14 $\pm$  0.21 & 0.802 & 0.16 & Swift XRT \\
2025-01-29T06:16:42  & 60704.262 & 0.064 & 5493 & 2keV & 24.65 $\pm$  0.23 & 0.503 & 0.10 & Swift XRT \\
2025-01-29T07:53:25  & 60704.329 & 0.131 & 11297 & 2keV & 24.99 $\pm$  0.23 & 0.367 & 0.08 & Swift XRT \\
2025-01-29T09:23:47  & 60704.392 & 0.194 & 16719 & 2keV & 24.64 $\pm$  0.28 & 0.505 & 0.13 & Swift XRT \\
2025-01-29T11:04:33  & 60704.461 & 0.263 & 22765 & 2keV & 24.39 $\pm$  0.23 & 0.637 & 0.13 & Swift XRT \\
2025-01-29T12:41:03  & 60704.529 & 0.330 & 28554 & 2keV & 24.80 $\pm$  0.25 & 0.436 & 0.10 & Swift XRT \\
2025-01-29T14:09:34  & 60704.590 & 0.392 & 33865 & 2keV & 25.24 $\pm$  0.31 & 0.292 & 0.08 & Swift XRT \\
2025-01-30T05:41:43  & 60705.237 & 1.039 & 89795 & 2keV & 27.52 $\pm$  0.20 & 0.036 & 0.01 & Swift XRT \\
2025-02-01T02:16:30  & 60707.095 & 2.897 & 250281 & 2keV & 28.92 $\pm$  0.23 & 0.010 & 0.00 & Swift XRT \\
2025-02-05T20:07:18  & 60711.838 & 7.640 & 660130 & 2keV & 30.77 $\pm$  0.33 & 0.002 & 0.00 & Swift XRT \\
2025-02-09T12:25:46  & 60715.518 & 11.320 & 978038 & 2keV & 31.41 $\pm$  0.51 & 0.001 & 0.00 & Swift XRT \\

\hline
\end{longtable}
\small
\begin{longtable}{|cc|cc|c|c|c|c|c|}
\caption{UV-optical-IR observations of GRB 250129A. In column (2),  $T_{\rm (s)}$ is the time delay between the start of the observation and the \textit{Swift} GRB trigger time (2025-01-29T04:45:09), all in days. Column (4) gives apparent magnitudes or 5$\sigma$ upper limits in the AB system, without any correction. The filters listed correspond to the reference-system (or reference-like) filters after cross-calibration and may differ from the instrumental filters used during acquisition. The transformation between systems is performed using color-term relations (see Section~\ref{sec:photometric_methods}).
In Column (7), a cross (x) means we did use this data point for the Bayesian analysis.} 
\label{tab:all_observations} \\
\hline
\multicolumn{2}{|c|}{$T_{\rm start}$} & \multicolumn{2}{c|}{$T-T_ {\rm GRB}$} & Filter & Magnitude & Corrected Magnitude & Telescope & Analysis \\
\multicolumn{1}{|c}{UT} & \multicolumn{1}{c|}{MJD} & Day & Minute & & & & & \\
\multicolumn{2}{|c|}{(1)} & \multicolumn{2}{|c|}{(2)} & (3) & (4) & (5) & (6) & (7) \\
\hline
\endfirsthead

\hline
\multicolumn{2}{|c|}{$T_{\rm start}$} & \multicolumn{2}{c|}{$T-T_ {\rm GRB}$} & Filter & Magnitude & Corrected Magnitude & Telescope & Analysis \\
\multicolumn{1}{|c}{UT} & \multicolumn{1}{c|}{MJD} & Day & Minute & & & & & \\
\multicolumn{2}{|c|}{(1)} & \multicolumn{2}{|c|}{(2)} & (3) & (4) & (5) & (6) & (7) \\
\hline
\endhead

\multicolumn{9}{|c|}{UV band} \\
\hline
2025-01-29T04:47:42 & 60704.200 & 0.002 & 2.557 & $v$ & 17.58$\pm$0.45 & 17.48$\pm$0.45 & $UVOT$  & x \\
2025-01-29T05:06:08 & 60704.213 & 0.015 & 20.987 & $white$ & 18.19$\pm$0.06 & 18.04$\pm$0.06 & $UVOT$  &  \\
2025-01-29T05:07:41 & 60704.214 & 0.016 & 22.535 & $v$ & 16.73$\pm$0.21 & 16.64$\pm$0.21 & $UVOT$  & x \\
2025-01-29T05:09:21 & 60704.215 & 0.017 & 24.216 & $b$ & 17.10$\pm$0.17 & 16.97$\pm$0.17 & $UVOT$  & x \\
2025-01-29T05:09:46 & 60704.215 & 0.017 & 24.625 & $white$ & 18.14$\pm$0.10 & 17.99$\pm$0.10 & $UVOT$  &  \\
2025-01-29T05:09:58 & 60704.215 & 0.017 & 24.828 & $uvw1$ & 19.55$\pm$0.35 & 19.34$\pm$0.35 & $UVOT$  &  \\
2025-01-29T05:10:23 & 60704.216 & 0.018 & 25.242 & $u$ & 18.03$\pm$0.14 & 17.87$\pm$0.14 & $UVOT$  & x \\
2025-01-29T05:10:35 & 60704.216 & 0.018 & 25.449 & $v$ & 16.97$\pm$0.27 & 16.88$\pm$0.27 & $UVOT$  & x \\
2025-01-29T05:12:15 & 60704.217 & 0.019 & 27.112 & $b$ & 16.86$\pm$0.16 & 16.73$\pm$0.16 & $UVOT$  & x \\
2025-01-29T05:12:39 & 60704.217 & 0.019 & 27.516 & $white$ & 18.17$\pm$0.12 & 18.01$\pm$0.12 & $UVOT$  &  \\
2025-01-29T05:15:07 & 60704.219 & 0.021 & 29.983 & $b$ & 17.25$\pm$0.23 & 17.13$\pm$0.23 & $UVOT$  & x \\
2025-01-29T05:15:33 & 60704.219 & 0.021 & 30.407 & $white$ & 18.10$\pm$0.12 & 17.95$\pm$0.12 & $UVOT$  &  \\
2025-01-29T05:15:45 & 60704.219 & 0.021 & 30.609 & $uvw1$ & 19.36$\pm$0.36 & 19.15$\pm$0.36 & $UVOT$  &  \\
2025-01-29T05:15:58 & 60704.219 & 0.021 & 30.829 & $uvw2$ & >21.06 & >20.81 & $UVOT$  &  \\
2025-01-29T05:16:09 & 60704.220 & 0.022 & 31.012 & $u$ & 17.76$\pm$0.16 & 17.61$\pm$0.16 & $UVOT$  & x \\
2025-01-29T05:16:23 & 60704.220 & 0.022 & 31.246 & $v$ & 17.45$\pm$0.44 & 17.36$\pm$0.44 & $UVOT$  & x \\
2025-01-29T06:11:33 & 60704.258 & 0.060 & 86.405 & $b$ & 17.56$\pm$0.32 & 17.43$\pm$0.32 & $UVOT$  & x \\
2025-01-29T06:14:13 & 60704.260 & 0.062 & 89.080 & $b$ & 17.56$\pm$0.06 & 17.43$\pm$0.06 & $UVOT$  & x \\
2025-01-29T06:17:38 & 60704.262 & 0.064 & 92.492 & $white$ & 18.42$\pm$0.06 & 18.27$\pm$0.06 & $UVOT$  &  \\
2025-01-29T07:56:09 & 60704.331 & 0.133 & 191.002 & $v$ & 17.97$\pm$0.08 & 17.87$\pm$0.08 & $UVOT$  & x \\
2025-01-29T09:24:31 & 60704.392 & 0.194 & 279.381 & $b$ & 17.48$\pm$0.06 & 17.35$\pm$0.06 & $UVOT$  & x \\
2025-01-29T11:00:21 & 60704.459 & 0.261 & 375.207 & $uvm2$ & >24.78 & >24.52 & $UVOT$  &  \\
2025-01-29T11:11:42 & 60704.466 & 0.268 & 386.551 & $uvw1$ & 20.14$\pm$0.13 & 19.93$\pm$0.13 & $UVOT$  &  \\
2025-01-29T12:34:14 & 60704.524 & 0.326 & 469.097 & $u$ & 18.42$\pm$0.06 & 18.27$\pm$0.06 & $UVOT$  & x \\
2025-01-29T12:49:19 & 60704.534 & 0.336 & 484.179 & $b$ & 18.05$\pm$0.06 & 17.93$\pm$0.06 & $UVOT$  & x \\
2025-01-29T15:42:35 & 60704.655 & 0.457 & 657.449 & $uvw1$ & 21.59$\pm$0.28 & 21.38$\pm$0.28 & $UVOT$  &  \\
2025-01-29T15:57:33 & 60704.665 & 0.467 & 672.403 & $u$ & 19.20$\pm$0.06 & 19.05$\pm$0.06 & $UVOT$  & x \\
2025-01-29T18:47:18 & 60704.783 & 0.585 & 842.152 & $b$ & 19.03$\pm$0.18 & 18.91$\pm$0.18 & $UVOT$  & x \\
2025-01-30T02:50:30 & 60705.118 & 0.920 & 1325.356 & $u$ & 20.08$\pm$0.17 & 19.92$\pm$0.17 & $UVOT$  & x \\
2025-01-30T02:55:56 & 60705.122 & 0.924 & 1330.799 & $white$ & 20.57$\pm$0.13 & 20.42$\pm$0.13 & $UVOT$  &  \\
2025-01-30T03:04:06 & 60705.128 & 0.930 & 1338.952 & $v$ & 19.20$\pm$0.32 & 19.11$\pm$0.32 & $UVOT$  & x \\
2025-01-30T12:18:01 & 60705.513 & 1.314 & 1892.873 & $white$ & 20.48$\pm$0.13 & 20.33$\pm$0.13 & $UVOT$  &  \\
2025-01-31T10:02:37 & 60706.418 & 2.220 & 3197.471 & $white$ & 21.17$\pm$0.12 & 21.01$\pm$0.12 & $UVOT$  &  \\
2025-01-31T21:02:59 & 60706.877 & 2.679 & 3857.843 & $white$ & 21.16$\pm$0.13 & 21.01$\pm$0.13 & $UVOT$  &  \\
2025-02-01T16:00:20 & 60707.667 & 3.469 & 4995.193 & $white$ & 21.38$\pm$0.12 & 21.23$\pm$0.12 & $UVOT$  &  \\
2025-02-01T19:48:07 & 60707.825 & 3.627 & 5222.977 & $white$ & 21.81$\pm$0.16 & 21.66$\pm$0.16 & $UVOT$  &  \\
2025-02-01T20:43:07 & 60707.863 & 3.665 & 5277.967 & $white$ & >23.00 & >22.85 & $UVOT$  &  \\
2025-02-06T13:16:46 & 60712.553 & 8.355 & 12031.617 & $u$ & 22.61$\pm$0.36 & 22.46$\pm$0.36 & $UVOT$  & x \\
2025-02-09T13:15:24 & 60715.552 & 11.354 & 16350.256 & $uvw1$ & >23.00 & 22.79 & $UVOT$  &  \\
2025-02-10T21:34:48 & 60716.899 & 12.701 & 18289.663 & $u$ & 22.32$\pm$0.35 & 22.17$\pm$0.35 & $UVOT$  & x \\
\hline
\multicolumn{9}{|c|}{$B$ band} \\
\hline
2025-01-29T05:13:55 & 60704.218 & 0.020 & 28.770 & $B$ & 17.27$\pm$0.07 & 17.14$\pm$0.07 & Skynet  & x \\
2025-01-29T05:16:31 & 60704.220 & 0.022 & 31.376 & $B$ & 17.23$\pm$0.05 & 17.11$\pm$0.05 & Skynet  & x \\
2025-01-29T05:20:13 & 60704.222 & 0.024 & 35.077 & $B$ & 17.30$\pm$0.06 & 17.18$\pm$0.06 & Skynet  & x \\
2025-01-29T05:26:09 & 60704.227 & 0.028 & 41.010 & $B$ & 17.34$\pm$0.04 & 17.21$\pm$0.04 & Skynet  & x \\
2025-01-29T05:27:38 & 60704.228 & 0.030 & 42.493 & $B$ & 17.29$\pm$0.04 & 17.16$\pm$0.04 & Skynet  & x \\
2025-01-29T05:35:46 & 60704.233 & 0.035 & 50.629 & $B$ & 17.39$\pm$0.04 & 17.26$\pm$0.04 & Skynet  & x \\
2025-01-29T05:36:01 & 60704.233 & 0.035 & 50.874 & $B$ & 17.27$\pm$0.04 & 17.14$\pm$0.04 & Skynet  & x \\
2025-01-29T05:45:38 & 60704.240 & 0.042 & 60.493 & $B$ & 17.40$\pm$0.04 & 17.27$\pm$0.04 & Skynet  & x \\
2025-01-29T05:47:59 & 60704.242 & 0.044 & 62.840 & $B$ & 17.47$\pm$0.03 & 17.34$\pm$0.03 & Skynet  & x \\
2025-01-29T05:56:54 & 60704.248 & 0.050 & 71.754 & $B$ & 17.39$\pm$0.03 & 17.26$\pm$0.03 & Skynet  & x \\
2025-01-29T06:02:16 & 60704.252 & 0.054 & 77.125 & $B$ & 17.44$\pm$0.03 & 17.32$\pm$0.03 & Skynet  & x \\
2025-01-29T06:10:05 & 60704.257 & 0.059 & 84.944 & $B$ & 17.44$\pm$0.04 & 17.32$\pm$0.04 & Skynet  & x \\
2025-01-29T06:18:39 & 60704.263 & 0.065 & 93.512 & $B$ & 17.59$\pm$0.03 & 17.47$\pm$0.03 & Skynet  & x \\
2025-01-29T06:29:53 & 60704.271 & 0.073 & 104.744 & $B$ & 17.64$\pm$0.03 & 17.52$\pm$0.03 & Skynet  & x \\
2025-01-29T06:38:09 & 60704.277 & 0.078 & 113.010 & $B$ & 17.66$\pm$0.03 & 17.54$\pm$0.03 & Skynet  & x \\
2025-01-29T07:51:06 & 60704.327 & 0.129 & 185.960 & $B$ & 18.32$\pm$0.06 & 18.20$\pm$0.06 & Skynet  & x \\
2025-01-29T08:33:24 & 60704.357 & 0.159 & 228.253 & $B$ & 17.87$\pm$0.04 & 17.75$\pm$0.04 & Skynet  & x \\
\hline
\multicolumn{9}{|c|}{$g'$ band} \\
\hline
2025-01-29T20:15:56 & 60704.844 & 0.646 & 930.783 & $g'$ & 19.21$\pm$0.03 & 19.09$\pm$0.03 & TNOT  & x \\
2025-01-30T01:05:00 & 60705.045 & 0.847 & 1219.853 & $g'$ & 19.40$\pm$0.06 & 19.28$\pm$0.06 & KAO  & x \\
2025-01-30T05:53:59 & 60705.246 & 1.048 & 1508.837 & $g'$ & 18.94$\pm$0.04 & 18.82$\pm$0.04 & KNC  & x \\
2025-01-30T06:04:01 & 60705.253 & 1.055 & 1518.877 & $g'$ & 19.00$\pm$0.05 & 18.89$\pm$0.05 & KNC  & x \\
2025-01-30T07:28:41 & 60705.312 & 1.114 & 1603.544 & $g'$ & 19.16$\pm$0.05 & 19.04$\pm$0.05 & KNC  & x \\
2025-01-30T07:50:29 & 60705.327 & 1.129 & 1625.337 & $g'$ & 19.11$\pm$0.05 & 18.99$\pm$0.05 & KNC  & x \\
2025-01-30T08:00:55 & 60705.334 & 1.136 & 1635.772 & $g'$ & 19.23$\pm$0.05 & 19.11$\pm$0.05 & KNC  & x \\
2025-01-31T05:39:41 & 60706.236 & 2.038 & 2934.540 & $g'$ & 20.63$\pm$0.14 & 20.52$\pm$0.14 & KNC  & x \\
2025-01-31T07:52:55 & 60706.328 & 2.130 & 3067.767 & $g'$ & 20.28$\pm$0.12 & 20.16$\pm$0.12 & KNC  & x \\
2025-02-01T06:37:04 & 60707.276 & 3.078 & 4431.930 & $g'$ & 20.52$\pm$0.12 & 20.41$\pm$0.12 & KNC  & x \\
2025-02-01T06:57:10 & 60707.290 & 3.092 & 4452.033 & $g'$ & 20.65$\pm$0.13 & 20.53$\pm$0.13 & KNC  & x \\
2025-02-01T20:18:13 & 60707.846 & 3.648 & 5253.081 & $g'$ & 20.79$\pm$0.15 & 20.68$\pm$0.15 & KNC  & x \\
2025-02-03T01:57:31 & 60709.082 & 4.884 & 7032.368 & $g'$ & 21.35$\pm$0.10 & 21.23$\pm$0.10 & NAO-2m  & x \\
2025-02-04T00:48:59 & 60710.034 & 5.836 & 8403.838 & $g'$ & 21.50$\pm$0.20 & 21.38$\pm$0.20 & KNC  & x \\
2025-02-04T02:42:14 & 60710.113 & 5.915 & 8517.089 & $g'$ & 22.30$\pm$0.06 & 22.18$\pm$0.06 & C2PU-O  & x \\
2025-02-04T04:58:11 & 60710.207 & 6.009 & 8653.038 & $g'$ & 22.47$\pm$0.08 & 22.35$\pm$0.08 & Artemis  & x \\
2025-02-05T04:57:09 & 60711.206 & 7.008 & 10092.016 & $g'$ & 22.62$\pm$0.09 & 22.50$\pm$0.09 & Artemis  & x \\
2025-02-05T05:31:52 & 60711.230 & 7.032 & 10126.717 & $g'$ & 22.60$\pm$0.25 & 22.48$\pm$0.25 & PicduMidi/T1M  & x \\
\hline
\multicolumn{9}{|c|}{$V$ band} \\
\hline
2025-01-29T05:14:45 & 60704.219 & 0.021 & 29.605 & $V$ & 17.02$\pm$0.04 & 16.93$\pm$0.04 & Skynet  & x \\
2025-01-29T05:17:36 & 60704.221 & 0.023 & 32.456 & $V$ & 16.96$\pm$0.03 & 16.86$\pm$0.03 & Skynet  & x \\
2025-01-29T05:21:19 & 60704.223 & 0.025 & 36.172 & $V$ & 17.01$\pm$0.04 & 16.92$\pm$0.04 & Skynet  & x \\
2025-01-29T05:27:42 & 60704.228 & 0.030 & 42.551 & $V$ & 17.03$\pm$0.03 & 16.93$\pm$0.03 & Skynet  & x \\
2025-01-29T05:28:57 & 60704.228 & 0.030 & 43.804 & $V$ & 17.00$\pm$0.03 & 16.91$\pm$0.03 & Skynet  & x \\
2025-01-29T05:37:33 & 60704.234 & 0.036 & 52.415 & $V$ & 17.07$\pm$0.03 & 16.98$\pm$0.03 & Skynet  & x \\
2025-01-29T05:37:39 & 60704.234 & 0.036 & 52.501 & $V$ & 17.04$\pm$0.02 & 16.94$\pm$0.02 & Skynet  & x \\
2025-01-29T05:47:28 & 60704.241 & 0.043 & 62.322 & $V$ & 17.13$\pm$0.02 & 17.04$\pm$0.02 & Skynet  & x \\
2025-01-29T05:50:17 & 60704.243 & 0.045 & 65.144 & $V$ & 17.09$\pm$0.02 & 16.99$\pm$0.02 & Skynet  & x \\
2025-01-29T05:59:02 & 60704.249 & 0.051 & 73.885 & $V$ & 17.24$\pm$0.02 & 17.15$\pm$0.02 & Skynet  & x \\
2025-01-29T06:05:07 & 60704.254 & 0.056 & 79.976 & $V$ & 17.26$\pm$0.02 & 17.17$\pm$0.02 & Skynet  & x \\
2025-01-29T06:12:37 & 60704.259 & 0.061 & 87.479 & $V$ & 17.25$\pm$0.02 & 17.16$\pm$0.02 & Skynet  & x \\
2025-01-29T06:22:06 & 60704.265 & 0.067 & 96.954 & $V$ & 17.34$\pm$0.02 & 17.25$\pm$0.02 & Skynet  & x \\
2025-01-29T06:32:59 & 60704.273 & 0.075 & 107.840 & $V$ & 17.45$\pm$0.02 & 17.36$\pm$0.02 & Skynet  & x \\
2025-01-29T06:42:17 & 60704.279 & 0.081 & 117.143 & $V$ & 17.51$\pm$0.02 & 17.42$\pm$0.02 & Skynet  & x \\
2025-01-29T07:56:15 & 60704.331 & 0.133 & 191.116 & $V$ & 18.01$\pm$0.04 & 17.92$\pm$0.04 & Skynet  & x \\
2025-01-29T08:39:52 & 60704.361 & 0.163 & 234.733 & $V$ & 17.61$\pm$0.02 & 17.52$\pm$0.02 & Skynet  & x \\
2025-01-30T05:15:39 & 60705.219 & 1.021 & 1470.512 & $V$ & 19.08$\pm$0.05 & 18.99$\pm$0.05 & Skynet  & x \\
2025-01-30T05:30:51 & 60705.230 & 1.032 & 1485.704 & $V$ & 19.00$\pm$0.04 & 18.91$\pm$0.04 & Skynet  & x \\
2025-01-30T05:46:04 & 60705.240 & 1.042 & 1500.925 & $V$ & 18.99$\pm$0.04 & 18.90$\pm$0.04 & Skynet  & x \\
2025-01-30T06:01:16 & 60705.251 & 1.053 & 1516.117 & $V$ & 18.91$\pm$0.03 & 18.82$\pm$0.03 & Skynet  & x \\
2025-01-30T06:16:28 & 60705.261 & 1.063 & 1531.324 & $V$ & 18.91$\pm$0.04 & 18.82$\pm$0.04 & Skynet  & x \\
2025-01-30T06:31:39 & 60705.272 & 1.074 & 1546.516 & $V$ & 18.93$\pm$0.04 & 18.83$\pm$0.04 & Skynet  & x \\
2025-01-30T06:46:52 & 60705.283 & 1.085 & 1561.722 & $V$ & 19.04$\pm$0.04 & 18.95$\pm$0.04 & Skynet  & x \\
2025-01-30T07:02:04 & 60705.293 & 1.095 & 1576.928 & $V$ & 19.17$\pm$0.04 & 19.08$\pm$0.04 & Skynet  & x \\
2025-01-30T07:17:16 & 60705.304 & 1.106 & 1592.120 & $V$ & 19.28$\pm$0.05 & 19.19$\pm$0.05 & Skynet  & x \\
2025-01-30T07:32:27 & 60705.314 & 1.116 & 1607.312 & $V$ & 19.23$\pm$0.05 & 19.14$\pm$0.05 & Skynet  & x \\
2025-01-30T07:45:08 & 60705.323 & 1.125 & 1619.999 & $V$ & 19.27$\pm$0.05 & 19.17$\pm$0.05 & Skynet  & x \\
2025-01-30T08:02:49 & 60705.335 & 1.137 & 1637.682 & $V$ & 19.26$\pm$0.05 & 19.16$\pm$0.05 & Skynet  & x \\
2025-01-30T08:18:02 & 60705.346 & 1.148 & 1652.888 & $V$ & 19.23$\pm$0.05 & 19.13$\pm$0.05 & Skynet  & x \\
2025-01-30T08:35:46 & 60705.358 & 1.160 & 1670.629 & $V$ & 19.20$\pm$0.04 & 19.10$\pm$0.04 & Skynet  & x \\
2025-02-01T05:56:31 & 60707.248 & 3.050 & 4391.380 & $V$ & 20.94$\pm$0.07 & 20.85$\pm$0.07 & Skynet  & x \\
2025-02-01T07:47:57 & 60707.325 & 3.127 & 4502.807 & $V$ & 21.17$\pm$0.07 & 21.08$\pm$0.07 & Skynet  & x \\
\hline
\multicolumn{9}{|c|}{$r'$ band} \\
\hline
2025-01-29T20:34:41 & 60704.857 & 0.659 & 949.550 & $r'$ & 18.96$\pm$0.05 & 18.88$\pm$0.05 & TNOT  & x \\
2025-01-30T01:17:48 & 60705.054 & 0.856 & 1232.666 & $r'$ & 19.17$\pm$0.05 & 19.09$\pm$0.05 & KAO  & x \\
2025-01-30T06:14:06 & 60705.260 & 1.062 & 1528.956 & $r'$ & 18.71$\pm$0.06 & 18.63$\pm$0.06 & KNC  & x \\
2025-01-30T07:40:07 & 60705.320 & 1.122 & 1614.977 & $r'$ & 18.90$\pm$0.05 & 18.82$\pm$0.05 & KNC  & x \\
2025-01-30T08:08:34 & 60705.339 & 1.141 & 1643.425 & $r'$ & 18.96$\pm$0.03 & 18.88$\pm$0.03 & COLIBRI-VIS  & x \\
2025-01-30T09:22:49 & 60705.391 & 1.193 & 1717.681 & $r'$ & 18.99$\pm$0.04 & 18.91$\pm$0.04 & COLIBRI-VIS  & x \\
2025-01-30T11:30:56 & 60705.480 & 1.282 & 1845.797 & $r'$ & 19.22$\pm$0.03 & 19.14$\pm$0.03 & COLIBRI-VIS  & x \\
2025-01-31T05:49:56 & 60706.243 & 2.045 & 2944.787 & $r'$ & 19.94$\pm$0.12 & 19.86$\pm$0.12 & KNC  & x \\
2025-01-31T08:08:01 & 60706.339 & 2.141 & 3082.871 & $r'$ & 20.33$\pm$0.05 & 20.25$\pm$0.05 & COLIBRI-VIS  & x \\
2025-01-31T08:24:30 & 60706.350 & 2.152 & 3099.361 & $r'$ & 20.47$\pm$0.08 & 20.39$\pm$0.08 & COLIBRI-VIS  & x \\
2025-01-31T10:16:49 & 60706.428 & 2.230 & 3211.681 & $r'$ & 20.71$\pm$0.09 & 20.63$\pm$0.09 & COLIBRI-VIS  & x \\
2025-01-31T11:59:50 & 60706.500 & 2.302 & 3314.691 & $r'$ & 20.51$\pm$0.05 & 20.43$\pm$0.05 & COLIBRI-VIS  & x \\
2025-02-01T07:21:03 & 60707.306 & 3.108 & 4475.900 & $r'$ & 20.43$\pm$0.10 & 20.35$\pm$0.10 & KNC  & x \\
2025-02-01T08:34:39 & 60707.357 & 3.159 & 4549.501 & $r'$ & 20.52$\pm$0.06 & 20.45$\pm$0.06 & COLIBRI-VIS  & x \\
2025-02-01T11:29:39 & 60707.479 & 3.281 & 4724.515 & $r'$ & 20.56$\pm$0.05 & 20.48$\pm$0.05 & COLIBRI-VIS  & x \\
2025-02-02T08:05:45 & 60708.337 & 4.139 & 5960.601 & $r'$ & 21.45$\pm$0.20 & 21.37$\pm$0.20 & COLIBRI-VIS  & x \\
2025-02-02T12:00:15 & 60708.500 & 4.302 & 6195.103 & $r'$ & 21.53$\pm$0.12 & 21.45$\pm$0.12 & COLIBRI-VIS  & x \\
2025-02-03T03:34:41 & 60709.149 & 4.951 & 7129.546 & $r'$ & 21.86$\pm$0.07 & 21.78$\pm$0.07 & NAO-2m  & x \\
2025-02-03T04:40:51 & 60709.195 & 4.997 & 7195.706 & $r'$ & 21.76$\pm$0.14 & 21.68$\pm$0.14 & OHP/MISTRAL  & x \\
2025-02-03T08:11:50 & 60709.342 & 5.144 & 7406.689 & $r'$ & 21.87$\pm$0.10 & 21.79$\pm$0.10 & COLIBRI-VIS  & x \\
2025-02-03T12:00:38 & 60709.500 & 5.302 & 7635.497 & $r'$ & 21.99$\pm$0.10 & 21.91$\pm$0.10 & COLIBRI-VIS  & x \\
2025-02-04T05:20:12 & 60710.222 & 6.024 & 8675.057 & $r'$ & 22.19$\pm$0.10 & 22.11$\pm$0.10 & Artemis  & x \\
2025-02-04T07:34:16 & 60710.315 & 6.117 & 8809.120 & $r'$ & 22.36$\pm$0.17 & 22.28$\pm$0.17 & COLIBRI-VIS  & x \\
2025-02-05T04:36:52 & 60711.192 & 6.994 & 10071.733 & $r'$ & 22.50$\pm$0.11 & 22.42$\pm$0.11 & PicduMidi/T1M  & x \\
2025-02-05T05:18:36 & 60711.221 & 7.023 & 10113.455 & $r'$ & 22.27$\pm$0.09 & 22.19$\pm$0.09 & Artemis  & x \\
2025-02-05T10:31:09 & 60711.438 & 7.240 & 10426.002 & $r'$ & 22.44$\pm$0.10 & 22.36$\pm$0.10 & COLIBRI-VIS  & x \\
2025-02-06T01:59:45 & 60712.083 & 7.885 & 11354.606 & $r'$ & 22.34$\pm$0.14 & 22.26$\pm$0.14 & PicduMidi/T1M  & x \\
2025-02-06T10:59:27 & 60712.458 & 8.260 & 11894.302 & $r'$ & 22.44$\pm$0.07 & 22.36$\pm$0.07 & COLIBRI-VIS  & x \\
2025-02-07T10:58:37 & 60713.457 & 9.259 & 13333.474 & $r'$ & 22.70$\pm$0.12 & 22.62$\pm$0.12 & COLIBRI-VIS  & x \\
2025-02-08T10:01:36 & 60714.418 & 10.220 & 14716.456 & $r'$ & 22.78$\pm$0.10 & 22.70$\pm$0.10 & COLIBRI-VIS  & x \\
2025-02-09T10:00:12 & 60715.417 & 11.219 & 16155.065 & $r'$ & 22.68$\pm$0.10 & 22.60$\pm$0.10 & COLIBRI-VIS  & x \\
2025-02-10T10:03:01 & 60716.419 & 12.221 & 17597.879 & $r'$ & 23.04$\pm$0.13 & >22.96$\pm$0.13 & COLIBRI-VIS  & x \\
2025-02-19T11:18:58 & 60725.472 & 21.273 & 30633.824 & $r'$ & 23.50$\pm$0.09 & 23.42$\pm$0.09 & COLIBRI-VIS  & x \\
2025-02-22T10:21:21 & 60728.431 & 24.233 & 34896.206 & $r'$ & 23.71$\pm$0.10 & 23.63$\pm$0.10 & COLIBRI-VIS  & x \\
\hline
\multicolumn{9}{|c|}{$R$ band} \\
\hline
2025-01-29T04:47:00 & 60704.199 & 0.001 & 1.866 & $R$ & >17.49$\pm$ & >17.42$\pm$ & TAROT/TCH  & x \\
2025-01-29T04:47:12 & 60704.199 & 0.001 & 2.067 & $R$ & >17.32$\pm$ & >17.25$\pm$ & TAROT/TCH  & x \\
2025-01-29T04:47:25 & 60704.200 & 0.002 & 2.267 & $R$ & 16.92$\pm$0.40 & 16.85$\pm$0.40 & TAROT/TCH  & x \\
2025-01-29T04:47:37 & 60704.200 & 0.002 & 2.467 & $R$ & 17.12$\pm$0.40 & 17.05$\pm$0.40 & TAROT/TCH  & x \\
2025-01-29T04:47:49 & 60704.200 & 0.002 & 2.667 & $R$ & 16.92$\pm$0.40 & 16.85$\pm$0.40 & TAROT/TCH  & x \\
2025-01-29T04:48:59 & 60704.201 & 0.003 & 3.844 & $R$ & 17.35$\pm$0.16 & 17.27$\pm$0.16 & TAROT/TCH  & x \\
2025-01-29T04:49:40 & 60704.201 & 0.003 & 4.520 & $R$ & 17.71$\pm$0.23 & 17.63$\pm$0.23 & TAROT/TCH  & x \\
2025-01-29T04:50:19 & 60704.202 & 0.004 & 5.168 & $R$ & 18.24$\pm$0.37 & 18.16$\pm$0.37 & TAROT/TCH  & x \\
2025-01-29T04:50:58 & 60704.202 & 0.004 & 5.831 & $R$ & 18.12$\pm$0.32 & 18.04$\pm$0.32 & TAROT/TCH  & x \\
2025-01-29T04:52:21 & 60704.203 & 0.005 & 7.213 & $R$ & 17.58$\pm$0.12 & 17.50$\pm$0.12 & TAROT/TCH  & x \\
2025-01-29T04:54:01 & 60704.204 & 0.006 & 8.869 & $R$ & 17.66$\pm$0.13 & 17.59$\pm$0.13 & TAROT/TCH  & x \\
2025-01-29T04:55:40 & 60704.205 & 0.007 & 10.525 & $R$ & 17.45$\pm$0.10 & 17.38$\pm$0.10 & TAROT/TCH  & x \\
2025-01-29T04:56:53 & 60704.206 & 0.008 & 11.735 & $R$ & 17.20$\pm$0.14 & 17.13$\pm$0.14 & FRAM-Auger  & x \\
2025-01-29T04:57:50 & 60704.207 & 0.009 & 12.700 & $R$ & 17.16$\pm$0.08 & 17.09$\pm$0.08 & TAROT/TCH  & x \\
2025-01-29T04:59:30 & 60704.208 & 0.010 & 14.356 & $R$ & 17.02$\pm$0.07 & 16.95$\pm$0.07 & TAROT/TCH  & x \\
2025-01-29T05:01:09 & 60704.209 & 0.011 & 16.012 & $R$ & 17.06$\pm$0.07 & 16.99$\pm$0.07 & TAROT/TCH  & x \\
2025-01-29T05:02:49 & 60704.210 & 0.012 & 17.668 & $R$ & 16.79$\pm$0.09 & 16.71$\pm$0.09 & FRAM-Auger  & x \\
2025-01-29T05:03:39 & 60704.211 & 0.013 & 18.503 & $R$ & 17.02$\pm$0.05 & 16.94$\pm$0.05 & TAROT/TCH  & x \\
2025-01-29T05:06:48 & 60704.213 & 0.015 & 21.656 & $R$ & 16.93$\pm$0.04 & 16.85$\pm$0.04 & TAROT/TCH  & x \\
2025-01-29T05:08:45 & 60704.214 & 0.016 & 23.600 & $R$ & 16.78$\pm$0.07 & 16.70$\pm$0.07 & FRAM-Auger  & x \\
2025-01-29T05:09:56 & 60704.215 & 0.017 & 24.796 & $R$ & 16.93$\pm$0.05 & 16.86$\pm$0.05 & TAROT/TCH  & x \\
2025-01-29T05:13:41 & 60704.218 & 0.020 & 28.540 & $R$ & 16.86$\pm$0.07 & 16.79$\pm$0.07 & Skynet  & x \\
2025-01-29T05:14:43 & 60704.219 & 0.021 & 29.576 & $R$ & 16.77$\pm$0.05 & 16.70$\pm$0.05 & Skynet  & x \\
2025-01-29T05:17:01 & 60704.220 & 0.022 & 31.880 & $R$ & 16.81$\pm$0.06 & 16.74$\pm$0.06 & Skynet  & x \\
2025-01-29T05:20:06 & 60704.222 & 0.024 & 34.962 & $R$ & 16.82$\pm$0.06 & 16.75$\pm$0.06 & Skynet  & x \\
2025-01-29T05:20:30 & 60704.223 & 0.025 & 35.351 & $R$ & 16.84$\pm$0.07 & 16.77$\pm$0.07 & FRAM-Auger  & x \\
2025-01-29T05:22:01 & 60704.224 & 0.026 & 36.877 & $R$ & 16.84$\pm$0.05 & 16.77$\pm$0.05 & Skynet  & x \\
2025-01-29T05:24:00 & 60704.225 & 0.027 & 38.864 & $R$ & 16.79$\pm$0.05 & 16.72$\pm$0.05 & Skynet  & x \\
2025-01-29T05:24:02 & 60704.225 & 0.027 & 38.893 & $R$ & 16.84$\pm$0.04 & 16.77$\pm$0.04 & TAROT/TCH  & x \\
2025-01-29T05:26:29 & 60704.227 & 0.029 & 41.341 & $R$ & 16.79$\pm$0.04 & 16.72$\pm$0.04 & Skynet  & x \\
2025-01-29T05:27:12 & 60704.227 & 0.029 & 42.061 & $R$ & 16.89$\pm$0.04 & 16.82$\pm$0.04 & TAROT/TCH  & x \\
2025-01-29T05:27:40 & 60704.228 & 0.030 & 42.522 & $R$ & 16.82$\pm$0.05 & 16.75$\pm$0.05 & Skynet  & x \\
2025-01-29T05:30:21 & 60704.229 & 0.031 & 45.215 & $R$ & 16.87$\pm$0.04 & 16.80$\pm$0.04 & TAROT/TCH  & x \\
2025-01-29T05:30:57 & 60704.230 & 0.032 & 45.805 & $R$ & 16.80$\pm$0.04 & 16.73$\pm$0.04 & Skynet  & x \\
2025-01-29T05:31:59 & 60704.231 & 0.033 & 46.842 & $R$ & 16.89$\pm$0.04 & 16.82$\pm$0.04 & Skynet  & x \\
2025-01-29T05:32:28 & 60704.231 & 0.033 & 47.317 & $R$ & 16.88$\pm$0.07 & 16.81$\pm$0.07 & FRAM-Auger  & x \\
2025-01-29T05:33:46 & 60704.232 & 0.034 & 48.628 & $R$ & 16.80$\pm$0.03 & 16.73$\pm$0.03 & TAROT/TCH  & x \\
2025-01-29T05:36:02 & 60704.233 & 0.035 & 50.888 & $R$ & 16.78$\pm$0.04 & 16.71$\pm$0.04 & Skynet  & x \\
2025-01-29T05:36:16 & 60704.234 & 0.036 & 51.133 & $R$ & 16.87$\pm$0.04 & 16.79$\pm$0.04 & Skynet  & x \\
2025-01-29T05:36:55 & 60704.234 & 0.036 & 51.781 & $R$ & 16.84$\pm$0.03 & 16.77$\pm$0.03 & TAROT/TCH  & x \\
2025-01-29T05:40:05 & 60704.236 & 0.038 & 54.935 & $R$ & 16.91$\pm$0.03 & 16.83$\pm$0.03 & TAROT/TCH  & x \\
2025-01-29T05:40:56 & 60704.237 & 0.039 & 55.799 & $R$ & 16.93$\pm$0.04 & 16.86$\pm$0.04 & Skynet  & x \\
2025-01-29T05:41:47 & 60704.237 & 0.039 & 56.648 & $R$ & 16.86$\pm$0.03 & 16.79$\pm$0.03 & Skynet  & x \\
2025-01-29T05:43:31 & 60704.239 & 0.041 & 58.376 & $R$ & 16.94$\pm$0.03 & 16.87$\pm$0.03 & TAROT/TCH  & x \\
2025-01-29T05:43:31 & 60704.239 & 0.041 & 58.376 & $R$ & 16.92$\pm$0.03 & 16.84$\pm$0.03 & TAROT/TCH  & x \\
2025-01-29T05:45:57 & 60704.240 & 0.042 & 60.810 & $R$ & 16.89$\pm$0.04 & 16.82$\pm$0.04 & Skynet  & x \\
2025-01-29T05:47:46 & 60704.242 & 0.043 & 62.624 & $R$ & 16.92$\pm$0.03 & 16.85$\pm$0.03 & Skynet  & x \\
2025-01-29T05:49:50 & 60704.243 & 0.045 & 64.684 & $R$ & 16.92$\pm$0.04 & 16.85$\pm$0.04 & TAROT/TCH  & x \\
2025-01-29T05:50:18 & 60704.243 & 0.045 & 65.159 & $R$ & 17.07$\pm$0.08 & 17.00$\pm$0.08 & FRAM-Auger  & x \\
2025-01-29T05:51:25 & 60704.244 & 0.046 & 66.282 & $R$ & 16.95$\pm$0.04 & 16.88$\pm$0.04 & Skynet  & x \\
2025-01-29T05:53:20 & 60704.245 & 0.047 & 68.197 & $R$ & 16.93$\pm$0.03 & 16.86$\pm$0.03 & TAROT/TCH  & x \\
2025-01-29T05:55:52 & 60704.247 & 0.049 & 70.732 & $R$ & 16.98$\pm$0.03 & 16.91$\pm$0.03 & Skynet  & x \\
2025-01-29T05:56:30 & 60704.248 & 0.050 & 71.365 & $R$ & 16.99$\pm$0.03 & 16.91$\pm$0.03 & TAROT/TCH  & x \\
2025-01-29T05:57:16 & 60704.248 & 0.050 & 72.128 & $R$ & 16.99$\pm$0.04 & 16.92$\pm$0.04 & Skynet  & x \\
2025-01-29T05:59:40 & 60704.250 & 0.052 & 74.519 & $R$ & 17.04$\pm$0.04 & 16.97$\pm$0.04 & TAROT/TCH  & x \\
2025-01-29T06:02:57 & 60704.252 & 0.054 & 77.816 & $R$ & 17.05$\pm$0.03 & 16.98$\pm$0.03 & TAROT/TCH  & x \\
2025-01-29T06:03:01 & 60704.252 & 0.054 & 77.874 & $R$ & 17.03$\pm$0.03 & 16.96$\pm$0.03 & Skynet  & x \\
2025-01-29T06:03:40 & 60704.253 & 0.055 & 78.522 & $R$ & 17.07$\pm$0.04 & 17.00$\pm$0.04 & Skynet  & x \\
2025-01-29T06:06:08 & 60704.254 & 0.056 & 80.984 & $R$ & 17.00$\pm$0.04 & 16.92$\pm$0.04 & TAROT/TCH  & x \\
2025-01-29T06:07:50 & 60704.255 & 0.057 & 82.684 & $R$ & 17.19$\pm$0.08 & 17.12$\pm$0.08 & FRAM-Auger  & x \\
2025-01-29T06:09:17 & 60704.256 & 0.058 & 84.138 & $R$ & 17.10$\pm$0.04 & 17.03$\pm$0.04 & TAROT/TCH  & x \\
2025-01-29T06:10:35 & 60704.257 & 0.059 & 85.434 & $R$ & 17.12$\pm$0.04 & 17.05$\pm$0.04 & Skynet  & x \\
2025-01-29T06:10:58 & 60704.258 & 0.060 & 85.823 & $R$ & 17.19$\pm$0.03 & 17.12$\pm$0.03 & Skynet  & x \\
2025-01-29T06:15:51 & 60704.261 & 0.063 & 90.704 & $R$ & 17.15$\pm$0.04 & 17.07$\pm$0.04 & TAROT/TCH  & x \\
2025-01-29T06:15:51 & 60704.261 & 0.063 & 90.704 & $R$ & 17.07$\pm$0.03 & 17.00$\pm$0.03 & TAROT/TCH  & x \\
2025-01-29T06:18:00 & 60704.263 & 0.064 & 92.864 & $R$ & 17.24$\pm$0.03 & 17.17$\pm$0.03 & Skynet  & x \\
2025-01-29T06:19:00 & 60704.263 & 0.065 & 93.858 & $R$ & 17.17$\pm$0.04 & 17.10$\pm$0.04 & TAROT/TCH  & x \\
2025-01-29T06:19:37 & 60704.264 & 0.066 & 94.477 & $R$ & 17.16$\pm$0.03 & 17.09$\pm$0.03 & Skynet  & x \\
2025-01-29T06:22:34 & 60704.266 & 0.068 & 97.429 & $R$ & 17.18$\pm$0.04 & 17.11$\pm$0.04 & TAROT/TCH  & x \\
2025-01-29T06:25:44 & 60704.268 & 0.070 & 100.597 & $R$ & 17.19$\pm$0.03 & 17.11$\pm$0.03 & TAROT/TCH  & x \\
2025-01-29T06:26:04 & 60704.268 & 0.070 & 100.928 & $R$ & 17.25$\pm$0.04 & 17.18$\pm$0.04 & Skynet  & x \\
2025-01-29T06:28:54 & 60704.270 & 0.072 & 103.751 & $R$ & 17.18$\pm$0.04 & 17.11$\pm$0.04 & TAROT/TCH  & x \\
2025-01-29T06:29:08 & 60704.270 & 0.072 & 103.996 & $R$ & 17.26$\pm$0.03 & 17.19$\pm$0.03 & Skynet  & x \\
2025-01-29T06:31:18 & 60704.272 & 0.074 & 106.156 & $R$ & 17.25$\pm$0.08 & 17.18$\pm$0.08 & FRAM-Auger  & x \\
2025-01-29T06:34:48 & 60704.274 & 0.076 & 109.655 & $R$ & 17.28$\pm$0.03 & 17.21$\pm$0.03 & Skynet  & x \\
2025-01-29T06:39:41 & 60704.278 & 0.080 & 114.536 & $R$ & 17.33$\pm$0.03 & 17.26$\pm$0.03 & Skynet  & x \\
2025-01-29T06:39:57 & 60704.278 & 0.080 & 114.810 & $R$ & 17.30$\pm$0.04 & 17.23$\pm$0.04 & TAROT/TCH  & x \\
2025-01-29T06:43:07 & 60704.280 & 0.082 & 117.978 & $R$ & 17.37$\pm$0.04 & 17.30$\pm$0.04 & TAROT/TCH  & x \\
2025-01-29T06:44:15 & 60704.281 & 0.083 & 119.101 & $R$ & 17.35$\pm$0.03 & 17.28$\pm$0.03 & Skynet  & x \\
2025-01-29T06:46:16 & 60704.282 & 0.084 & 121.132 & $R$ & 17.33$\pm$0.04 & 17.25$\pm$0.04 & TAROT/TCH  & x \\
2025-01-29T06:54:59 & 60704.288 & 0.090 & 129.844 & $R$ & 17.51$\pm$0.10 & 17.43$\pm$0.10 & FRAM-Auger  & x \\
2025-01-29T07:17:48 & 60704.304 & 0.106 & 152.653 & $R$ & 17.51$\pm$0.04 & 17.44$\pm$0.04 & Skynet  & x \\
2025-01-29T07:21:14 & 60704.306 & 0.108 & 156.095 & $R$ & 17.59$\pm$0.05 & 17.51$\pm$0.05 & TAROT/TCH  & x \\
2025-01-29T07:24:23 & 60704.309 & 0.111 & 159.248 & $R$ & 17.59$\pm$0.04 & 17.52$\pm$0.04 & TAROT/TCH  & x \\
2025-01-29T07:27:33 & 60704.311 & 0.113 & 162.402 & $R$ & 17.63$\pm$0.04 & 17.55$\pm$0.04 & TAROT/TCH  & x \\
2025-01-29T07:40:50 & 60704.320 & 0.122 & 175.693 & $R$ & 17.79$\pm$0.05 & 17.72$\pm$0.05 & TAROT/TCH  & x \\
2025-01-29T07:44:00 & 60704.322 & 0.124 & 178.861 & $R$ & 17.68$\pm$0.04 & 17.61$\pm$0.04 & TAROT/TCH  & x \\
2025-01-29T07:47:09 & 60704.324 & 0.126 & 182.015 & $R$ & 17.72$\pm$0.05 & 17.65$\pm$0.05 & TAROT/TCH  & x \\
2025-01-29T08:00:14 & 60704.334 & 0.135 & 195.090 & $R$ & 17.87$\pm$0.05 & 17.80$\pm$0.05 & TAROT/TCH  & x \\
2025-01-29T08:00:16 & 60704.334 & 0.135 & 195.119 & $R$ & 17.94$\pm$0.04 & 17.87$\pm$0.04 & Skynet  & x \\
2025-01-29T08:03:23 & 60704.336 & 0.138 & 198.244 & $R$ & 17.85$\pm$0.05 & 17.77$\pm$0.05 & TAROT/TCH  & x \\
2025-01-29T08:06:32 & 60704.338 & 0.140 & 201.397 & $R$ & 17.86$\pm$0.05 & 17.79$\pm$0.05 & TAROT/TCH  & x \\
2025-01-29T08:44:55 & 60704.365 & 0.167 & 239.773 & $R$ & 17.43$\pm$0.02 & 17.36$\pm$0.02 & Skynet  & x \\
2025-01-29T09:27:18 & 60704.394 & 0.196 & 282.152 & $R$ & 17.13$\pm$0.05 & 17.05$\pm$0.05 & KAIT  & x \\
2025-01-29T09:32:03 & 60704.397 & 0.199 & 286.904 & $R$ & 16.78$\pm$0.05 & 16.71$\pm$0.05 & KAIT  & x \\
2025-01-29T09:36:46 & 60704.401 & 0.203 & 291.628 & $R$ & 16.63$\pm$0.05 & 16.56$\pm$0.05 & KAIT  & x \\
2025-01-29T09:36:49 & 60704.401 & 0.203 & 291.671 & $R$ & 16.65$\pm$0.05 & 16.58$\pm$0.05 & Skynet  & x \\
2025-01-29T09:41:31 & 60704.404 & 0.206 & 296.380 & $R$ & 16.48$\pm$0.05 & 16.41$\pm$0.05 & KAIT  & x \\
2025-01-29T09:46:17 & 60704.407 & 0.209 & 301.146 & $R$ & 16.47$\pm$0.05 & 16.40$\pm$0.05 & KAIT  & x \\
2025-01-29T09:52:18 & 60704.411 & 0.213 & 307.151 & $R$ & 16.62$\pm$0.05 & 16.55$\pm$0.05 & Skynet  & x \\
2025-01-29T09:55:49 & 60704.414 & 0.216 & 310.679 & $R$ & 16.45$\pm$0.05 & 16.38$\pm$0.05 & KAIT  & x \\
2025-01-29T10:00:33 & 60704.417 & 0.219 & 315.402 & $R$ & 16.47$\pm$0.05 & 16.40$\pm$0.05 & KAIT  & x \\
2025-01-29T10:05:45 & 60704.421 & 0.223 & 320.600 & $R$ & 16.58$\pm$0.05 & 16.51$\pm$0.05 & KAIT  & x \\
2025-01-29T10:10:28 & 60704.424 & 0.226 & 325.324 & $R$ & 16.71$\pm$0.05 & 16.64$\pm$0.05 & KAIT  & x \\
2025-01-29T10:12:56 & 60704.426 & 0.228 & 327.786 & $R$ & 16.87$\pm$0.06 & 16.80$\pm$0.06 & Skynet  & x \\
2025-01-29T10:15:12 & 60704.427 & 0.229 & 330.061 & $R$ & 16.82$\pm$0.05 & 16.75$\pm$0.05 & KAIT  & x \\
2025-01-29T10:18:06 & 60704.429 & 0.231 & 332.956 & $R$ & 16.92$\pm$0.06 & 16.85$\pm$0.06 & Skynet  & x \\
2025-01-29T10:20:00 & 60704.431 & 0.233 & 334.856 & $R$ & 16.94$\pm$0.05 & 16.87$\pm$0.05 & KAIT  & x \\
2025-01-29T10:24:45 & 60704.434 & 0.236 & 339.608 & $R$ & 16.99$\pm$0.05 & 16.92$\pm$0.05 & KAIT  & x \\
2025-01-29T10:29:32 & 60704.437 & 0.239 & 344.389 & $R$ & 17.03$\pm$0.05 & 16.96$\pm$0.05 & KAIT  & x \\
2025-01-29T10:34:17 & 60704.440 & 0.242 & 349.141 & $R$ & 17.03$\pm$0.05 & 16.96$\pm$0.05 & KAIT  & x \\
2025-01-29T10:41:38 & 60704.446 & 0.248 & 356.500 & $R$ & 17.05$\pm$0.05 & 16.98$\pm$0.05 & KAIT  & x \\
2025-01-29T10:51:22 & 60704.452 & 0.254 & 366.220 & $R$ & 17.09$\pm$0.05 & 17.02$\pm$0.05 & KAIT  & x \\
2025-01-29T11:00:52 & 60704.459 & 0.261 & 375.724 & $R$ & 17.16$\pm$0.05 & 17.09$\pm$0.05 & KAIT  & x \\
2025-01-29T11:10:25 & 60704.466 & 0.268 & 385.271 & $R$ & 17.16$\pm$0.05 & 17.09$\pm$0.05 & KAIT  & x \\
2025-01-29T11:20:06 & 60704.472 & 0.274 & 394.962 & $R$ & 17.20$\pm$0.05 & 17.13$\pm$0.05 & KAIT  & x \\
2025-01-29T11:29:49 & 60704.479 & 0.281 & 404.682 & $R$ & 17.26$\pm$0.05 & 17.19$\pm$0.05 & KAIT  & x \\
2025-01-29T11:40:31 & 60704.486 & 0.288 & 415.367 & $R$ & 17.32$\pm$0.05 & 17.25$\pm$0.05 & KAIT  & x \\
2025-01-29T11:52:31 & 60704.495 & 0.297 & 427.376 & $R$ & 17.38$\pm$0.05 & 17.31$\pm$0.05 & KAIT  & x \\
2025-01-30T00:20:54 & 60705.015 & 0.816 & 1175.759 & $R$ & 19.17$\pm$0.06 & 19.10$\pm$0.06 & Skynet  & x \\
2025-01-30T01:03:57 & 60705.044 & 0.846 & 1218.800 & $R$ & 19.26$\pm$0.06 & 19.19$\pm$0.06 & Skynet  & x \\
2025-01-30T02:36:01 & 60705.108 & 0.910 & 1310.874 & $R$ & 19.28$\pm$0.06 & 19.21$\pm$0.06 & Skynet  & x \\
2025-01-30T03:14:31 & 60705.135 & 0.937 & 1349.380 & $R$ & 19.32$\pm$0.09 & 19.25$\pm$0.09 & Skynet  & x \\
2025-01-30T03:50:28 & 60705.160 & 0.962 & 1385.322 & $R$ & 19.30$\pm$0.06 & 19.23$\pm$0.06 & Skynet  & x \\
2025-01-30T04:16:09 & 60705.178 & 0.980 & 1411.012 & $R$ & 19.46$\pm$0.07 & 19.39$\pm$0.07 & Skynet  & x \\
2025-01-30T05:30:50 & 60705.230 & 1.032 & 1485.690 & $R$ & 18.71$\pm$0.03 & 18.64$\pm$0.03 & Skynet  & x \\
2025-01-30T05:45:56 & 60705.240 & 1.042 & 1500.796 & $R$ & 18.71$\pm$0.03 & 18.64$\pm$0.03 & Skynet  & x \\
2025-01-30T06:31:13 & 60705.272 & 1.074 & 1546.069 & $R$ & 18.82$\pm$0.02 & 18.75$\pm$0.02 & Skynet  & x \\
2025-01-30T07:01:24 & 60705.293 & 1.095 & 1576.252 & $R$ & 18.93$\pm$0.03 & 18.86$\pm$0.03 & Skynet  & x \\
2025-01-30T07:46:41 & 60705.324 & 1.126 & 1621.540 & $R$ & 18.99$\pm$0.03 & 18.92$\pm$0.03 & Skynet  & x \\
2025-01-30T08:34:28 & 60705.357 & 1.159 & 1669.319 & $R$ & 18.88$\pm$0.02 & 18.81$\pm$0.02 & Skynet  & x \\
2025-01-30T09:22:02 & 60705.390 & 1.192 & 1716.896 & $R$ & 18.86$\pm$0.05 & 18.79$\pm$0.05 & KAIT  & x \\
2025-01-30T09:39:54 & 60705.403 & 1.205 & 1734.752 & $R$ & 18.97$\pm$0.05 & 18.90$\pm$0.05 & KAIT  & x \\
2025-01-30T09:58:05 & 60705.415 & 1.217 & 1752.940 & $R$ & 18.99$\pm$0.05 & 18.92$\pm$0.05 & KAIT  & x \\
2025-01-30T10:15:55 & 60705.428 & 1.230 & 1770.767 & $R$ & 19.03$\pm$0.05 & 18.96$\pm$0.05 & KAIT  & x \\
2025-01-30T10:34:07 & 60705.440 & 1.242 & 1788.983 & $R$ & 19.03$\pm$0.05 & 18.96$\pm$0.05 & KAIT  & x \\
2025-01-30T10:52:04 & 60705.453 & 1.255 & 1806.925 & $R$ & 19.06$\pm$0.05 & 18.99$\pm$0.05 & KAIT  & x \\
2025-01-30T11:10:12 & 60705.465 & 1.267 & 1825.055 & $R$ & 19.09$\pm$0.05 & 19.02$\pm$0.05 & KAIT  & x \\
2025-01-30T11:28:13 & 60705.478 & 1.280 & 1843.069 & $R$ & 19.11$\pm$0.05 & 19.04$\pm$0.05 & KAIT  & x \\
2025-01-30T11:46:14 & 60705.490 & 1.292 & 1861.098 & $R$ & 19.16$\pm$0.05 & 19.09$\pm$0.05 & KAIT  & x \\
2025-01-30T12:04:19 & 60705.503 & 1.305 & 1879.170 & $R$ & 19.21$\pm$0.05 & 19.14$\pm$0.05 & KAIT  & x \\
2025-01-30T12:22:19 & 60705.516 & 1.317 & 1897.170 & $R$ & 19.28$\pm$0.05 & 19.21$\pm$0.05 & KAIT  & x \\
2025-01-30T12:40:28 & 60705.528 & 1.330 & 1915.328 & $R$ & 19.24$\pm$0.05 & 19.17$\pm$0.05 & KAIT  & x \\
2025-01-30T12:58:25 & 60705.541 & 1.343 & 1933.271 & $R$ & 19.38$\pm$0.05 & 19.31$\pm$0.05 & KAIT  & x \\
2025-01-30T13:16:35 & 60705.553 & 1.355 & 1951.444 & $R$ & 19.42$\pm$0.05 & 19.35$\pm$0.05 & KAIT  & x \\
2025-01-30T13:16:34 & 60705.553 & 1.355 & 1951.417 & $R$ & 19.43$\pm$0.15 & 19.36$\pm$0.15 & KNC & \\
2025-01-30T13:34:28 & 60705.566 & 1.368 & 1969.328 & $R$ & 19.47$\pm$0.05 & 19.40$\pm$0.05 & KAIT  & x \\
2025-01-30T13:52:44 & 60705.578 & 1.380 & 1987.588 & $R$ & 19.48$\pm$0.05 & 19.41$\pm$0.05 & KAIT  & x \\
2025-01-30T14:11:14 & 60705.591 & 1.393 & 2006.092 & $R$ & 19.64$\pm$0.05 & 19.57$\pm$0.05 & KAIT  & x \\
2025-01-30T17:27:26 & 60705.727 & 1.529 & 2202.300 & $R$ & 19.98$\pm$0.15 & 19.90$\pm$0.15 & KNC  & x \\
2025-01-30T22:16:50 & 60705.928 & 1.730 & 2491.683 & $R$ & 20.00$\pm$0.10 & 19.93$\pm$0.10 & AbAO-T70  & x \\
2025-01-31T05:19:45 & 60706.222 & 2.024 & 2914.602 & $R$ & 20.32$\pm$0.06 & 20.25$\pm$0.06 & Skynet  & x \\
2025-01-31T06:20:07 & 60706.264 & 2.066 & 2974.967 & $R$ & 20.44$\pm$0.07 & 20.37$\pm$0.07 & Skynet  & x \\
2025-01-31T07:20:28 & 60706.306 & 2.108 & 3035.332 & $R$ & 20.37$\pm$0.07 & 20.30$\pm$0.07 & Skynet  & x \\
2025-01-31T08:25:52 & 60706.351 & 2.153 & 3100.722 & $R$ & 20.32$\pm$0.06 & 20.25$\pm$0.06 & Skynet  & x \\
2025-01-31T23:10:05 & 60706.965 & 2.767 & 3984.933 & $R$ & 20.44$\pm$0.11 & 20.37$\pm$0.11 & AbAO-T70  & x \\
2025-02-01T05:57:27 & 60707.248 & 3.050 & 4392.301 & $R$ & 20.25$\pm$0.04 & 20.18$\pm$0.04 & Skynet  & x \\
2025-02-01T07:30:44 & 60707.313 & 3.115 & 4485.587 & $R$ & 20.55$\pm$0.05 & 20.48$\pm$0.05 & Euler  & x \\
2025-02-01T07:48:04 & 60707.325 & 3.127 & 4502.922 & $R$ & 20.49$\pm$0.04 & 20.42$\pm$0.04 & Skynet  & x \\
2025-02-02T07:20:42 & 60708.306 & 4.108 & 5915.562 & $R$ & 21.06$\pm$0.10 & 20.99$\pm$0.10 & Euler  & x \\
2025-02-03T08:26:27 & 60709.352 & 5.154 & 7421.317 & $R$ & 21.72$\pm$0.10 & 21.65$\pm$0.10 & Euler  & x \\
2025-02-05T08:58:33 & 60711.374 & 7.176 & 10333.416 & $R$ & 22.79$\pm$0.11 & 22.72$\pm$0.11 & Euler  & x \\
2025-02-08T08:09:32 & 60714.340 & 10.142 & 14604.396 & $R$ & 22.81$\pm$0.28 & 22.74$\pm$0.28 & Euler  & x \\
\hline
\multicolumn{9}{|c|}{$i'$ band} \\
\hline
2025-01-29T05:40:14 & 60704.236 & 0.038 & 55.083 & $i'$ & 16.55$\pm$0.03 & 16.49$\pm$0.03 & OHP/MISTRAL  & x \\
2025-01-30T01:30:37 & 60705.063 & 0.865 & 1245.483 & $i'$ & 18.92$\pm$0.03 & 18.86$\pm$0.03 & KAO  & x \\
2025-01-30T06:34:13 & 60705.274 & 1.076 & 1549.074 & $i'$ & 18.53$\pm$0.11 & 18.48$\pm$0.11 & KNC  & x \\
2025-01-30T08:13:54 & 60705.343 & 1.145 & 1648.755 & $i'$ & 18.75$\pm$0.12 & 18.69$\pm$0.12 & KNC  & x \\
2025-02-01T07:44:08 & 60707.322 & 3.124 & 4498.999 & $i'$ & 20.17$\pm$0.05 & 20.11$\pm$0.05 & Euler  & x \\
2025-02-02T07:34:06 & 60708.315 & 4.117 & 5928.956 & $i'$ & 21.00$\pm$0.06 & 20.94$\pm$0.06 & Euler  & x \\
2025-02-03T02:04:30 & 60709.086 & 4.888 & 7039.353 & $i'$ & 21.30$\pm$0.10 & 21.24$\pm$0.10 & NAO-2m  & x \\
2025-02-03T08:39:54 & 60709.361 & 5.163 & 7434.761 & $i'$ & 21.75$\pm$0.10 & 21.69$\pm$0.10 & Euler  & x \\
2025-02-04T03:18:10 & 60710.138 & 5.940 & 8553.025 & $i'$ & 21.80$\pm$0.11 & 21.74$\pm$0.11 & C2PU-O  & x \\
2025-02-05T09:10:24 & 60711.382 & 7.184 & 10345.261 & $i'$ & 22.45$\pm$0.20 & 22.39$\pm$0.20 & Euler  & x \\
2025-02-08T08:33:00 & 60714.356 & 10.158 & 14627.866 & $i'$ & >22.25$\pm$ & >22.19$\pm$ & Euler  & x \\
\hline
\multicolumn{9}{|c|}{$I$ band} \\
\hline
2025-01-29T05:15:46 & 60704.219 & 0.021 & 30.628 & $I$ & 16.69$\pm$0.07 & 16.64$\pm$0.07 & Skynet  & x \\
2025-01-29T05:22:34 & 60704.224 & 0.026 & 37.424 & $I$ & 16.56$\pm$0.05 & 16.51$\pm$0.05 & Skynet  & x \\
2025-01-29T05:30:34 & 60704.230 & 0.032 & 45.431 & $I$ & 16.63$\pm$0.05 & 16.58$\pm$0.05 & Skynet  & x \\
2025-01-29T05:39:17 & 60704.236 & 0.038 & 54.143 & $I$ & 16.74$\pm$0.05 & 16.69$\pm$0.05 & Skynet  & x \\
2025-01-29T05:49:29 & 60704.243 & 0.045 & 64.338 & $I$ & 16.70$\pm$0.04 & 16.65$\pm$0.04 & Skynet  & x \\
2025-01-29T06:01:24 & 60704.251 & 0.053 & 76.261 & $I$ & 16.85$\pm$0.04 & 16.80$\pm$0.04 & Skynet  & x \\
2025-01-29T06:15:21 & 60704.261 & 0.063 & 90.215 & $I$ & 16.98$\pm$0.04 & 16.93$\pm$0.04 & Skynet  & x \\
2025-01-29T06:36:20 & 60704.275 & 0.077 & 111.196 & $I$ & 17.19$\pm$0.03 & 17.14$\pm$0.03 & Skynet  & x \\
2025-01-29T21:35:47 & 60704.900 & 0.702 & 1010.650 & $I$ & 19.14$\pm$0.10 & 19.08$\pm$0.10 & KNC  & x \\
2025-01-30T01:19:13 & 60705.055 & 0.857 & 1234.067 & $I$ & 18.97$\pm$0.11 & 18.91$\pm$0.11 & KNC  & x \\
2025-02-02T19:02:22 & 60708.793 & 4.595 & 6617.230 & $I$ & >20.69$\pm$ & >20.64$\pm$ & Generic  & x \\
\hline
\multicolumn{9}{|c|}{$z$ band} \\
\hline
2025-02-04T02:59:50 & 60710.125 & 5.927 & 8534.685 & $z'$ & 20.97$\pm$0.20 & 20.93$\pm$0.20 & C2PU-O  & x \\
2025-02-04T05:42:14 & 60710.238 & 6.040 & 8697.085 & $z'$ & 21.81$\pm$0.16 & 21.77$\pm$0.16 & Artemis  & x \\
2025-02-05T05:41:13 & 60711.237 & 7.039 & 10136.076 & $z'$ & 22.20$\pm$0.20 & 22.16$\pm$0.20 & Artemis  & x \\
\hline
\end{longtable}

\clearpage
\twocolumn

\label{lastpage}
\end{document}